\documentclass[reprint,superscriptaddress,aps,onecolumn,nofootinbib,notitlepage,showpacs]{revtex4-1}

\usepackage[intlimits]{amsmath}
\usepackage{scalerel}
\usepackage{stackengine,wasysym}
\usepackage{amssymb}
\usepackage[scr=boondox,scrscaled=1.03]{mathalfa}
\usepackage{placeins}
\usepackage{amsbsy}
\usepackage{bbold}
\usepackage[T1]{fontenc}
\usepackage[utf8]{inputenc}
\usepackage[portuguese,english]{babel}
\usepackage{multirow}
\usepackage{microtype}
\usepackage{amsfonts}
\usepackage{amssymb}
\usepackage{setspace}
\usepackage{graphicx}
\usepackage{wrapfig}
\usepackage{textcomp}         
\usepackage{array}
\usepackage{slashed}
\usepackage{url}
\usepackage[usenames,table]{xcolor}
\usepackage[colorlinks=true,linkcolor=blue,citecolor=blue,urlcolor=blue]{hyperref}
\usepackage{natbib}
\usepackage{mathtools}

\DeclareMathSizes{10.95}{10.95}{7}{7}
\setcitestyle{numbers,sort&compress}

\def\dj{d\kern-0.4em\char"16\kern-0.1em}

\def\be{\begin{equation}}
\def\ee{\end{equation}}
\def\bea{\begin{eqnarray}}
\def\eea{\end{eqnarray}}
\def\bfl{\begin{flushleft}}
\def\efl{\end{flushleft}}
\def\bfr{\begin{flushright}}
\def\efr{\end{flushright}}
\def\bc{\begin{center}}
\def\ec{\end{center}}
\def\ben{\begin{enumerate}}
\def\een{\end{enumerate}}
\def\bit{\begin{itemize}}
\def\eit{\end{itemize}}

\def\dzn{,\kern-0.1em,}

\def\graph{\includegraphics}
\def\tew{\textwidth}
\def\er{\eqref}

\def\d#1{{#1\kern-0.4em\char"16\kern-0.1em}}
\def\D#1{{\raise0.2ex\hbox{-}\kern-0.4em 31}}

\def\e{\mbox{e}}
\def\d{\mbox{d}}

\def\D{{\cal{D}}}

\newcommand {\apgt} {\ {\raise-.5ex\hbox{$\buildrel>\over\sim$}}\ }
\newcommand {\aplt} {\ {\raise-.5ex\hbox{$\buildrel<\over\sim$}}\ }

\begin{document}

\title{Tensor representations of lattice vertices \\ from hypercubic symmetry}

\author{Milan Vujinovi\'c}
\affiliation{Institute of Physics, NAWI Graz, University of Graz, \\ Universit\"atsplatz 5, 8010 Graz, Austria}
\email{milan.vujinovic@uni-graz.at}

\begin{abstract}

We present a symmetry-based method for obtaining suitable tensor descriptions of lattice vertex functions without spinor components.~The approach is based on
finding the polynomial functions of vertex momenta, which satisfy the appropriate tensor transformation laws under hypercubic symmetry transformations.~We use 
the method to find the most general possible (up to finite volume effects) basis decompositions for lattice vectors and second-rank tensors.~The leading-order
non-continuum versions of these representations are then applied to the Landau gauge gluon propagator and ghost-gluon vertex of Monte Carlo simulations, to
reveal two interesting insights.~First, it is demonstrated numerically and analytically that there exist special kinematic configurations where the basis 
descriptions of both functions reduce to their continuum analogues.~Second, for the gluon two-point correlator it is shown that the rate at which the function
approaches its continuum form in the infrared is independent of the lattice gauge coupling $\beta$ (when working in lattice units)\,:~the said rate depends on 
kinematics alone and is ultimately dictated by the numerical gauge-fixing procedure.~We also comment on how this reflects on the lattice investigations of the
anomalous magnetic moment of the muon.~Finally, we argue how our findings can be used to directly test some of the continuum extrapolation methods.

\end{abstract}          

\maketitle

\section{Introduction}\label{sec: intro}

Many investigations of quantum chromodynamics (QCD) and similar strongly-interacting models concentrate on a direct extraction of physical observables present 
in these theories, see e.\,g.~\cite{Aoki:2016frl, Eichmann:2016yit} and references therein.~However, during the last few decades a fair amount of attention was
also dedicated to the analysis of the elementary degrees of freedom of these mathematical frameworks, which can generically be labelled as the quark and gluon 
fields.~While not being directly detectable in experiments, the quark, gluon and ghost propagators (wherein ghosts arise from gauge-fixing), as well as their 
corresponding interaction vertices still attract considerable interest among researchers.~This is because of the supposed connection of these objects to the 
phenomena of confinement \cite{Gribov:1977wm, Zwanziger:1993dh, Zwanziger:2001kw, Zwanziger:2003cf, Kugo:1979gm} and dynamical chiral symmetry breaking 
\cite{Alkofer:2008tt}, as well as the pivotal role they play in functional studies of bound states \cite{Eichmann:2016yit, Alkofer:2008tt, Vujinovic:2014ioa, 
Sanchis-Alepuz:2015qra, Williams:2015cvx, Binosi:2016rxz, Eichmann:2016hgl, Sanchis-Alepuz:2017jjd, Rodriguez-Quintero:2018wma, Vujinovic:2018nko, Eichmann:2019tjk}.

The non-perturbative calculations of the elementary correlators of strongly-interacting theories roughly consist of two complimentary methods.~These two groups are the 
various continuum approaches \cite{Alkofer:2008tt, Vujinovic:2014ioa, Sanchis-Alepuz:2015qra, Williams:2015cvx, Binosi:2016rxz, Eichmann:2016hgl, Rodriguez-Quintero:2018wma,
Vujinovic:2018nko, Schleifenbaum:2004id, Pawlowski:2005xe, Kellermann:2008iw, Alkofer:2008dt,  Huber:2012zj, Huber:2012kd, Aguilar:2013xqa, Aguilar:2013vaa, Pelaez:2013cpa,
Blum:2014gna, Eichmann:2014xya, Cyrol:2014kca, Binosi:2014kka, Mitter:2014wpa, Aguilar:2014lha, Pelaez:2015tba, Binosi:2016wcx, Cyrol:2016tym, Aguilar:2016lbe, Cyrol:2017ewj, 
Huber:2017txg, Oliveira:2018fkj, Corell:2018yil, Aguilar:2018csq}, as well as those based on lattice Monte Carlo simulations \cite{Parrinello:1994wd, Alles:1996ka, Boucaud:1998bq,
Skullerud:2002ge, Skullerud:2003qu, Cucchieri:2004sq, Cucchieri:2006tf, Ilgenfritz:2006he, Maas:2007uv, Cucchieri:2008qm, Maas:2011se, Sternbeck:2012mf,Boucaud:2013jwa, 
Maas:2013aia, Duarte:2016jhj, Athenodorou:2016oyh, Sternbeck:2016ltn, Boucaud:2017obn, Sternbeck:2017ntv, Vujinovic:2018nqc, Maas:2018ska, Maas:2019tnm}.~While both frameworks 
have their specific advantages and disadvantages, the lattice method features a particular issue which we think has not been adequately addressed so far.~Namely, for continuum
theories there exist unique and well-known ``recipes'' for obtaining complete tensor descriptions for virtually any correlator of interest, see e.\,g.~\cite{Hassani:1999sny}.~On 
the lattice, such recipes are sorely lacking, and in some situations this leads to systematic errors which are not easy to quantify.~For instance, many lattice investigations of 
the quark-gluon and three-gluon vertex employ the corresponding tensor bases from the continuum theory \cite{Parrinello:1994wd, Alles:1996ka, Boucaud:1998bq, Skullerud:2002ge, 
Skullerud:2003qu, Boucaud:2013jwa, Duarte:2016jhj, Athenodorou:2016oyh, Boucaud:2017obn,Sternbeck:2017ntv}, and it is hard to say what is the systematic uncertainty associated
with such an approximation. 

Here we attempt to resolve some of these matters by invoking symmetry-based arguments.~On symmetric (hyper)cubic lattices in $d$ dimensions, the rotational symmetry of an Euclidean
continuum theory reduces to the hypercubic group $H(d\,)$, whose elements can all be generated from parity transformations and $\pi/2$ rotations around the coordinate axes \cite{
Morty:1962prc}.~This affects the basis decompositions of lattice correlators, since the hypercubic operators induce far weaker constraints on the allowed tensor structures, compared 
to the full continuum symmetry group.~A detailed account on how exactly this modifies the tensor representations of certain lattice vertices will be provided later.~To the best of
our knowledge, no systematic investigation of this kind has been performed before, although the issue was approached from different sides in the past, in the context of lattice 
perturbation theory \cite{Kawai:1980ja}, improved gauge actions \cite{Weisz:1983bn} and lattice investigations of the anomalous magnetic moment of the muon \cite{Aubin:2015rzx}. 

The biggest asset of our approach to vertex tensor bases is its generality.~Since we only employ the constraints coming from hypercubic symmetry, the applicability 
of our method does not depend on a particular choice of lattice action or gauge-fixing method, as long as an equal treatment of all the coordinates is maintained in 
all aspects of the calculation.~This means that any computations done with our bases can be taken from e.\,g.~the case of the Wilson gauge action \cite{Wilson:1974sk} 
to the $\mathcal{O}(a^2)$ tree-level improved one \cite{Weisz:1983bn, Weisz:1982zw, Symanzik:1983gh}, without any alterations in parts of the code which deal only with
correlator form factors.~Apart from this, we will argue that our framework also allows one to (more-less) directly quantify the rotational symmetry breaking effects in 
vertex dressing functions, perform tests of continuum extrapolation procedures, and identify special kinematic configurations where the lattice-modified bases get reduced 
to their continuum form.~However, in the course of this work it will become clear that all of this comes at a cost:~when deriving the tensor structures of the lattice theory
using symmetry arguments alone, one may easily end up with so many elements that actual calculations with the full basis become very challenging, if not downright impossible.
\!Thus, for any particular problem at hand one has to judge if the potential gains provided by our framework outweight the considerable rise in algebraic difficulty. 

Our paper is organised as follows. In section \ref{sec: basics} we discuss the basic ideas behind our method, and show how scalar and vector quantities get modified on the
lattice, as compared to their continuum counterparts.~In section \ref{sec: vertex_n_propag} we use the same principles to derive the most general tensor decompositions (up to
finite volume effects) for the lattice ghost-gluon vertex and gluon propagator.~In section \ref{sec: numerics} we apply the obtained tensor bases, in their lowest-order 
(non-continuum) versions to the gluon and ghost-gluon correlators, as evaluated in numerical Monte Carlo simulations in Landau gauge.~We point to some interesting insights
which come out of these applications, including the fact that the gluon propagator approaches its continuum form at low energies, at a rate which is independent of the 
parameters of the numerical lattice implementation.~Based on this observation we also comment on how the lattice studies of the anomalous magnetic moment of the muon may 
(not) get affected by discretisation artifacts.~We conclude in section \ref{sec: conclude}.~Most of the purely technical discussions have been relegated to the four 
detailed appendices.     

\section{Continuum and hypercubic tensors:~basic ideas}\label{sec: basics}

 \subsection{Tensor bases in the continuum theory}\label{sec: tens_cont}
 
We begin the discussion of our method by briefly reviewing some basic facts about tensor descriptions of certain vertex functions in the 
continuum.~Most of the points we will cover here are well-known from elementary textbooks, and some of them could even be considered rather 
trivial.~Nonetheless, we think it is important to go through these ``trivialities'' , in order to fully understand how the arguments change when 
going from continuum to discretised spacetimes.~We emphasise that throughout this paper, we shall \textit{not} be using the Einstein summation 
convention, since we will frequently encounter non-covariant objects and expressions.~Thus, in relations which feature summations over indices, 
the sum symbol will always be explicitly written out.

A continuous, $d$-dimensional Euclidean space is often said to possess an $O(d\,)$ symmetry, meaning that the distances, or scalar products of
vectors in the space, are all preserved under an action of arbitrary orthogonal $d \times d$ matrices.~For orthogonal operators, it holds that the
operation of matrix transposition is equivalent to inversion, or explicitly
\begin{align}\label{eqn: ortho_define}
O_{\mu\nu} = O^{-1}_{\nu\mu} \, , \qquad \quad \mu,\, \nu = 1 \ldots d \, , 
\end{align}  

\noindent
where indices $\mu$ and $\nu$ stand for operator components.~The fact that the matrices $O$ represent symmetry transformations of a continuous space means 
that all of the quantities in the space (i.\,e.~scalars, vectors, second-rank tensors, etc.) have to be defined with respect to the orthogonal group.~As
an example, take a set of numbers which constitute the components of a vector $v$.~This means that, under arbitrary orthogonal transformations, these 
numbers/components satisfy a particular transformation law, namely ($v_\mu$ denotes the $\mu$-th component of $v$):  
\begin{align}\label{eqn: cont_trans_law}
v'_\mu = \sum_{\nu = 1}^d O_{\mu\nu} \, v_\nu \, , \qquad \quad \mu = 1 \ldots d . 
\end{align}

In the above relation, prime ($'$) signifies the vector components in the transformed system.~Transformation laws like \er{eqn: cont_trans_law} put stringent 
constraints on the possible momentum dependencies of tensor quantities of various rank.~Take as an example a tensor of rank zero i.\,e.~a scalar function
$S$ which depends on a single momentum variable $p$.~Being a scalar, or an invariant quantity, means that $S(p)$ does not change under arbitrary orthogonal 
transformations of $p$.~In other words, for a general orthogonal matrix $O$, it holds that 
\begin{align}\label{eqn: scalar_p_trans}
p'_\mu = \sum_{\nu = 1}^d O_{\mu\nu} \, p_\nu \, , \qquad \text{and} \qquad S'(p') = S(p) \, ,
\end{align}

\noindent
where index $\mu$ runs from 1 to $d$, the number of dimensions.~It is well known (see e.\,g.~\cite{Hassani:1999sny}) that the invariance of $S$ implies that it
can only depend on $p$ through the scalar product $p^2$, which is defined in $d$ dimensions as 
\begin{align}\label{eqn: p2_cont}
p^2 = \sum_{\mu = 1}^d p^2_\mu = p_1^2 + p_2^2 + \ldots + p_d^2 \, .
\end{align}

Invariance of $p^2$ under arbitrary $d$-dimensional orthogonal transformations follows directly from the property \er{eqn: ortho_define}.~Going back to the
function $S$, one sees that the ``demand'' that it remain unchanged under general $O$ matrices leads to the conclusion that it can depend solely on the product
$p^2$.~Similar restrictions follow for tensors of arbitrary rank.~As an example, instead of a scalar quantity, one might be working with some vector $\Gamma$,
which is a function of momentum $p$.~Being a vector, $\Gamma$ has to obey a transformation law akin to \er{eqn: cont_trans_law}, meaning that 
\begin{align}\label{eqn: gamma_p_trans}
p'_\mu = \sum_{\nu = 1}^d O_{\mu\nu} \, p_\nu \, , \qquad \text{and} \qquad \Gamma'_\mu(p') = \sum_{\nu = 1}^d O_{\mu\nu} \, \Gamma_\nu(p) \, .
\end{align}

Now, even if one had no prior knowledge on the way that $\Gamma$ depends on $p$, a careful consideration of \er{eqn: gamma_p_trans} would quickly lead one to
the deduction that $\Gamma$ has to have the form 
\begin{align}
\Gamma_\mu(p) =  A(p) \, p_\mu \, ,
\end{align}

\noindent
with dressing function (or form factor) $A(p)$ being an orthogonal invariant, i.\,e.~depending on $p^2$ alone.~In words, the vector $\Gamma$ has to be strictly
linear in \textit{components} of $p$, since any non-linear terms with an open vector index (e.\,g.~$p^2_\mu$) would not obey \er{eqn: gamma_p_trans}.~For
instance, a structure quadratic in $p$ components, the aforementioned object $p^2_\mu$, would transform under general $O$ matrices as 
\begin{align}
p_\mu^2 \rightarrow p'^{\, 2}_\mu = \sum_{\nu = 1}^d \sum_{\rho = 1}^d O_{\mu\nu} O_{\mu\rho} \, p_\nu \, p_\rho \, , 
\end{align}   

\noindent
which is clearly incompatible with the vector-like transformation law for $\Gamma$ itself, see \er{eqn: gamma_p_trans}.

We conclude this section with comments on how some of the above observations change when going from continuum spaces to discretised ones.~As stated in the 
Introduction, on standard cubic lattices the orthogonal group $O(d\,)$ gets broken down to its hypercubic\footnote{In the context of our work, the term 
``hypercubic'' is not really correct since it implies a four-dimensional setting, whereas most of our arguments will not depend on the number of dimensions.
\!Nonetheless, as we do not wish to keep switching between different group names for different dimensions, we will continue this mild abuse of terminology 
throughout the rest of this paper.} subgroup $H(d\,)$, which is comprised of $d$-dimensional $\pi/2$ rotations and parity transformations.~We shall see soon 
that, when represented as matrices, the hypercubic symmetry operations have a somewhat special structure, which makes the equations like \er{eqn: scalar_p_trans}
and \er{eqn: gamma_p_trans} far less restrictive than in the case of general orthogonal operators.~For scalar functions depending on momentum $p$, it is by now 
well known that the hypercubic group has more invariants than just $p^2$ \cite{Weyl:1939prc}, a fact to which we shall return later.~We will show in this paper 
that similar considerations apply for tensors of higher rank as well:~taking again vectors as an example, it will turn out that there are open-index objects 
which are non-linear in momentum components (i.\,e.~$p^{\,n}_\mu$, with integer $n > 1$), which despite their non-linearity, still satisfy the adequate vector 
transformation law \er{eqn: cont_trans_law} under hypercubic symmetry transformations.  
 
\subsection{Hypercubic group as permutations and inversions of coordinates}\label{sec: hyper_perms}

As already mentioned, the group $H(d\,)$ consists of (powers of) $\pi/2$ rotations around the coordinate axes, and parity transformations (sometimes also called 
inversions).~Here, we want to show that the hypercubic group can equally well be represented with permutations and inversions of coordinates, since a $\pi/2$ rotation
in an arbitrary plane can always be written as a composition of permutation and inversion transformations.~Demonstrating the aforementioned equivalence is important 
since we wish to adopt the ``permutations + inversions'' viewpoint in this paper, because it makes much of the forthcoming analysis easier.

We start with the simplest possible example, that of ``hypercubic'' symmetry transformations in two dimensions.~First, we will need a matrix representation of 
a clockwise $\pi/2$ rotation for $d=2$.~If we take some vector $p = (p_1, p_2)$, and denote its clockwise $\pi/2$-rotated version with $p' = (p'_1, p'_2)$, the 
operation of rotation can be written down as  
\begin{align}
p'_\mu = \sum_{\nu = 1}^2 L^{\pi/2}_{\mu\nu} \, p_\nu \, . 
\end{align}  

\begin{figure}[!t]
\begin{center}
\graph[width = 0.64\tew]{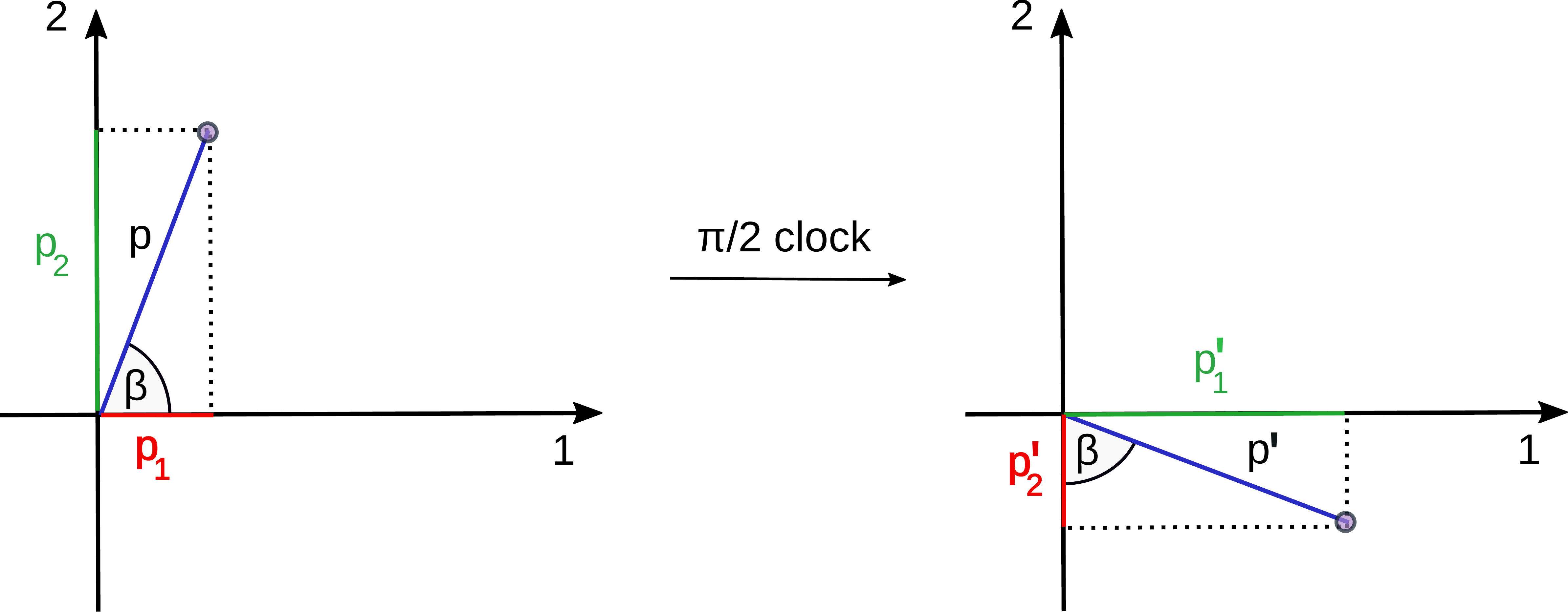}\\
\caption{\textit{Left:}~A graphical representation of a clockwise $\pi/2$ rotation of a two-dimensional vector $p$.~$(p_1, p_2)$ and $(p'_1, p'_2)$ denote, respectively, 
components of the vector before and after the rotation.} 
\label{fig: rotation}
\end{center}
\end{figure}

The explicit form of the matrix $L^{\pi/2}$ can be easily deduced with a bit of visual help, shown in Figure \ref{fig: rotation}.~From the Figure it should be 
relatively clear that the primed components $p'_\mu$ are related to the un-primed ones $p_\mu$ via 
\begin{align}
p'_1 = p_2 \, , \quad \text{and} \quad p'_2 = - p_1 \, ,
\end{align}

\noindent
from which it immediately follows that $L^{\pi/2}$ has a matrix representation
\begin{align}\label{eqn: pi_half_matrix}
L^{\pi/2} = \left[ \begin{array}{cc} 0 & \;\; 1 \\ -1 & \;\; 0  \end{array} \right] \, .
\end{align}

Note that it makes no essential difference if one were to look at counter-clockwise rotations, instead of clockwise ones:~the only difference between the two kinds of
matrices is an overall minus sign, which is unimportant for the upcoming arguments.~Also note that the fourth power of the matrix $L^{\pi/2}$ is an indentity element:
\begin{align}\label{eqn: roto_2pi}
\left(L^{\pi/2}\right)^4 \, = \, L^{2\pi} \, = \, + \left[ \begin{array}{cc} 1 & \;\; 0 \\ 0 & \;\; 1  \end{array} \right] \, .
\end{align}

The above is to be expected, since a rotation by $2\pi$ is equivalent to making no change to the system at all\,\footnote{A $\pi/2$ rotation in a plane generates a 
group isomorphic to the cyclic group $Z_4$, i.\,e.~an Abelian group with four elements (say, $a, b, c$ and $e$, with $e$ the identity element), and the multiplication
law(s) $a^2 = b , \; ab = a^3 = c$, and $a^4 = b^2 = e$.~In general, planar rotations by $2\pi/n$ degrees, with integer $n$, generate groups isomorphic to cyclic groups
$Z_n$ \cite{Morty:1962prc}.}.~Besides the operator $L^{\pi/2}$ and its powers (modulo 4), there are two remaining elementary operations which leave a two-dimensional
hypercube (i.\,e.~a square) intact, and these are the parity transformations.~Their matrix representations are \cite{Morty:1962prc}

\begin{align}\label{eqn: 2d_inver_def}
\rho_1 =  \left[ \begin{array}{cc} -1 & \;\; 0 \\ 0 & \;\; 1  \end{array} \right] \, , \qquad 
\rho_2 =  \left[ \begin{array}{cc} 1 & \; 0 \\ 0 & \; -1  \end{array} \right] \, . 
\end{align}

Now, a key observation to be made here is that the matrix $L^{\pi/2}$ of \er{eqn: pi_half_matrix} can itself be written as a composition of a parity transformation
and a permutation operator $\Pi^{12}$, or explicitly 
\begin{align}\label{eqn: 2d_rot_example}
L^{\pi/2} = \Pi^{12} \cdot \rho_1 = \left[ \begin{array}{cc} 0 & \;\; 1 \\ 1 & \;\; 0  \end{array} \right] \cdot \left[ \begin{array}{cc} -1 & \;\; 0 \\
0 & \;\; 1  \end{array} \right] \, .
\end{align}

$L^{\pi/2}$ can also be obtained as $L^{\pi/2} = \rho_2 \cdot \Pi^{12}$.~In these expressions, $\Pi^{12}$ stands for a permutation which exchanges the first and
second momentum components, i.\,e.~
\begin{align}\label{eqn: 2d_perm_def}
\Pi^{12} \cdot p =  \left[ \begin{array}{cc} 0 & \;\; 1 \\ 1 & \;\; 0  \end{array} \right] \cdot  \left( \begin{array}{c} p_1 \\ p_2 \end{array} \! \right) = 
\left( \begin{array}{c} p_2 \\ p_1  \end{array} \! \right) \, .
\end{align}

Since the matrix $L^{\pi/2}$ and its powers (modulo 4) can all be written as compositions of elementary transformations $\rho_1$, $\rho_2$ and $\Pi^{12}$, one concludes
that these last three operators, and their combinations, exhaust all of the possible symmetry operations of a two-dimensional hypercube.~The argument can be straightforwardly
extended to an arbitrary number of dimensions, by analysing each of the available rotation planes separately.~In three dimensions, for instance, there are three spatial 
planes, which one might denote as (12), (13) and (23).~Taking the clockwise $\pi/2$ rotation in the plane (23) as an example, one has (we will leave out the `$\pi/2$' 
designation in the following, since it should be clear that we are not working with any other angles of rotation):
\begin{align}\label{eqn: 3d_rot_example} 
L^{23} = \left[ \begin{array}{ccc} 1 & \; 0 & \; 0 \\ 0 & \; 0 & \; 1 \\ 0 & -1 & \; 0 \end{array} \right] = \, \Pi^{\, 23} \cdot \rho_2 = \left[ \begin{array}{ccc}
1 & \; 0 & \; 0 \\ 0 & \; 0 & \; 1 \\ 0 & 1 & \; 0 \end{array} \right] \cdot \left[ \begin{array}{ccc} 1 & \; 0 & \; 0 \\ 0 & -1 & \; 0 \\ 0 & \; 0 & \; 1 \end{array}
\right] \, .
\end{align}

Thus, the rotation $L^{23}$ can be represented as a product of two elementary operations, where $\Pi^{\,23}$ permutes the second and third momentum components,
and $\rho_2$ turns momentum $p = (p_1, p_2, p_3)$ into $p' = (p_1, - p_2, p_3)$.~In the same vein, $\pi/2$ rotations in the other two planes follow as adequate 
compositions of exchange operators $\Pi^{12}$ and $\Pi^{13}$, with one of the elementary inversions $\rho_k \, (k = 1\ldots 3)$.~It should be relatively clear
that by combining various powers (modulo 4) of rotation matrices $L^{12}$, $L^{13}$ and $L^{23}$, one can generate any symmetry rotation of a three-dimensional 
hypercube\,\footnote{In all these proceedings, we have ignored the symmetry rotations around the (hyper)cube diagonals.~The fact that these too can be decomposed
into permutations and parity transformations (or equivalently, into inversions and $\pi/2$ rotations about the coordinate axes) is demonstrated in some detail in
Appendix \ref{sec: diagonals}.}.~And since these three rotation operators can themselves be decomposed into simpler permutation and inversion transformations, it 
follows that combinations of coordinate permutations and inversions exhaust all possible operations which leave a hypercube unchanged, in $d = 3$.~A generalisation
of these arguments to higher dimensions is straightforward.       
 
We wish to conclude this section by briefly going back to the ``point of the whole exercise'', i.\,e.~to why it was important to show that combinations of parity 
transformations and permutations cover all of the elements of the symmetry group $H(d\,)$.~Suppose that one wished to find general expressions for hypercubic vectors,
meaning the most general possible functions of given momenta, which transform as vectors under the hypercubic group.~From the above analysis, it should be clear that
it is enough to identify those functions, which transform as vectors under $both$ permutations and inversions:~the said quantity will surely constitute a vector under 
an arbitrary hypercubic symmetry transformation.~In these matters, it will turn out to be very useful to analyse the two kinds of alterations (parity and permutation) 
independently from each other, as they obviously have different effects on momentum components.

\subsection{Hypercubic scalars}  

Assume that one is working in a theory with a single momentum $p$, in an arbitrary number of dimensions $d$.~We already mentioned that in the continuum, the only available
scalar quantity in this scenario would be the product $p^2$ defined in \er{eqn: p2_cont}.~The said product is invariant under general orthogonal operators.~Now, on a 
$d$-dimensional hypercube, the very definition of a scalar gets generalised:~instead of being a function which is left unchanged by arbitrary orthogonal matrices, it ``only''
needs to be unvarying under the effects of parity transformations and permutations, in accordance with the analysis of the previous section.~For a $d$-dimensional vector $p$
with components $p = (p_1, p_2, \ldots p_d)$, all of the following functions would constitute hypercubic invariants:
\begin{align}\label{eqn: hyper_scalars}
p^{\,[2n]} = \sum_{\mu = 1}^d p^{\,2n}_\mu = p_1^{\,2n} + p_2^{\,2n} + \ldots + p_d^{\,2n} \, , \qquad n \in N , 
\end{align}

\noindent
with $N$ the set of positive integers.~The bracketed superscript notation (i.\,e.~$[2n]$) was taken from \cite{Becirevic:1999uc, deSoto:2007ht}.~Note that $p^{
[2]} = p^2$ is an invariant of the continuum theory.~Throughout the rest of this paper, we will use the notation $p^2$ instead of $p^{[2]}$, but only for this 
particular scalar product:~all the other hypercubic invariants of a single momentum $p$ will follow the convention \er{eqn: hyper_scalars}.~The fact that the
quantities \er{eqn: hyper_scalars} do not change under arbitrary permutations and inversions of momentum components should be relatively self-evident.~What is
perhaps not so obvious, is that not all functions of the form \er{eqn: hyper_scalars} can be algebraically independent from each other.~In \cite{Weyl:1939prc}
it is shown that the number of independent hypercubic scalars, with a single momentum $p$ at ones disposal, does not exceed the number of dimensions of the
theory under consideration.~This is best illustrated with an example, for which we (again) turn to the simplest case of two dimensions.~For $d = 2$, and momentum
$p = (p_1,p_2)$, there are only two independent invariants of the form \er{eqn: hyper_scalars}, which one might choose to be (say) $p^2$ and $p^{\,[4]}$.~All the
other hypercubic scalars follow as polynomial functions of these two, for an example
\begin{align}
&p^{\,[6]} = p_1^6 + p_2^6 = \frac{1}{2} \left( 3 \, p^2\cdot p^{\,[4]} - \big(p^2\big)^3\right) \, , \nonumber \\ 
&p^{\,[8]} = p_1^8 + p_2^8 = \frac{1}{6} \left( 3 \, \big(p^{\,[4]}\big)^2  - 3 \, \big(p^2\big)^4 + 6 \, \big(p^2\big)^2 \cdot p^{\,[4]} \right) \, , 
\end{align}

\noindent
and similarly for invariants with higher mass dimensions.~In the same manner, a three-dimensional theory would contain three independent hypercubic scalars (say, 
$p^2, p^{\,[4]}$ and $p^{\,[6]}$) and so on.~An elegant proof, for an arbitrary dimension number, can be found in \cite{Weyl:1939prc}, while \cite{deSoto:2007ht} 
treats a more specific case of four dimensions.~Note that, for a $d$-dimensional theory, one can choose any $d$ functions of the form \er{eqn: hyper_scalars}, and
use them as a ``basis'' for calculations:~the symmetry itself does not dictate which invariants should be chosen.~We shall see soon that similar ambiguities arise 
for tensor bases of lattice vertex functions.~To at least partially address the ambiguity, we will always choose the bases according to the ascending order of mass 
dimension, meaning that the preference will be given to elements which feature the smallest powers of momentum components. 

To conclude, we want to mention some practical implications of the observations made in this section.~Suppose that one is studying some lattice vertex function,
which depends on momentum $p$, and that one has obtained the corresponding data for vertex form factors.~For reasons discussed above, the said form factors will not
be functions of the product $p^2$ alone, but will also depend on other hypercubic scalars like $p^{\,[4]},p^{\,[6]}$ etc.~In this context, the presence of additional
invariants is an unwanted lattice artifact, which one would generally like to mitigate as much as possible.~To this end, a powerful computational tool has been
developed, the so-called hypercubic extrapolation, where one attemps to extrapolate the available lattice data towards the limit where some of the ``extra'' lattice
invariants $(\text{e.\,g.~}p^{\,[4]})$ vanish.~For some examples on the use of this method, see e.\,g.~\cite{Becirevic:1999uc, deSoto:2007ht, Becirevic:1999hj, 
Blossier:2014kta, Boucaud:2018xup} and references therein.         
   
\subsection{Hypercubic vectors}\label{sec: hyper_vector}

As was argued at the end of section \ref{sec: hyper_perms}, the task of finding general expressions for hypercubic vectors amounts to finding the functions which 
transform as vectors under both permutations and inversions of momentum components.~We shall split this task into two parts, wherein we analyse the two kinds of
tranformations separately, since they have different effects on vector components. 

We shall start with permutations.~We consider a situation with a single momentum variable $p$, in an arbitrary number of dimensions $d$.~Let $\Pi^{\sigma\tau}$ 
denote a permutation which exchanges the $\sigma$-th and $\tau$-th momentum components, where each index can run from 1 to $d$,  and $\sigma = \tau$ corresponds 
to an indentity matrix.~The operators which swap only two elements at a time are sometimes called transpositions, and the fact that we consider only such matrices
does not diminish the generality of our upcoming results.~This is because an arbitrary permutation can always be broken down into a product of transpositions, in 
infinitely many different ways \cite{Hassani:1999sny}:~thus a quantity which transforms as a vector under arbitrary transpositions will also constitute a vector 
under any one permutation.~Now, an operator $\Pi^{\sigma\tau}$ is obtained from a $d$-dimensional unity matrix $\mathbb{1}$, by swapping
the identity elements $\sigma$-th and $\tau$-th rows \cite{Hassani:1999sny}.~As an example, the matrix $\Pi^{14} = \Pi^{41}$ in (say) four dimensions follows as
\begin{align}
\mathbb{1}_{d=4} = \left[ \begin{array}{cccc} 1 & \; 0 & \; 0 & \; 0 \\ 0 & \; 1 & \; 0 & \; 0 \\ 0 & 0 & \; 1 & \; 0 \\ 0 & \; 0 & \; 0 & \; 1 \end{array} \right]
\overset{\text{1st row $\leftrightarrow$ 4th row}}{\longrightarrow} \left[ \begin{array}{cccc} 0 & \; 0 & \; 0 & \; 1 \\ 0 & \; 1 & \; 0 & \; 0 \\ 0 & 0 & \; 1 & \; 0 \\ 
1 & \; 0 & \; 0 & \; 0 \end{array} \right] = \Pi^{14}
\, .
\end{align}

It is straightforward to check that the operator $\Pi^{14}$ permutes the first and fourth components of a four-dimensional vector $p$.~The above construction 
principle implies that, in terms of matrix components, a transposition $\Pi^{\sigma\tau}$ can be written as  
\begin{align}\label{eqn: transpos_one}
\Pi^{\sigma\tau}_{\mu\nu} \, = \, \delta_{\mu\nu} \, , \qquad \text{if} \quad \mu \neq \sigma, \, \tau \, , 
\end{align}  

\noindent 
whereas for the $\sigma$-th and $\tau$-th rows of $\Pi^{\sigma\tau}$ it holds that 
\begin{align}\label{eqn: transpos_two}
\Pi^{\sigma\tau}_{\sigma\tau} \, = \, 1 \, =  \, \Pi^{\sigma\tau}_{\tau\sigma} \, ,
\end{align}

\noindent
with all the other elements in the aforementioned rows being zero.~With the help of the above component-wise representation for $\Pi^{\sigma\tau}$, it is 
easy to see how an arbitrary $d$-dimensional vector $p$ changes under transpositions.~By plugging in the equations \er{eqn: transpos_one} and \er{eqn: 
transpos_two} into the vector-like transformation law 
\begin{align}\label{eqn: vector_trans}
p_\mu \rightarrow p'_\mu = \sum_{\nu = 1}^d \Pi^{\sigma\tau}_{\mu\nu} \, p_\nu \, , 
\end{align}

\noindent
one notes that, regardless of the value of the index $\mu$, the above sum is always ``killed'' i.\,e.~there is always only a single momentum component that 
survives the summation.~As an example, for $\mu = \sigma$, one has
\begin{align}\label{eqn: perm_example}
p'_\sigma = \sum_{\nu = 1}^d \Pi^{\sigma\tau}_{\sigma\nu} \, p_\nu \, =  \Pi^{\sigma\tau}_{\sigma\tau} \, p_\tau = p_\tau \, , 
\end{align}

\noindent
wherein we used the fact that, in the $\sigma$-th row of $\Pi^{\sigma\tau}$, only the element $\Pi^{\sigma\tau}_{\sigma\tau} = 1$ is non-vanishing.~The full 
change of vector $p$ under this transposition is 
\begin{align}\label{eqn: perm_vect}
&p'_\mu =  \, p_\mu \, , \qquad \text{if} \quad \mu \neq \sigma, \, \tau \, , \nonumber \\
&p'_\sigma = p_\tau \, , \qquad \: p'_\tau = p_\sigma \, .  
\end{align}

With the above small machinery set up, it is very little additional effort to show that the vector-like modifications akin to \er{eqn: perm_vect} are obeyed 
by arbitrary polynomial functions of $p$, with an open vector index $\mu$.~In other words, any expression of the form $p_\mu^{\,m}$, with $m \in N$, will 
transform as a vector with respect to permutations of momentum components.~Under the action of $\Pi^{\sigma\tau}$, one gets 
\begin{align}\label{eqn: multi_trans}
p_\mu^{\,m} \rightarrow \left(\,p'_\mu\,\right)^m \: = \: \overbracket{p'_\mu \cdot p'_\mu \cdot p'_\mu \cdot \ldots \cdot p'_\mu}^{m \: \text{terms}} \: = \:   
\underbracket{\sum_{\rho = 1}^d \, \Pi^{\sigma\tau}_{\mu\rho} \, p_\rho \, \sum_{\lambda = 1}^d \, \Pi^{\sigma\tau}_{\mu\lambda} \, p_\lambda \ldots \sum_{
\xi = 1}^d \,\Pi^{\sigma\tau}_{\mu\xi} \, p_\xi}_{m \: \text{terms}} \, .
\end{align}  

The above alteration rule obviously involves multiple sums and looks quite dissimilar to the way that the momentum $p$ itself changes.~Nonetheless, there is no
actual difference between \er{eqn: vector_trans} and \er{eqn: multi_trans}.~To show this, we again concentrate on the case $\mu = \sigma$, for which it holds 
\begin{align}\label{eqn: fin_example}
\left(\,p'_\sigma\,\right)^m \: = \: \underbracket{\sum_{\rho = 1}^d \, \Pi^{\sigma\tau}_{\sigma\rho} \, p_\rho \, \sum_{\lambda = 1}^d \, \Pi^{\sigma\tau}_{\sigma\lambda}
\, p_\lambda \ldots \sum_{\xi = 1}^d \,\Pi^{\sigma\tau}_{\sigma\xi} \, p_\xi}_{m \: \text{terms}} \: = \: \overbrace{\Pi^{\sigma\tau}_{\sigma\tau} \, p_\tau \, \Pi^{\sigma
\tau}_{\sigma\tau} \, p_\tau \ldots \Pi^{\sigma\tau}_{\sigma\tau} \, p_\tau}^{m \: \text{terms}} \: = \: \Pi^{\sigma\tau}_{\sigma\tau} \, p_\tau^{\,m} \: = \: \sum_{\nu = 
1}^d \,\Pi^{\sigma\tau}_{\sigma\nu} \, p_\nu^{\,m} \,\, . 
\end{align} 

In obtaining the final result in \er{eqn: fin_example}, we again made use of the fact that $\Pi^{\sigma\tau}_{\sigma\tau} = 1$ is the only non-zero entry in the $\sigma$-th 
row of the operator $\Pi^{\sigma\tau}$, as well as the fact that $\left(\Pi^{\sigma\tau}_{\sigma\tau}\right)^m = \Pi^{\sigma\tau}_{\sigma\tau}$.~The final form in \er{eqn:
fin_example} was written in a suggestive way, to make it clear that $p_\mu^{\,m}$ indeed behaves as a vector under arbitrary transpositions, for any integer value $m$ and 
for a fixed index $\mu = \sigma$.~The full tranformation rule for the functions $p_\mu^{\,m}$ is
\begin{align}
&\left(\,p'_\mu\,\right)^m =  \, p^{\,m}_\mu \, , \qquad \text{if} \quad \mu \neq \sigma, \, \tau \, , \nonumber \\
&\left(\,p'_\sigma\,\right)^m = p^{\,m}_\tau \, , \qquad \: \left(\,p'_\tau\,\right)^m = p^{\,m}_\sigma \, ,
\end{align} 

\noindent
and it matches the vector-like change of momentum $p$ itself, see equation \er{eqn: perm_vect}.~Thus, when it comes to finding general expressions for hypercubic
vectors, permutations alone offer very few restrictions, as any open-indexed polynomial in $p$ will change in a vector-like fashion under these kinds of operators. 

The argument is not done however, as we also have to take into account the parity transformations.~We will start with the functions $p^{\,m}_\mu$, where $m \in N$,
and see if we can impose some additional constraints on them, by invoking their vector-like nature under inversions.~As an example, let us take the $d$-dimensional 
momentum $p = (p_1, p_2,\ldots,p_d)$ and perform a parity transformation on its $\eta$-th component, so that $p'_\eta = - p_\eta$.~Here the index $\eta$ can take on
any value between 1 and $d$.~Our prospective lattice vector $p^{\,m}_\mu$ should transform in exactly the same way as momentum $p$ itself, meaning that  
\begin{align}
p^{\,m}_\eta \rightarrow p'^{\,m}_\eta \: = \: - p^{\,m}_\eta  \, ,  \qquad \quad \eta = 1, \ldots d \, ,
\end{align}

\noindent
with all the other components (with $\mu \neq \eta$) of $p^{\,m}_\mu$ being intact.~The above considerations lead us to a conclusion that $m$ has to be an \textit{odd}
integer.~If $m$ were even, the polynomial functions $p^{\,m}_\eta$ would be completely indifferent to inversions of momentum components, while any combination of even 
and odd factors (e.\,g.~$p^{\,2}_\mu + p^{\,3}_\mu$) would have no definitive symmetry properties under parity changes.~To conclude, any polynomial expression of the form 
\begin{align}\label{eqn: latt_vector}
t_\mu(p) = p_\mu^{\,2k+1} \, , \quad k \in N_0 \, ,   
\end{align}

\noindent
will constitute a lattice vector, i.\,e.~it will transform as a vector under coordinate permutations and inversions.~In the above relation, $N_0$ stands for
a set of non-negative integers.~An immediate corollary is that any function which can be expanded in an odd Taylor series in $p_\mu$, would also comprise the 
components of a lattice vector.~As an example of this, in lattice perturbation theory one often encounters expressions where the standard continuum momentum 
$p_\mu$ is replaced with the following function \cite{Rothe:1992nt, Capitani:2002mp}
\begin{align}\label{eqn: def_hatp}
\hat{p}_\mu = 2 \sin\left(\frac{p_\mu}{2}\right) \, .  
\end{align}

It is obvious that the quantity $\hat{p}$ doesn't change in a vector-like fashion, under general orthogonal transformations of $p$, but it does transform as a
vector with respect to inversions and permutations of momentum components.~In other words, $\hat{p}$ is a lattice vector (as it arguably should be), and its 
Taylor series expansion would result in a summation over infinitely many terms of the form \er{eqn: latt_vector}.

This finally brings us to the question of some practical relevance.~Given some lattice vector $\Gamma$, which is a function of momentum $p$, what is the suitable 
tensor representation of $\Gamma(p)$?~Based on our preceding arguments, any object like \er{eqn: latt_vector} could be used as a tensor element when describing
$\Gamma(p)$, since they all behave as vectors with respect to lattice symmetry operations.~At first, this kind of ``infinite freedom'' of choice might seem rather
absurd, especially if one considers the fact that basis decompositions of continuum functions are unique.~However, such ambiguities are nothing new in the world
of lattice calculations, as it is well known that any field theory can be discretised in infinitely many different ways, all of which have the same continuum 
limit.~Concerning the vertex tensor parametrisations, we shall partially resolve the ambiguity by adhering to the order of ascending mass dimension, meaning that
tensors with smaller powers in $p$ will be preferred.~Then, a tensor description of a vector-like quantity $\Gamma$ would be 
\begin{align}
\Gamma_\mu(p) = \sum_{k = 1}^d \mathcal{F}^k \, p_\mu^{\,2k + 1} \, , 
\end{align}

\noindent
with $\mathcal{F}^k$ a form factor of the $k$-th tensor element.~Note that the sum in the above relation does not include infinitely many terms, but rather 
terminates at the $d$-th contribution, with $d$ the number of dimensions.~This is because, in a $d$-dimensional space, there can be no more than $d$ linearly 
independent basis vectors.~In fact, the notion of dimension is often \textit{defined} as the number of linearly independent vectors needed to cover the space 
\cite{Hassani:1999sny, Morty:1962prc}.~For concreteness, let us assume a three-dimensional setting, so that a complete tensor description of a lattice vector
$\Gamma_\mu$ should be given by   
\begin{align}\label{eqn: basis_primer}
\Gamma_\mu(p) = A(p) \, p_\mu + B(p) \, p_\mu^3 + C(p) \, p_\mu^5 \, . 
\end{align}

Now, one might notice an apparent problem with the above arguments.~In $d$ dimensions, \textit{any} collection of $d$ linearly independent elements will constitute a 
complete tensor representation, for vector-like quantities.~If basis completeness was the only relevant criterion, then a decomposition like \er{eqn: basis_primer}
should not be favoured, for $d=3$, over any other choices of three linearly independent objects with a vector index $\mu$.~This might even include the previously mentioned 
structures of the kind $p_\mu^{\,m}$, with an even integer $m$.~However, one ought to remember that the vertex form factors [\,i.\,e.~the functions like $A(p), \, B(p)$
and $C(p)$ of \er{eqn: basis_primer}\,] should be hypercubic invariants, and this will not happen for arbitrary choices of tensor parametrisation.~To see why, take the 
particular example of the basis \er{eqn: basis_primer} and suppose that one has obtained the corresponding projectors $P_\mu$.~Then, the dressing (say) $A(p)$ would follow
as 
\begin{align}\label{eqn: a_project}
A(p) = P_\mu^{\,A} \cdot \Gamma_\mu = \sum_{\mu = 1}^d \, P_\mu^{\,A} \, \Gamma_\mu \, .
\end{align}

In order for the contraction \er{eqn: a_project} to be a hypercubic invariant, both the projector ($P_\mu^{\,A}$) and the vertex $(\Gamma_\mu)$ itself ought to transform
as hypercubic vectors.~Since the correlator $\Gamma_\mu$ is assumed to be a lattice vector from the outset, the symmetry (or lack thereof) of $A(p)$ is determined completely
by the projector $P_\mu^{\,A}$.~In turn, the transformation properties of $P_\mu^{\,A}$ follow directly from the choice of basis, since the projectors are always linear 
combinations of basis elements themselves, see e.\,g.~Appendix \ref{sec: projectors}.~This also means that it takes only a single wrong (non-vector) basis structure
to ruin the symmetry properties for \textit{all} of the involved coefficient functions.~In the case of the parametrisation like \er{eqn: basis_primer}, one should feel 
somewhat ``safe'' since all the tensor elements behave as hypercubic vectors.~These claims regarding the symmetry features of correlator form factors will be addressed 
directly in our Monte Carlo simulations, where we shall compare the values of the dressing functions before and after averaging over all possible permutations and 
inversions of momentum components:~we shall see that (within statistical uncertainties) all of the relevant dressings pass the test of hypercubic invariance.            

Equipped with these basic facts on how scalar and vector functions get modified on the lattice, compared to their continuum counterparts, we may proceed towards 
some practical applications for the knowledge we've acquired.~Namely, we wish to deduce the tensor representations of two concrete vertex functions of lattice 
Yang Mills theory, the ghost-gluon vertex and gluon propagator.~We shall see if we can learn something interesting about these lattice correlators along the way. 

\section{Tensor bases for lattice ghost-gluon vertex and gluon propagator}\label{sec: vertex_n_propag}

\subsection{Ghost-gluon vertex:~continuum basis}

The ghost-gluon vertex is the lowest-order correlation function which encodes the interaction of ghost and gluon fields.~It plays a pivotal role in many truncations of
functional equations of motion (see e.\,g.~\cite{Schleifenbaum:2004id, Kellermann:2008iw, Alkofer:2008dt, Huber:2012zj, Huber:2012kd, Aguilar:2013xqa, Aguilar:2013vaa, 
Binosi:2014kka, Huber:2017txg, Alkofer:2004it}), due to its non-renormalisation in Landau gauge \cite{Taylor:1971ff}.~Here, we wish to see how the tensor description of
the function changes when going from continuum to discretised spacetimes, and what this can tell us about the relation between the lattice and continuum versions of the 
correlator.  

Let us start with the tensor basis in a continuum setting.~In this section, we will keep the discussion independent of the number of dimensions:~a definitive 
value for $d$ will be chosen only when we start considering the lattice vertex.~The momenta pertaining to the ghost, antighost and the gluon leg of the function 
will be denoted, respectively,  with $p,\, q \, \text{and} \, k$.~Due to momentum conservation at the vertex, with $p + q + k = 0$, only two out of these three
momenta are linearly independent, and any two can be chosen for constructing the vertex tensor elements.~We opt to work in terms of vectors $q$ and $p$:~with this
choice, the continuum correlator can be represented as  
\begin{align}\label{eqn: vert_cont}
\Gamma_\mu(q, k) = A(q,p) \, q_\mu + B(q,p) \, p_\mu \, .
\end{align}

The projectors for the above basis can be found straightforwardly, with standard methods of linear algebra.~Their construction is explained in Appendix \ref{sec: projectors},
and here we simply cite the final answer:
\begin{align}\label{eqn: ab_project}
P_\mu^{\,A} = \frac{-p^2 \, q_\mu + q\cdot p \, p_\mu}{(q\cdot p)^2 - q^2 \, p^2} \, , \qquad \qquad \: P_\mu^{\,B} = \frac{ q\cdot p \, q_\mu -  \, q^2 \, p_\mu}{(q\cdot p)^2
- q^2 \, p^2} \, .
\end{align}
 
Note that both of the above functions are ill-defined for a kinematic configuration 
\begin{align}\label{eqn: paral_momenta}
q_\mu = c\cdot p_\mu \, , \qquad c = \text{const.}
\end{align}

Namely, for the momentum setup of \er{eqn: paral_momenta}, the projectors in \er{eqn: ab_project} reduce to undefined expressions of the form ``$0/0$''.~For the
purposes of latter discussion, it is worthwhile to dwell on the origin of this problem.~Let us thus look at (say) the function $P_\mu^{\, A}$, and its two defining
equations: 
\begin{align}
\sum_{\mu = 1}^d P_\mu^{\,A} \, q_\mu = 1 \, , \qquad \text{and} \qquad  \sum_{\mu = 1}^d P_\mu^{\,A} \, p_\mu = 0 \, .
\end{align}

For the kinematic choice \er{eqn: paral_momenta}, the above set of equations becomes contradictory, as one gets
\begin{align}\label{eqn: contradict}
c \, \sum_{\mu = 1}^d P_\mu^{\,A} \, p_\mu = 1 \, , \qquad \text{and} \qquad  \sum_{\mu = 1}^d P_\mu^{\,A} \, p_\mu = 0 \, .
\end{align}

It should be clear that no well-defined object $P_\mu^{\, A}$ can obey the constraints of \er{eqn: contradict}.~The same holds for $P_\mu^{\, B}$.~These issues are closely 
related to the concept of linear (in)dependence of basis elements.~For the particular configuration \er{eqn: paral_momenta}, the vectors $q$ and $p$ are evidently not linearly
independent, since one of the momenta is proportional to the other one.~It is a rather general statement of linear algebra, that no well-defined projectors can be constructed
for basis descriptions with feature linearly dependent elements, see e.\,g.~\cite{Hassani:1999sny} or Appendix \ref{sec: projectors}.

The solution for the above problems is simple, and it amounts to using a reduced basis, where needed.~For the kinematic choice of \er{eqn: paral_momenta}, any one of the
following descriptions would work
\begin{align}\label{eqn: cont_reduce}
\Gamma_\mu(q, k) = A(q,p) \, q_\mu  \, , \qquad \text{or} \qquad \Gamma_\mu(q, p) = B(q,p) \, p_\mu  \, .
\end{align}

Basis completeness of these reduced decompositions follows from the kinematics \er{eqn: paral_momenta} itself.~Namely, for parallel momenta $q$ and $p$, any one 
of the vectors will contain the full information about the vertex, since the other element has no ``new information'' to add.~This constitutes a general rule,
concerning the tensor representations of vertex functions (both continuum and lattice ones):~if a given basis becomes redundant, for a particular kinematic choice,
one is allowed to ``throw away'' the basis elements, until a non-redundant description is reached.~Here, by a redundant decomposition we mean the one where some of 
the basis structures can be expressed as linear combinations of other tensor elements.~In the next section, we will see that there exist special kinematic choices
on the lattice, where for reasons of linear dependence, the continuum tensor description of \er{eqn: vert_cont} determines the lattice correlator fully. 

\subsection{Ghost-gluon vertex:~lattice-modified basis}\label{sec: ghost_latt}
   
In section \ref{sec: hyper_vector} we already discussed possible tensor representations for lattice vector-valued functions, which depend on a single momentum
variable.~For the ghost-gluon correlator, these arguments need to be generalised to a situation with two independent momenta, in order to capture the full
kinematic dependence of the vertex.~Such a generalisation is rather straightforward, and we shall not provide the details here.~We merely state without proof,
that functions of two momenta (say, $q$ and $p$), which transform as vectors under permutations and parity transformations, will neccessarily have one of the
following two forms:
\begin{align}\label{eqn: taus_lattice_vertex}
&\tau^{1, \, rs}_\mu(q,p) = p_\mu^{\, 2r} q_\mu^{\, 2s + 1} \, , \nonumber \\ 
&\tau^{2, \, rs}_\mu(q,p) = q_\mu^{\, 2r} p_\mu^{\, 2s + 1} \, , \qquad \text{with} \qquad r, \, s \in N_0 \, .
\end{align}    

Of course, any linear combinations of the above structures are also allowed.~At the risk of overstating the obvious, we highlight that for functions of multiple 
momenta, the same symmetry transformation (permutation, inversion) always has to be applied to \textit{all} of the vectors involved.~Thus, if one wished to change
the sign on a (say) second momentum component, it has to be done to both vectors $q$ and $p$, so that (in three dimensions)
\begin{align}
q' = (q_1, - q_2, q_3) \, , \qquad p' = (p_1, - p_2, p_3) \, .
\end{align}  

The above comes from the fact that the symmetry operations relate to the whole lattice, not just to individual momentum vectors.~For instance, if one wanted to rotate
the space by $\pi/2$ degrees in a certain plane, there is no way to perform this transformation without affecting all of the lattice momenta equally.~Also, in the absence 
of the aforementioned rule the scalar products like $q\cdot p$ would not be invariant under the supposed lattice symmetry operations.

If we now go back to the tensor structures of \er{eqn: taus_lattice_vertex}, we see that the lowest-order terms give us the continuum basis elements $q_\mu$ and
$p_\mu$, while the leading-order lattice-induced structures would be $q_\mu^{\,3}, \: p_\mu^{\,3}, \: q_\mu^{\,2} \, p_\mu$ and $p_\mu^{\,2} \,q_\mu$.~For concreteness,
let us say that we work in three spatial dimensions, since this is the case we will consider in our numerical simulations.~For $d=3$, any three linearly independent 
vectors would suffice to describe the vertex fully, which means that we only need to add one more element to the continuum representation of \er{eqn: vert_cont}.~Any one of 
the leading-order lattice modifications would fit equally well, from the symmetry perspective, and we simply choose the additional vector to be $q^{\,3}_\mu$.~This brings
us to the following basis for the lattice correlator 
\begin{align}\label{eqn: vert_latt}
\Gamma_\mu(q, p) = E(q,p) \, q_\mu + F(q,p) \, p_\mu \, + G(q,p) \, q^{\,3}_\mu \, .
\end{align}

Again, the construction of the corresponding projectors in explained in Appendix \ref{sec: projectors}, and here we will abstain from providing the full expressions 
for these objects, due to their considerable length and complexity.~For the parallel configuration \er{eqn: paral_momenta}, one of the continuum basis vectors (either
$q_\mu$ or $p_\mu$) can be neglected in the overall tensor representation, as it contains no information which is not already present in some of the other elements.
\!However, what is arguably more interesting is to identify those kinematic choices where the entire lattice correlator collapses onto the continuum structures
of \er{eqn: vert_cont}.~In other words, one may attempt to find such momentum points where \textit{all} of the tensor elements \er{eqn: taus_lattice_vertex} become 
proportional to the continuum vectors.~While this might seem like an impossible task at first, it is in fact very easy to think of at least one situation where this must
happen:~if both momenta $p$ and $q$ point along the lattice diagonal, with $p = (m,m,m)$ and $q = (n,n,n)$ (where in general $m \neq n$), all of the lattice-induced tensor
elements will become parallel to either $q$ or $p$, with 
\begin{align}\label{eqn: taus_diagonal}
&\tau^{1, \, rs}_\mu(q,p) = m^{\, 2r} n^{\, 2s} q_\mu \, , \nonumber \\ 
&\tau^{2, \, rs}_\mu(q,p) = n^{\, 2r} m^{\, 2s} p_\mu \, , \qquad \text{with} \qquad r, \, s \in N_0 \, .
\end{align} 

The fully diagonal setup is also a special case of the parallel kinematics \er{eqn: paral_momenta}, meaning that the actual tensor parametrisation shrinks even further to 
\er{eqn: cont_reduce}.~The fact that the lattice function $\Gamma_\mu(q,p)$ is described exactly by a continuum tensor basis (within statistical errors), for fully diagonal 
kinematics, will be demonstrated in our numerical Monte Carlo simulations.~Another example where the lattice modifications of the basis \er{eqn: vert_cont} become redundant, 
is the one where at least one momentum is diagonal [\,say, $q = (n,n,n)$\,], while the other vector is either on-axis [\,with $p = (m,0,0)$, plus permutations thereof\:] or 
is of the form $p = (m,0,m)$ (plus permutations of $p$ components).~Some other interesting cases will be discussed in due course.~What is important to note here is that the 
completeness of the continuum tensor description does not imply that the lattice vertex is ``equal'' to its continuum counterpart, and that the two functions could/should be 
directly compared to each other.~In all special kinematic cases, the discretisation-induced tensor structures become effectively degenerate with the continuum basis elements,
but this does not mean that the vertex cannot host a myriad of finite-spacing artifacts within its ``continuum'' dressing functions.~Ideally, elaborate continuum extrapolations 
should be performed before any serious comparisons between the continuum and lattice functions are made.~The only kinematic choices where the lattice correlators could be 
regarded as being truly continuum-like are those in the deep infrared (IR) energy region:~since most lattice corrections have a comparatively high mass dimension [\,see e.\,g.
\!\er{eqn: hyper_scalars} and \er{eqn: basis_primer}\,], one can naturally expect them to be suppressed at low momenta, thus bringing about the dominance of the continuum terms
(barring the finite volume effects). 

\subsection{Gluon propagator:~lattice-modified basis}\label{sec: gluon_lattice} 

In this section we want to deduce, using hypercubic symmetry considerations, the tensor description of the lattice gluon propagator.~The arguments we will cover 
here also apply to any other (lattice) second-rank tensors which depend on single momentum $p$, such as the photon two-point function, or the hadronic 
electromagnetic current $\Pi_{\mu\nu}$, see e.\,g.~\cite{Aubin:2015rzx}.~To keep things simple, we will refer only to the ``gluon propagator'' in the following
text, since this is the one function which we will study in some detail in numerical Monte Carlo simulations.~We will also employ the corresponding standard notation
$D_{\mu\nu}(p)$.

The tensor parametrisation of the gluon two-point function is well known on the lattice, and is determined unambiguously by the gauge-fixing procedure \cite{Alles:1996ka}.
\!Thus, one might wonder why we are investing effort to tackle a problem which already has a satisfying solution.~We can provide two justifications in this regard.~First, 
deriving a basis description which follows purely from symmetry can potentially provide some interesting insights which would otherwise remain hidden.~Second, the calculations
to be carried out here can be seen as a preparation for obtaining symmetry-based decompositions of other lattice correlators of higher rank, like the three-gluon vertex, whose 
tensor representations are not fully constrained by gauge-fixing \cite{Vujinovic:2018nqc}.~We start the discussion with the continuum propagator:~the corresponding basis
decomposition is 
\begin{align}\label{eqn: gluon_cont}
D_{\mu\nu}(p) = A(p) \, \delta_{\mu\nu} + B(p) \, p_\mu p_\nu ,
\end{align}    

\noindent
where indices $\mu$ and $\nu$ run from 1 to $d$.~The above two basis elements are the most general structures which satisfy the appropriate tensor 
transformation law 
\begin{align}\label{eqn: second_tensor_law}
\tau_{\mu\nu}(p) \rightarrow  \tau''_{\mu\nu}(p) = \sum_{\sigma = 1}^d \sum_{\rho = 1}^d O_{\mu\sigma} O_{\nu\rho} \, \tau_{\sigma\rho} ,
\end{align}

\noindent
with $O$s being arbitrary $d$-dimensional orthogonal matrices.~The projectors for the above basis will be provided later.~We now wish to see how the
representation \er{eqn: gluon_cont} may be generalised in a discretised theory.~Similarly to hypercubic scalars and vectors, one needs to look for most 
general possible functions which satisfy the transformation rule \er{eqn: second_tensor_law}, with matrices $O$ belonging to to the hypercubic symmetry
group.~The obvious course of action would be to look for second-rank extensions of the equation \er{eqn: latt_vector}, i.\,e.~to find operators with higher
powers in $p_\mu$ and $p_\nu$, which can be added to the continuum tensor description.~The addition of such new terms is indeed possible on a lattice, but
right now we want to discuss a different type of modification of the continuum basis, which is reminiscent of how rotational symmetry breaking manifests 
itself in QCD and Yang-Mills studies at finite temperature, see e.\,g.~\cite{Fischer:2018sdj} and references therein.~In particular, we wish to argue that 
the tensor structure \er{eqn: gluon_cont} generalises to 
\begin{align}\label{eqn: latt_glue_2d}          
&D_{\mu\mu} = E(p)\, \delta_{\mu\mu} \: + \: F(p) \, p_\mu^{\,2} \, , \qquad \quad \mu = 1, \ldots d \nonumber \\ 
&D_{\nu\mu} = G(p) \, p_\nu p_\mu \, , \qquad \qquad \qquad \quad \: \:  \mu, \, \, \nu = 1, \ldots d \, , \qquad \mu \neq \nu \, ,
\end{align}  

\noindent
on discretised spacetimes.~Put in words, the diagonal and off-diagonal components of the lattice gluon propagator are parametrised by different dressing 
functions, in contrast to \er{eqn: gluon_cont}.~The above splitting comes from the fact that, unlike general orthogonal operators, permutations and inversions
cannot mix diagonal and off-diagonal components of second-rank tensors, and the two kinds of terms transform independently from each other, under hypercubic
symmetry operations.~To demonstrate this, we start (yet again) with an example of a two-dimensional theory.~In section \ref{sec: hyper_perms} we argued that
the three operators $\rho_1, \rho_2$ and $\Pi^{12}$, and their matrix compositions, exhaust all of the symmetry transformations of a square, see equations
\er{eqn: 2d_inver_def} and \er{eqn: 2d_perm_def}.~Now, under the action of a permutation $\Pi^{12}$, the 2\,$\times$\,2 gluon propagator transforms in the
following fashion
\begin{align}
D\,'' = &\left( \begin{array}{cc} D\,''_{11} & D\,''_{12} \\ D\,''_{21} & D\,''_{22}  \end{array} \right) \: = \: \Pi^{12} \cdot D \cdot \left(\Pi^{12}
\right)^T = \nonumber \\[0.12cm]
\left[ \begin{array}{cc} 0 & \;\; 1 \\ 1 & \;\; 0  \end{array} \right] \cdot & \left( \begin{array}{cc} D_{11} & D_{12} \\ D_{21} & D_{22}  \end{array}\right) 
\cdot \left[ \begin{array}{cc} 0 & \;\; 1 \\ 1 & \;\; 0  \end{array} \right] = \left( \begin{array}{cc} D_{22} & D_{21} \\ D_{12} & D_{11}  \end{array}\right) \, .
\end{align} 

In the above relations, $T$ stands for a matrix transpose.~The overall effect of the exchange operator $\Pi^{12}$ is thus  
\begin{align}\label{eqn: gluon_permuted_2d}
&D\,''_{11} = D_{22} \, , \qquad \: D\,''_{12} = D_{21} \, , \nonumber \\
&D\,''_{21} = D_{12} \, , \qquad \: D\,''_{22} = D_{11} \, .
\end{align} 

In the transformation rule \er{eqn: gluon_permuted_2d}, a diagonal component (i.\,e.~a term of the form $D_{\mu\mu}$) always changes into another diagonal 
component, while an off-diagonal factor (a quantity like $D_{\mu\nu}$, with $\mu \neq \nu$) always changes into another off-diagonal factor.~Thus, the 
transformation itself does not combine the diagonal and off-diagonal terms, and the two kinds of contributions change separately from each other, under 
the action of $\Pi^{12}$.~The same remark holds also for the operators $\rho_1$ and $\rho_2$ of \er{eqn: 2d_inver_def}.~For instance, the $\rho_1$ matrix
changes the propagator as follows  
\begin{align}\label{eqn: gluon_reflected_2d}
&D\,''_{11} = D_{11} \, , \qquad \: D\,''_{12} = - D_{12} \, , \nonumber \\
&D\,''_{21} = -D_{21} \, , \qquad \: D\,''_{22} = D_{22} \, . 
\end{align} 

In \er{eqn: gluon_reflected_2d}, there is again no mixing between off-diagonal and diagonal pieces of the gluon two-point function.~The same observation is 
also true for the inversion element $\rho_2$.~One concludes that none of the elementary transformations $\rho_1, \rho_2$ and $\Pi^{12}$, and thus also none 
of their compositions, combines the propagator terms of the kind $D_{\mu\mu}$ with those of the kind $D_{\mu\nu}$ (with $\mu \neq \nu$).~This is the origin 
of the splitting given in equation \er{eqn: latt_glue_2d}, for $d=2$.~Of course, one would like to check if the argument generalises naturally to higher 
dimensions as well.~The most direct way of testing this would be to take the three- or higher-dimensional permutation and inversion matrices, apply them to
the gluon propagator of appropriate dimensionality, and deduce the corresponding constraints on the correlators tensor decomposition.~Such a procedure is 
however very tedious, as one has to check the overall effect of every elementary permutation and parity operator.~To make matters simpler, we will now try to
formulate relatively general and dimensionality-independent arguments on why hypercubic symmetry transformations cannot combine the off-diagonal and diagonal 
components of second rank tensors.~In the process, we will also attempt to extend these considerations to other correlators of interest, like the lattice 
three-gluon vertex.  

As usual, we will look at the hypercubic group as a composition of permutation and inversion transformations, and will shape our line of reasoning for each 
of the two symmetry operations separately.~We start with the easier case of parity changes.~For these matrices, it is somewhat obvious that they cannot 
induce mixing of different components of second-rank tensors, or indeed tensors of arbitrary rank.~Namely, the inversion operators always have the overall
structure of a unity matrix $\mathbb{1}$:~the ``only'' difference between parity transformations and the identity element comes from the minus signs, see 
e.\,g.~\er{eqn: 2d_inver_def} and \er{eqn: 3d_rot_example}.~While the minuses are evidently important, the general unity-like composition of these operators
means that they cannot rearrange the components of arbitrary tensors in any non-trivial way, see \er{eqn: gluon_reflected_2d} as an example.~This also implies
that inversions cannot mix the diagonal and off-diagonal terms of second-rank tensors, independent of the number of dimensions.

This brings us to permutations.~To understand why permutations cannot mix contributions of the kind $D_{\mu\mu}$ with those of the kind $D_{\mu\nu}$ (with
$\mu \neq \nu$), we will go back to the example of a two-dimensional theory and the transformation rule \er{eqn: gluon_permuted_2d}.~The change \er{eqn: 
gluon_permuted_2d} can be written in a more abstract and concise manner as 
\begin{align}\label{eqn: abstract_permute}
  11 \leftrightarrow 22 \: , \qquad 12 \leftrightarrow 21 \: .
\end{align} 

The above is a symbolic way to say that under a permutation $\Pi^{12}$, the gluon component $D_{11}$ gets exchanged with $D_{22}$, while $D_{12}$ 
exchanges places with $D_{21}$:~these swaps constitute the full content of the equation \er{eqn: gluon_permuted_2d}.~One now notes that the rule \er{eqn:
abstract_permute} matches the way in which an ordered set of numbers `$jk$' (with $j\, , k = 1, 2$) transforms under a permutation $1 \leftrightarrow2$.~With
this abstraction, it becomes somewhat obvious why $\Pi^{12}$ cannot mix diagonal and off-diagonal components of second-rank tensors:~there is no possible 
permutation of numbers which can turn configurations of the form ``11'' and ``22'' into those of the form ``12'' or ``21'', and vice versa.~The reasoning
straightforwardly extends to an arbitrary dimension number.~In three dimensions, for instance, there are three elementary permutations (these are $1 
\leftrightarrow 2 \, , 1 \leftrightarrow 3$ and $ 2 \leftrightarrow 3 $), none of which can turn any of the diagonal configurations (11, 22 and 33) into any
of the off-diagonal ones (12, 13, 23 + permutations).~Thus, the splitting between off-diagonal and diagonal components of the gluon propagator persists also
for $d = 3$, and indeed for any dimension number.~A mathematically more formal version of this heuristic reasoning is given in Appendix \ref{sec: append_permutes}.

We wish to point out that the different treatment of diagonal and off-diagonal tensor terms, as in equation \er{eqn: latt_glue_2d}, was also noted in the lattice
study of the hadronic vacuum polarisation contribution to the anomalous magnetic moment of the muon \cite{Aubin:2015rzx}.~However, no direct comparison between our 
approach and that of \cite{Aubin:2015rzx} is possible, since there an asymmetric four-dimensional lattice was used, with different temporal and spatial extensions.
\!Also, the authors of \cite{Aubin:2015rzx} eventually abandon the explicit consideration of discretisation artifacts in favour of a careful analysis of finite volume 
effects, which are here ignored.~We will make a qualitative/semi-quantitative argument about the validity of their approximation later in this paper.  
   
The arguments which had led us to the decomposition \er{eqn: latt_glue_2d} can also be applied to other correlators, like the three-gluon vertex.~With a bit of 
work, one may quickly deduce that the lattice three-gluon correlator contains five independent ``cycles'', which cannot combine with each other under either
parity or permutation transformations.~The five cycles are 1) $\Gamma_{\mu\mu\mu}$ 2) $\Gamma_{\nu\mu\mu}$ 3) $\Gamma_{\mu\nu\mu}$ 4) $\Gamma_{\mu\mu\nu}$ 5) 
$\Gamma_{\mu\nu\rho}$:~for cycles 2) to 4), it holds that $\mu \neq \nu$, while in cycle 5) all the indices $\mu, \, \nu$ and $\rho$ are different from each 
other .~In practice, this means that a single tensor entity of the continuum theory, like (say) $p_\mu \, p_\nu \, p_\rho$, will break into five independent 
pieces on the lattice, each with its own dressing function.~We leave explicit calculations concerning this vertex function for future studies.  

Going back to the gluon propagator, the equation \er{eqn: latt_glue_2d} does not exhaust all the possibilities concerning the correlators tensor representations 
on the lattice.~Namely, with the same arguments as employed in section \ref{sec: hyper_vector}, it can be easily shown that any functions of the form 
\begin{align}\label{eqn: gluon_higher}
2 \,\tau_{\mu\nu}(p) \, = \, p_\mu^{\,2k + 1}  p_\nu^{\,2n + 1} \: + \:\: p_\nu^{\,2k + 1} \, p_\mu^{\,2n + 1} \, , \qquad \: \: k, \, n \in N_0 \, , \qquad \: \: 
\mu \, , \nu  = 1
\ldots d \, , 
\end{align} 

\noindent
will satisfy the transformation laws adequate for a second-rank tensor, under permutations and inversions\,\footnote{In principle, the Kronecker tensor $\delta_{\mu\nu}
$ can also receive higher-order lattice corrections, like e.\,g.~$\delta_{\mu\nu}\,p_\mu^{\,2}$ \cite{Kawai:1980ja, Weisz:1983bn}.~However, the only non-vanishing part 
of such a term is $\delta_{\mu\mu} \, p^{\,2}_\mu = p_\mu^{\,2}$, which is already present in \er{eqn: latt_glue_2d}.~We thus do not consider such contributions separately,
as they are in fact indistinguishable from the diagonal parts of \er{eqn: gluon_higher}.}.~The symmetrisation in \er{eqn: gluon_higher} was carried out to comply with the 
symmetry property of the propagator itself, namely $D_{\mu\nu}(p) = D_{\nu\mu}(p)$.~In \er{eqn: gluon_higher}, we did not explicitly indicate a split between the diagonal
and off-diagonal contributions, for reasons of simplicity.~It should be understood that this kind of separate treatment is applicable to the above higher-order tensors, 
just as it is for the decomposition \er{eqn: latt_glue_2d}.~Among the elements \er{eqn: gluon_higher}, the leading-order correction to the continuum term $p_\nu p_\mu$
has the form 
\begin{align}\label{eqn: gluon_lead}
2 \, \tau^\text{\,lead}_{\mu\nu}(p) \, = \, p_\mu \, p_\nu^{3} \: + \: p_\nu \, p_\mu^3 \, = \, p_\nu \, p_\mu \, (p_\mu^2 \: + \: p_\nu^2) \, , \qquad \mu \, , \nu  = 1
\ldots d \, ,
\end{align} 

\noindent
and it appears in gluon propagator representations involving the $\mathcal{O}(a^2)$ improved lattice gauge action \cite{Weisz:1983bn}.~Now, while it is important
to keep in mind that the decomposition \er{eqn: latt_glue_2d} can be augmented with higher-order corrections, throughout this paper we will work \textit{only} with
the tensor structures of \er{eqn: latt_glue_2d}.~In two dimensions, it actually turns out that this basis is complete, i.\,e.~that it describes the gluon propagator
without any loss of information.~This follows from the fact that, being a symmetric $d \times d$ matrix in $d$ dimensions, the gluon two-point function cannot
contain more that $N_d$ free parameters (for fixed momentum $p$), where \cite{Morty:1962prc}
\begin{align}\label{eqn: nd_def}
N_d = \frac{d\,(d + 1)}{2} \, . 
\end{align}

For a two-dimensional theory, $N_d$ equals three, which is exactly the number of free parameters present in \er{eqn: latt_glue_2d}.~To solidify the case for 
completeness of this basis in two dimensions, in Appendix \ref{sec: projectors} we show that the leading-order correction \er{eqn: gluon_lead} can be described
exactly as a linear combination of the elements in \er{eqn: latt_glue_2d}.~In three dimensions, $N_d$ is equal to six, and the decomposition \er{eqn: latt_glue_2d}
is no longer complete.~In our Monte Carlo simulations, we will show that even for $d = 3$, the lattice-modified representation \er{eqn: latt_glue_2d} describes the 
propagator rather well, and certainly significantly better than the continuum one.~Of course, showing an (approximate) completeness of a given basis is not enough, 
as we argued at the end of section \ref{sec: hyper_vector}\,:~it is always possible to find basis decompositions which are ``trivially'' complete, by virtue of 
exhausting all of the free parameters of a correlator at hand.~The real issue is whether the basis in question features form factors which have adequate symmetry
properties.~Therefore the hypercubic invariance of dressing functions pertaining to the decomposition \er{eqn: latt_glue_2d} will be tested numerically, and they
will be shown to perform quite well in this regard.~Explicit formulas for calculating the coefficients of \er{eqn: latt_glue_2d} will be given later.      

To conclude, we want to point to an interesting notion concerning the lattice propagator basis.~Naively, one would expect that the gluon two-point function 
becomes more ``continuum-like'' as one approaches the infrared energy region.~In the context of the parametrisation  \er{eqn: latt_glue_2d}, this suggests 
that the dressing functions $F(p)$ and $G(p)$ should become equal to each other, so that the form \er{eqn: gluon_cont} is recovered, as one considers smaller 
and smaller values for momentum components $p_\mu$.~Indeed, such a behaviour will be confirmed in our numerical calculations.~However, it will also turn out 
that the scenario ``$F \approx G$'' is not tied exclusively to the infrared limit, and that there exist alternative lattice kinematics, some at rather high 
momenta, where the decomposition \er{eqn: latt_glue_2d} effectively reduces to the continuum basis \er{eqn: gluon_cont}.

\section{Numerical calculations with lattice-adjusted bases}\label{sec: numerics}

\subsection{General setup and vertex reconstruction}\label{sec: general_setup}

We now wish to perform lattice Monte Carlo calculations with the gluon propagator and ghost-gluon vertex, using both the continuum and lattice-modified tensor
bases for these functions.~Our aim in the following will be roughly threefold.~The first goal is to show that, for general kinematics, the lattice-adjusted
bases are ``more complete'' than their continuum counterparts.~Details on how this (approximate) completeness is tested will be given shortly.~Our second aim 
is to demonstrate numerically and analytically that there exist such kinematics on the lattice, where the continuum bases describe the examined correlation 
functions without any loss of information.~For the ghost-gluon vertex, the analytic part of this problem was already partly discussed in \ref{sec: ghost_latt},
while the appropriate calculations for the gluon propagator have been postponed since they are more involved.~Our third goal is to show numerically, that the 
lattice-modified tensor bases for these $n$-point functions $(n = 2, 3)$ feature form factors which are invariant under arbitrary permutations and inversions
of momentum components, i.\,e.~that the dressing functions are actual hypercubic invariants.~Since we are mostly concerned with proof-of-principle evaluations 
here, in our numerics we will only consider two- and three-dimensional lattices.~While this obviously does not correspond to the physical situation, it still 
captures many of the essential feautures which should be present in higher-dimensional settings. 

To begin, we shall provide some details on the setup of our Monte Carlo calculations.~We consider equally-sided lattices in two and three dimensions, with
periodic boundary conditions.~The gauge field configurations are thermalised and subsequently updated for measurements using the standard gauge action of
Wilson \cite{Wilson:1974sk}:
\begin{align}\label{eqn: wilson_action}
S = \frac{\beta}{N_c} \sum_\text{plaq} \text{Re}  \left[ \, \text{Tr} \left(\mathbb{1} - U_\text{plaq} \right)\right] \, , 
\end{align}   

\noindent
where $N_c$ is the number of colours (in our case $N_c = 2$), and $U_\text{plaq}$ is the Wilson plaquette operator:
\begin{align}\label{eqn: plaquette}
U_\text{plaq}(x) \, = & \,\,\, U_\mu(x)\,U_\nu(x+\hat{\mu})\,U^\dagger_\mu(x+\hat{\nu})\,U^\dagger_\nu(x) \, .
\end{align} 

The operators $U_\sigma$ in the above equation belong to the $SU(2)$ gauge group, and are parametrised as $U \equiv U_0\,\mathbb{1} + i\,\vec{U}
\cdot\vec{\sigma}$, with $\mathbb{1}$ standing for a $2 \times 2$ unity element, and $\vec{\sigma} = (\sigma^1,\,\sigma^2,\,\sigma^3)$ being the
vector of Pauli matrices.~The coefficients $(U_0, \, \vec{U})$ are real numbers satisfying $U_0^2 + \vec{U}^2 = 1$.~The symbol $\beta$ in \er{eqn:
wilson_action} denotes the bare lattice gauge coupling.  

The gauge field configurations are updated by means of the hybrid-over-relaxation algorithm (HOR), consisting of three over-relaxation \cite{Adler:1981sn, 
Adler:1987ce} and one heat-bath step:~for the heat-bath sweep, we use the Kennedy-Pendleton procedure \cite{Kennedy:1985nu}.~Starting from a cold initial 
guess, we perform 5000 HOR sweeps for thermalisation, while in actual measurements we discard a certain number of updated configurations, to lessen the 
effect of autocorrelations.~Concretely, we perform 1.5\,$N$\,HOR updates before each measurement, with $N$ denoting the linear extent of the lattice in one
direction:~as an example, for lattices with $N=32$, we perform 48 HOR steps prior to measurement.~In the end, we use 9600 configurations for evaluations
of the gluon propagator, and 480 configurations for the ghost-gluon vertex, for each $(N,\beta)$ pair considered in this work.~We obtain estimates for 
statistical errors via an integrated autocorrelation time analysis, according to the automatic windowing procedure outlined in section 3.3 of \cite{Wolff:2003sm},
with parameter $S = 2.5$.~For all of the calculated quantities in this work, the integrated autocorrelation time was always estimated to be less than 0.75 
(we remind that $\tau_\text{int}$ = 0.5 means no autocorrelations), but this might be an underestimation caused by gauge-fixing, which can ``artificially'' 
decrease autocorrelations.

One of the basic quantities needed in the upcoming simulations is the lattice gluon potential $A_\nu$, which is defined in terms of the link variables $U_\nu(x)$
as 
\begin{align}\label{eqn: intro_latt_glue}
A_\nu(x) \, \equiv \, \frac{1}{2} \left[ \, U_\nu(x) - U^\dagger_\nu(x) \, \right] = i \, \vec{U} \cdot \vec{\sigma} \, .
\end{align} 
  
The colour components of $A_\nu(x)$ can be projected out with appropriate Pauli matrices, where one has $A^b_\nu(x) = (1/2i) \cdot \text{Tr}\,[A_\nu(x) \, \sigma^b]$,
with $b = 1, \, 2, \, 3$.~Some other ingredients, needed in calculations of specific lattice correlation functions, will be discussed in due course.~Concerning our 
general numerical setup, there are two remaining issues to clarify.~One is the gauge-fixing procedure:~since we are interested in gauge-dependent quantitites, we have 
to specify a gauge to work in, lest all our Monte Carlo averages end up being zero.~Here, we shall concentrate exclusively on (lattice) Landau gauge, as it is 
computationally by far the easiest one to implement.~Certain other choices will be discussed only briefly in due time.~For gauge-fixing to Landau gauge, we choose the
so-called over-relaxation method \cite{Mandula:1990vs, Cucchieri:1995pn}:~the corresponding iterative steps are explained in detail in e.\,g.~section 3.3 of \cite{
Cucchieri:1995pn}.~The algorithm features a free parameter $\omega \in (1,2)$, which may be tuned to improve convergence.~The ``optimal'' values of $\omega$, for each 
set of considered gauge field configurations, can be found in Table \ref{tab: config_details}.~The gauge-fixing process is stopped when $e_6 \leq 10^{-14}$, where \cite{
Cucchieri:1995pn}:
\begin{align}\label{eqn: e6_def}
e_6 \equiv \frac{1}{3\,N\,d} \sum_{\nu = 1}^d \sum_{b = 1}^3 \sum_{x_\nu = 1}^N \frac{\left[ Q_v(x_v) - \hat{Q}_\nu \right]_b^2}{\left[\hat{Q}_\nu\right]_b^2} \, ,
\end{align}
\noindent
with
\begin{align}\label{eqn: def_q}
Q_\nu(x_\nu) \equiv \sum_{\mu \neq \nu} \sum_{x_\mu} A_\nu(x) \, , \quad \text{and} \quad \hat{Q}_\nu \equiv \frac{1}{N} \sum_{x_\nu = 1}^N Q_\nu(x_\nu) \, .
\end{align}

In \er{eqn: def_q}, the index $\nu$ runs from 1 to $d$, and $A_\nu(x)$ is the gluon potential introduced in \er{eqn: intro_latt_glue}.~Also, the index $b = 1,\,2,
\,3$ in \er{eqn: e6_def} stands for the colour components of the bracketed expressions.~$e_6$ essentially measures the spatial fluctuations of the quantity $Q_\nu$,
defined in \er{eqn: def_q}:~according to \cite{Mandula:1987rh}, the functions $Q_\nu$ should be independent of $x_\nu$, for periodic lattices and for gauge field 
configurations fixed to Landau gauge. 

This brings us to one of the final notions we will need for the upcoming analysis, and this is the vertex reconstruction procedure.~The method is discussed in some
detail in \cite{Vujinovic:2018nqc}, but here we wish to repeat the main ideas.~Vertex reconstruction is a way of quantifying how (un)well some tensor basis describes
a given correlation function.~Suppose that one is working with some generic lattice $n$-point function $\Gamma$, and that one wishes to test if a tensor basis $\tau$,
with the appropriate quantum numbers, describes the correlator $\Gamma$ well.~One approach to doing this would be to assume that the elements $\tau$ form a complete 
basis, and that $\Gamma$ can thus be written as a linear combination of these tensor structures:
\begin{align}\label{eqn: recon_rep}
\Gamma = \sum_{j} \: \mathcal{F}^{\,j} \, \tau^{\,j} \, , 
\end{align}    
\noindent
with $\tau^{j\,}$ denoting the $j$-th basis element, and $\mathcal{F}^{\,j}$ the corresponding form factor.~The first step in the procedure is to calculate the
dressing functions $\mathcal{F}^{\,j}$ of the lattice vertex in the usual way.~One then reconstructs the correlator via equation \er{eqn: recon_rep}, by using
the obtained form factors and the basis $\tau$ itself.~The final part is to compare the reconstructed and the original lattice vertex, in whatever way seems 
appropriate.~The main idea behind this method is that, if the basis $\tau$ is truly complete, then no information will be lost when computing the coefficients
$\mathcal{F}^{\,j}$.~Thus, the original and the reconstructed correlator should exactly match.~Any difference between the two correlation functions suggests that
the structures $\tau$ do not contain the full information about $\Gamma$, and the ``size'' of the difference can be seen as a measure of the (un)suitability of
the basis, for given kinematics.~This strategy will be used to test the (approximate) completeness of tensor bases to be considered in the following. 

\newpage

Concerning the above procedure, there is one more issue of practical importance to be discussed.~Namely, we will look at vertex/propagator functions with Lorentz indices,
and comparing the original and the reconstructed correlator for each value of these indices would be highly impractical for the presentation of results.~To address
this, in our plots we will always give the results for ratios of index-averaged quantities.~In the case of the gluon two-point function, the said ratio would look like  
\begin{align}\label{eqn: recon_ratio}
\mathcal{R} =  \frac{D^\text{\,origo}_{|\left\langle \mu\nu \right\rangle|}}{D^\text{\,recon}_{|\left\langle \mu\nu \right\rangle|}} \equiv \frac{\sum_{\mu} 
\sum_{\nu} | D_{\mu\nu}^\text{\,origo} |}{\sum_{\mu} \sum_{\nu} | D_{\mu\nu}^\text{\, recon} |}.
\end{align}

In the above relation, superscripts ``origo'' and ``recon'' denote, respectively, the original and the reconstructed correlator, while $|.|$ stands for a (complex 
number) absolute value.~The reasons for using the absolute value when evaluating $\mathcal{R}$ are discussed in some detail in section III of \cite{Vujinovic:2018nqc},
and will not be repeated here.~Note that, in all these proceedings, the original correlation functions (``origo'') are the only ones for wich statistical uncertainties 
are calculated directly, by means of the aforementioned integrated autocorrelation time analysis.~For all the other quantities, like the reconstructed correlators and
ratios akin to \er{eqn: recon_ratio}, the corresponding errors are estimated from those of the original function, via error propagation.~Regarding the propagation
of uncertainty itself, we always consider only the leading-order (variance) formulas, meaning that all of the involved variables are treated as if being statistically
independent from each other.~With these important computational details clarified, we may finally proceed towards some actual results.    

\subsection{Gluon propagator in two dimensions}\label{sec: 2d_glue}

In lattice Monte Carlo simulations, the gluon two-point function can be calculated as
\begin{align}\label{eqn: d_propagator}
D^{\,ab}_{\mu\nu}(p) = \frac{1}{V} \left\langle \tilde{A}^{a}_\mu(p) \,  \tilde{A}^{b}_\nu(-p) \right\rangle \, , 
\end{align} 

\noindent 
with $V = N^d$ being the lattice volume, and $\tilde{A}^{a}_\mu(p)$ is the Fourier tranform of $A^{a}_\mu(x)$:
\begin{align}\label{eqn: fourier_glue}
\tilde{A}^a_\mu(p) & \equiv  \sum_x \, A^a_\mu(x) \, \exp \left[ 2\pi i (p\cdot x + p_\mu/2) \right] \, , \quad \text{where} \nonumber \\  
& p_\mu \equiv \frac{2\pi \, n_\mu}{aN} \, , \quad n_\mu \in [0, \, N-1] \, .
\end{align}

Note that all of the momenta in our plots and text will be given in terms of the vector $n_\mu$ defined above, with one exception:~components lying exactly half-way on 
the lattice sides (corresponding to $p_\mu = \pi$) will be written in the text as `$\pi$'.~One may also observe that the equation \er{eqn: fourier_glue} contains an 
additional term $p_\mu/2$, as opposed to the standard definition of a discrete Fourier transform:~the purpose of this modification is to make the lattice gluon potential
obey the continuum Landau gauge condition with $\mathcal{O}(a^2)$ corrections, instead of $\mathcal{O}(a)$ ones \cite{Alles:1996ka}.~With lattice gauge field configurations
fixed to Landau gauge, and the Fourier transform of the gluon potential defined according to \er{eqn: fourier_glue}, the gluon propagator of Monte Carlo simulations should 
have the following colour and tensor structure \cite{Alles:1996ka}: 
\begin{align}\label{eqn: gauge_tensor}
D^{\,ab}_{\mu\nu}(p) = \left( \delta_{\mu\nu} - \frac{\hat{p}_\mu \, \hat{p}_\nu}{\hat{p}^2} \right) \, \delta^{ab} D(p) \, , 
\end{align}   

\noindent
with the lattice vector $\hat{p} = 2\, \sin(p/2)$.~Henceforth, we shall assume that this function is diagonal in colour space, as indicated above, and will work only with 
colour-averaged quantities $D_{\mu\nu} = \frac{1}{3}\sum_{a} D^{aa}_{\mu\nu}$.~The tensor representation \er{eqn: gauge_tensor} will not be used for vertex reconstruction 
in the upcoming analysis, but it should still be kept in mind since many of the results we will obtain can only be properly understood with the help of \er{eqn: 
gauge_tensor}.~Also, for comparison purposes, we will plot the results for the dressing function $D(p)$ of \er{eqn: gauge_tensor}, which is easily evaluated in $d$ dimensions 
as 
\begin{align}\label{eqn: dressing_d}
D(p) = \frac{1}{d - 1} \sum_{\mu = 1}^d \, D_{\mu\mu}(p) \, .
\end{align} 

The above formula does not apply for vanishing momentum $p$, but since the case $p = 0$ will not be considered in our numerics, this is of 
no concern to us.~This finally brings us to the two tensor representations to be actively explored in this and the next section:~we shall 
repeat the corresponding definitions for convenience.~The first is the continuum parametrisation for the gluon propagator, given by 
\begin{align}\label{eqn: cont_glue}
D_{\mu\nu}(p) = A(p) \, \delta_{\mu\nu} + B(p) \, p_\mu p_\nu ,
\end{align} 
   
\noindent
with the appropriate projectors (assuming $p \neq 0$ and a $d$-dimensional space):
\begin{align}\label{eqn: cont_project_glue}
P^{\,A}_{\mu\nu} = \frac{1}{d-1}\left(\delta_{\mu\nu} \: - \: \frac{p_\mu \, p_\nu}{p^2}\right) \, , \qquad \qquad
P^{\,B}_{\mu\nu} = \frac{1}{d-1}\left(-\frac{\delta_{\mu\nu}}{p^2} \: + \: \frac{ d \, p_\mu \, p_\nu}{(p^2)^2}\right) \, .
\end{align}

The above projectors are constructed explicitly in Appendix \ref{sec: projectors}.~Note that $P^A_{\mu\nu}$ is the standard transverse projector in $d$ 
dimensions.~The second basis to be scrutinised in detail is the lattice-modified version of \er{eqn: cont_glue}, with 
\begin{align}\label{eqn: lattice_glue}         
&D_{\mu\mu} = E(p)\, \delta_{\mu\mu} \: + \: F(p) \, p_\mu^{\,2} \, , \qquad \quad \mu = 1, \ldots d \nonumber \\ 
&D_{\nu\mu} = G(p) \, p_\nu p_\mu \, , \qquad \qquad \qquad \quad \: \:  \mu, \, \, \nu = 1, \ldots d \, , \qquad \mu \neq \nu \, , 
\end{align}  

The dressing functions of the above decomposition can be calculated in $d$ dimensions as [\,equations \er{eqn: diag_ffs} and \er{eqn: off_diag_ff} of Appendix \ref{sec: 
projectors}\,]\,:
\begin{align}\label{eqn: latt_dress_glue}
&E(p) \: = \: \frac{ p^{\,[4]} \sum_{\mu} D_{\mu\mu} \:\: - \:\: p^2 \sum_{\mu} p_\mu^2 \, D_{\mu\mu} }{d \, p^{\,[4]} \, - \, (p^2)^2} \, , \nonumber \\[0.2cm] 
&F(p) \: = \: \frac{ - p^2 \sum_{\mu} D_{\mu\mu} \:\: + \:\: d \sum_{\mu} p_\mu^2 \, D_{\mu\mu} }{d \, p^{\,[4]} \, - \, (p^2)^2} \, , \nonumber \\[0.2cm]
&G(p) \: = \: \frac{ \sum_{\substack{\mu, \nu \\ \mu \neq \nu}} \: p_\nu p_\mu  D_{\mu\nu}}{(p^2)^2 \, - \, p^{\,[4]}} \, .
\end{align}

\begin{figure}[!t]
\begin{center}
\graph[width = 0.41\tew]{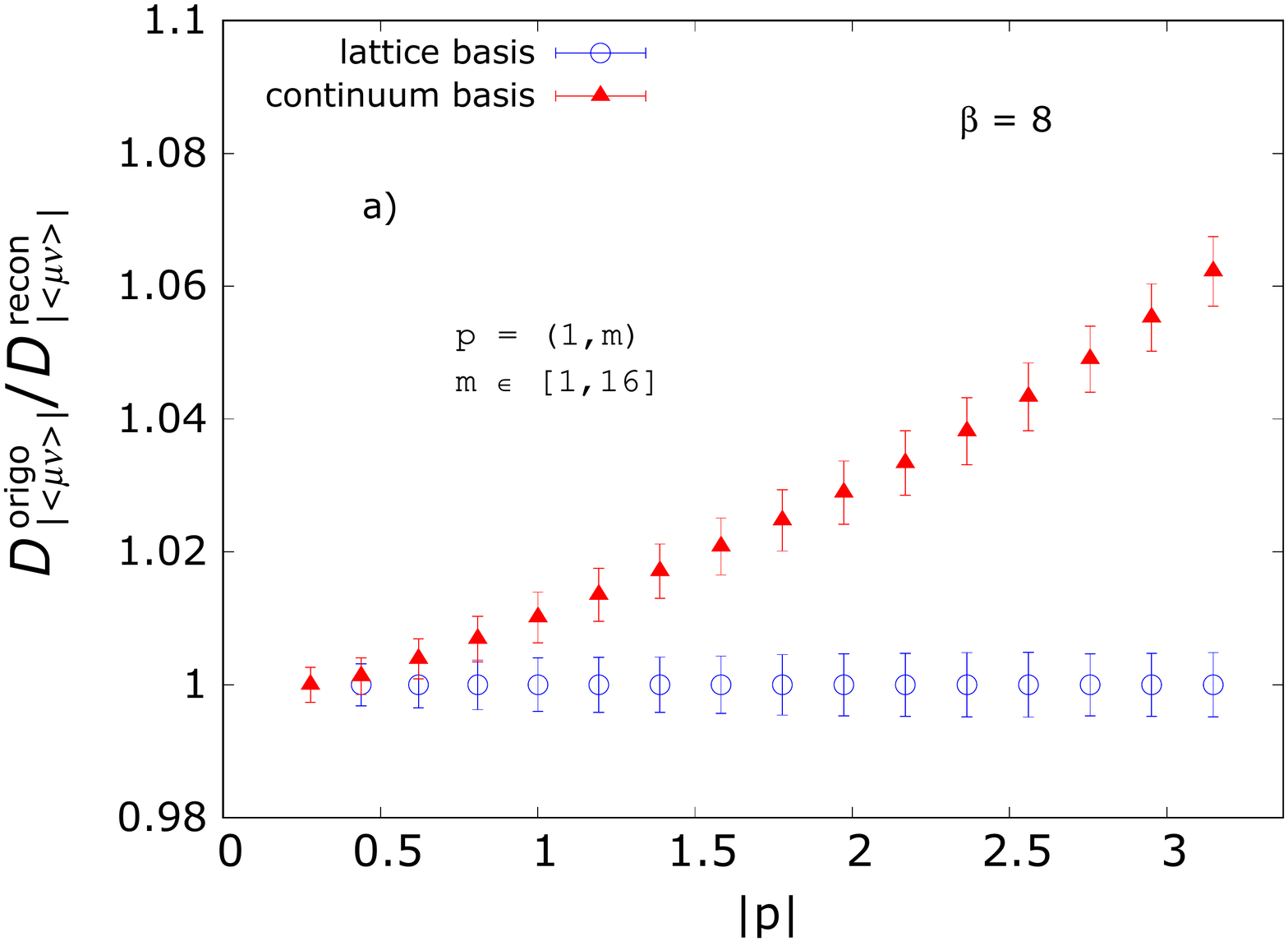}\graph[width = 0.41\tew]{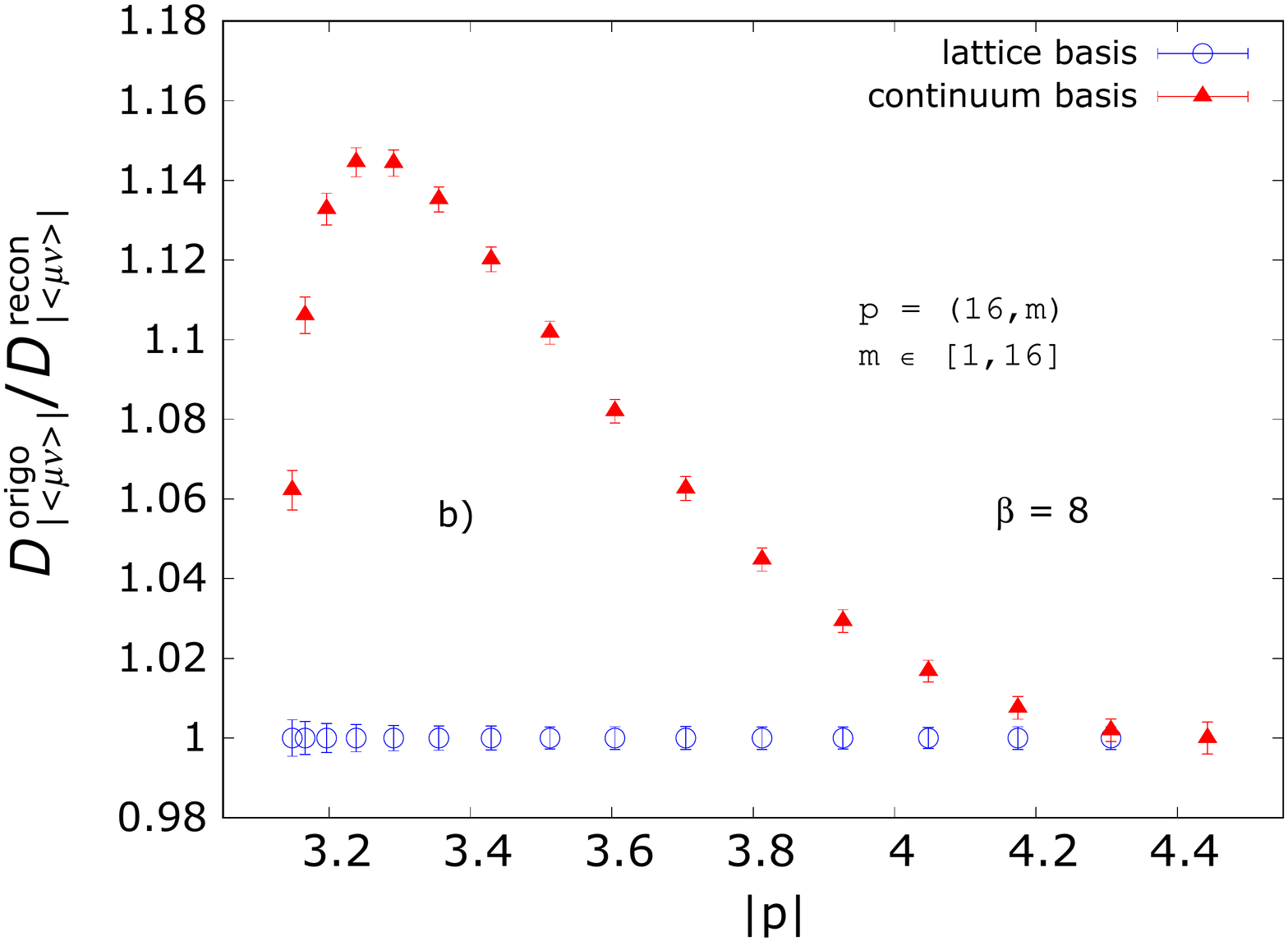} 
\graph[width = 0.41\tew]{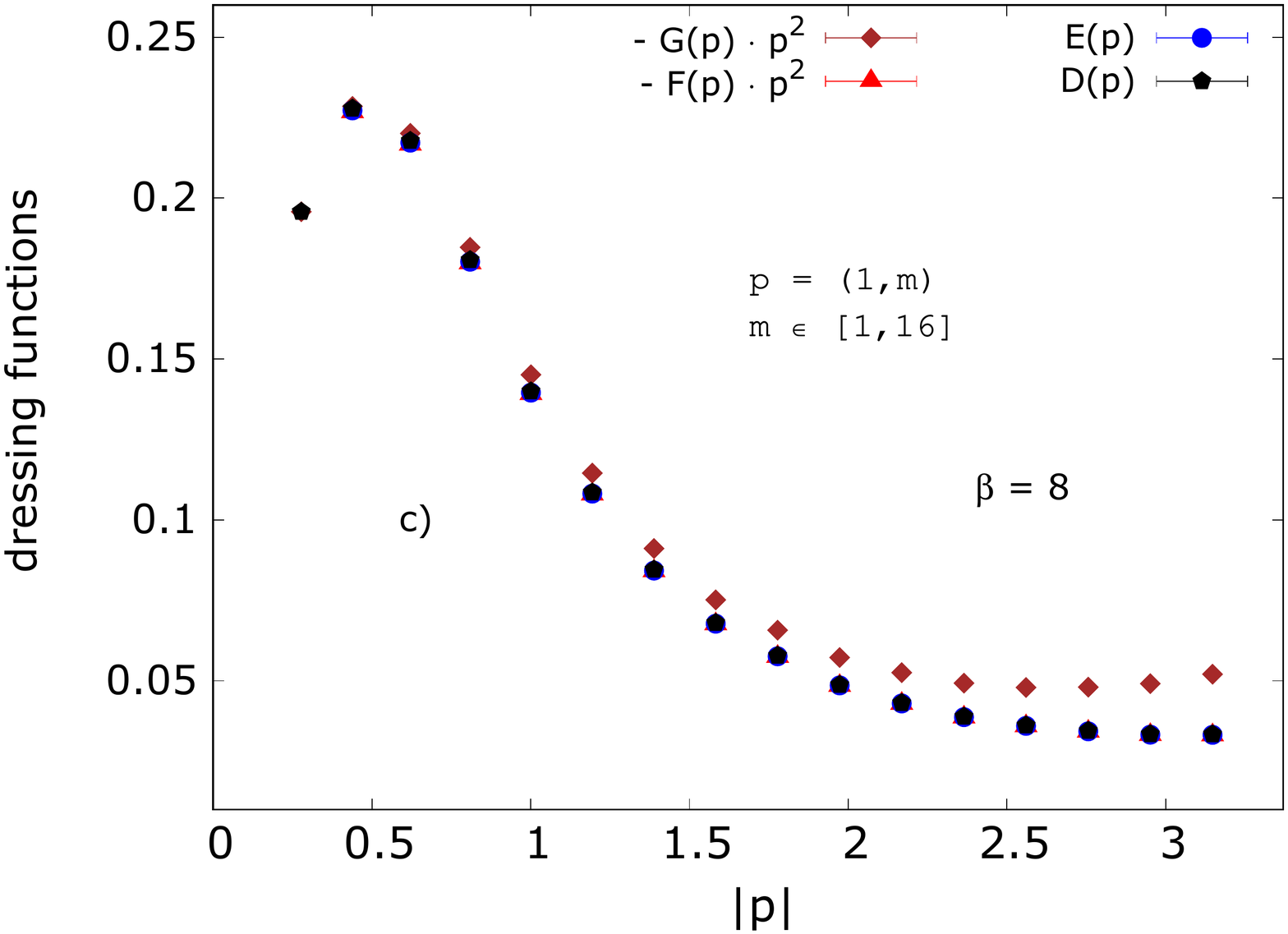}\graph[width = 0.41\tew]{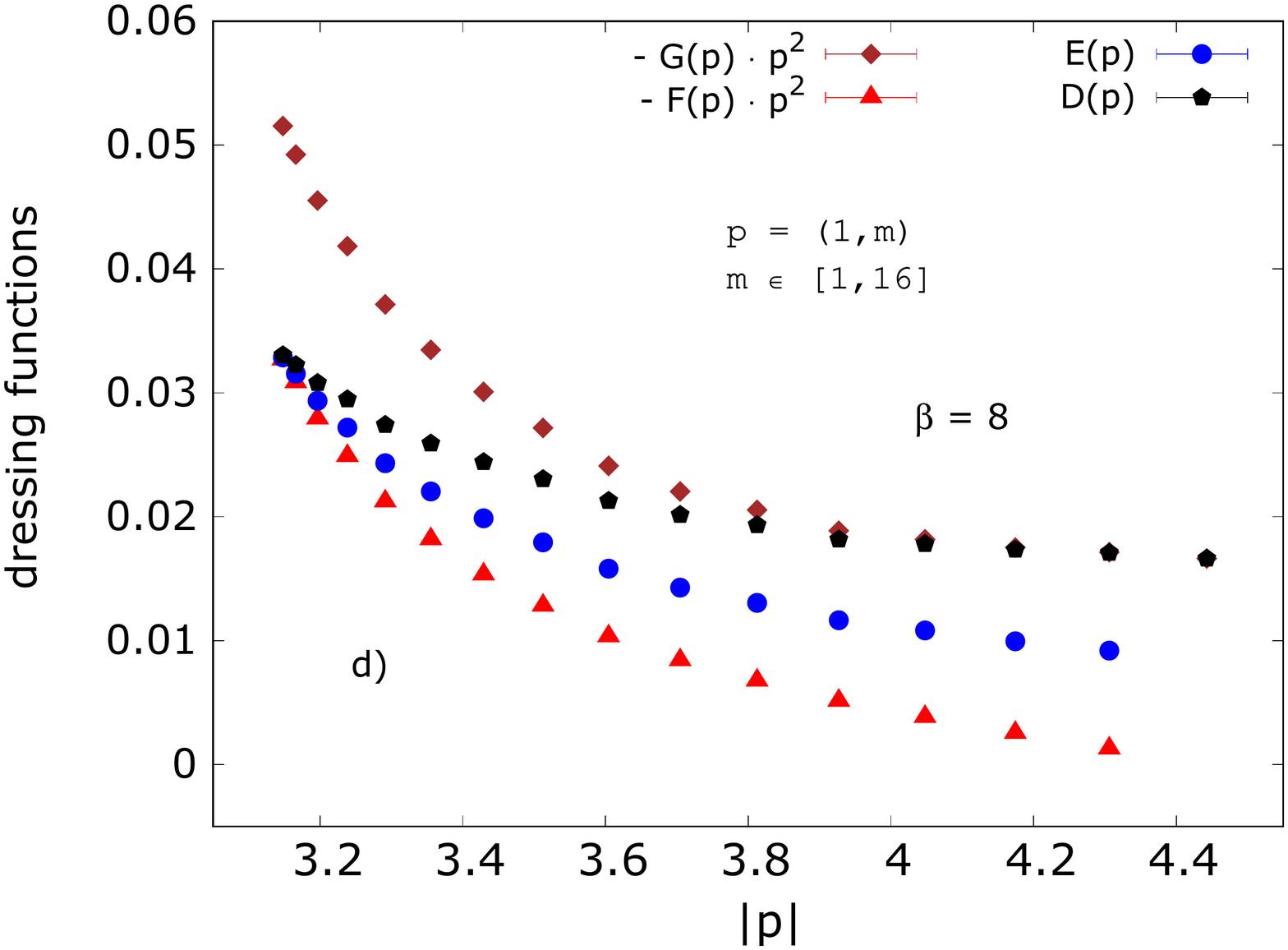}
\caption{Upper panel:~Results of vertex reconstruction on a $32^2$ lattice, according to continuum [\,equations \er{eqn: cont_glue} and \er{eqn: 
cont_project_glue}\,]  and lattice [\,equations \er{eqn: lattice_glue} and \er{eqn: latt_dress_glue}\,] decompositions.~Lower panel:~Data for form
factors of \er{eqn: dressing_d} and \er{eqn: latt_dress_glue}.~Results are plotted as functions of $|p| = \sqrt{p^2}$ (in lattice units), with momenta
defined in terms of vector $n_\mu$ in \er{eqn: fourier_glue}.~$\beta$ is the bare coupling of \er{eqn: wilson_action}.} 
\label{fig: 2d_glue}
\end{center}
\end{figure}

In the above expressions, all of the sums run from 1 to $d$ [\,with the appropriate restriction $\mu \neq \nu$ in the case of the function $G(p)$\,], and $
p^{\,[4]}$ is a hypercubic invariant $p^{\,[4]} = \sum_{\mu = 1}^{d} \, p_\mu^{\,4}$.~Vertex reconstruction results according to the basis descriptions of
\er{eqn: cont_glue} and \er{eqn: lattice_glue} are given in Figure \ref{fig: 2d_glue}, for the gluon propagator of two-dimensional lattice Monte Carlo 
simulations.~In the same Figure we show the data for the dressing functions of \er{eqn: dressing_d} and \er{eqn: latt_dress_glue}.~More accurately, instead 
of form factors $F(p)$ and $G(p)$ in \er{eqn: latt_dress_glue}, in the plots we provide the results for functions $-p^2 \cdot F(p)$ and $-p^2 \cdot G(p)
$\,:~the reason for this choice will become clear shortly.

Let us first discuss the data points for vertex reconstruction.~As might be expected, use of the continuum basis \er{eqn: cont_glue} leads to appreciable
differences between the reconstructed and the original propagator, with the deviations peaking at about 15 percent, for the considered kinematics.~On the
other hand, within statistical uncertainties there are no discrepancies present for the lattice-modified basis of \er{eqn: lattice_glue}.~This is in accord
with the arguments made towards the end of section \ref{sec: gluon_lattice}, wherein we claimed that the basis \er{eqn: lattice_glue} is complete, in a 
two-dimensional setting.~Some further analytic calculations that support this claim can be found in Appendix \ref{sec: projectors}.~From Figure \ref{fig: 
2d_glue}\,b) one also notes that the diagonal momentum point, corresponding to $p = (\pi, \pi)$, is somewhat special as the continuum decomposition describes 
the lattice correlator fully.~However, we do not yet want to elaborate on diagonal kinematics in detail, and instead turn our attention to the results of Figure 
\ref{fig: 2d_glue}\,c). 

The arguably most interesting thing about the said Figure is that the displayed data points for functions $D(p)$, $E(p)$ and $-p^2 \cdot F(p)$ seem to lie on top
of each other, i.\,e.~these functions seem to have the same values.~The momenta examined in the plot are all of the form $p = (1,m)$ [\,with $m \in (1,16) $\,], 
which is ``very close'' to the kinematic choice $p = (0,m)$.~For momentum vectors of the kind $p = (0,m)$, with non-vanishing $m$, one can easily demonstrate that
the exact equalities $D(p) = E(p) = -p^2 \cdot F(p)$ hold.~For instance, by plugging the vector $p = (0,m)$ into the function $E(p)$ of \er{eqn: latt_dress_glue} 
(with $d = 2$), one gets  
\begin{align}\label{eqn: e_axis_one}
E(p) = \frac{m^4\cdot(D_{11} + D_{22}) - m^4 \, D_{22}}{2 \, m^4 - m^4} = \frac{ m^4 \, D_{11}}{m^4} = D_{11} \, .
\end{align}

To fully evaluate the above expression, we turn to the relation \er{eqn: gauge_tensor}.~For on-axis momentum $p = (0,m)$, the representation \er{eqn: gauge_tensor} 
states that
\begin{align}\label{eqn: e_axis_two}
D_{11} = \left( \delta_{\,11} - \frac{ \sin^2(0) }{\hat{m}^2} \right) \cdot D(p) = D(p) \, ,
\end{align}

\noindent
where $\hat{m} = 2 \, \sin(m/2)$.~Combining the equations \er{eqn: e_axis_one} and \er{eqn: e_axis_two} gives $E(p) = D(p)$.~In the same manner, one can show that
the relation $D(p) = -p^2 \cdot F(p)$ holds, for on-axis momentum $p$.~In Figure \ref{fig: 2d_glue}\,c), we purposefully do \textit{not} look at the situation $p =
(0,m)$, choosing instead the kinematic points $p = (1,m)$.~This is because we wanted to be able to include also the form factor $G(p)$ of \er{eqn: latt_dress_glue} 
in the same graph.~Namely, for kinematic configurations like $p = (0,m)$, the function $G(p)$ evaluates to an ambiguous expression ``0/0'', or explicitly
\begin{align}\label{eqn: off_diag_axis}
G(p) = \frac{ p_1 \, p_2 \cdot (D_{12} + D_{21}) }{m^4 - m^4} = \frac{0}{0} \, ,
\end{align}

\noindent
wherein we used the fact that $p_1 \cdot p_2 = 0$, if $p = (p_1, p_2) = (0,m)$.~The dressing $G(p)$ is indeterminate here because the off-diagonal part of the propagator vanishes 
for on-axis momentum $p$, since $p_\mu \, p_\nu = 0$ (if $\mu \neq \nu$) .~With the help of the representation \er{eqn: gauge_tensor}, these results [\,the equalities $D(p) = E(p) =
 -p^2 \cdot F(p)$, ill-defined function $G(p)$\,] can be extended to any $d$-dimensional vectors with only a single non-zero component.~Put differently, for on-axis momenta, and for 
lattices of arbitrary  dimension, the full information about the gluon correlator is contained in its diagonal part $D_{\mu\mu}$, as the off-diagonal terms are anyway zero.~This 
reasoning can be taken even further, to an eventual conclusion that the on-axis propagator should be described fully with a continuum basis of \er{eqn: cont_glue}.~We will discuss
this last point in detail in the next section, where we analyse the gluon two-point function on a three-dimensional lattice.  

From data in Figure \ref{fig: 2d_glue}\,c) it may also be observed that, as one goes deeper into the infrared (IR) energy region, the off-diagonal dressing $-p^2 \cdot G(p)$ becomes
almost equal to the diagonal form factors $E(p)$ and $-p^2 \cdot F(p)$.~This is what one would expect, because it means that the continuum (Landau gauge) form of the propagator
is recovered at low energies.~However, for the decomposition \er{eqn: lattice_glue} it is not so obvious why the relations like $F(p) \approx G(p)$ should hold at small momentum 
values.~To fully explore this issue, we first need to discuss diagonal lattice kinematics, with non-zero momenta of the kind $p = (m, m)$.  

In terms of the representation \er{eqn: lattice_glue}, diagonal kinematics are special in two ways.~First, the dressing functions $E(p)$ and $F(p)$ take an ill-defined
form ``0/0'' in such cases:~this is shown in Appendix \ref{sec: projectors}, for an arbitrary number of dimensions.~This ambiguity in the definitions of $E(p)$ and $F(p)$ is 
the reason that some results for the basis \er{eqn: lattice_glue} are missing in Figure \ref{fig: 2d_glue}.~The issue has to do with linear dependence of basis elements:~for
diagonal momenta $p = (m,m)$ it holds that $p^{\,2}_\mu = m^2 \, \delta_{\mu\mu} \, (\mu = 1, 2)$, and so the tensor structures of the propagator $D_{\mu\mu}$ become degenerate.
\!This suggests that a reduced tensor description is needed for such momentum points.~The second interesting feature concerning diagonal kinematics is the fact that the 
off-diagonal form factor $- p^2 \cdot G(p)$ becomes equal to $D(p)$.~To see this, one may put the momentum $p = (m,m)$ into the definition of $G(p)$ in \er{eqn: latt_dress_glue},
and get  
\begin{align}\label{eqn: glue_diag}
G(p) = \frac{m^2 \cdot (D_{12} + D_{21})}{ 4 \, m^4 - 2 \, m^4} = \frac{- m^2 \cdot 2 \, \hat{m}^2 D(p) }{(2 \, m^4) \cdot (2 \, \hat{m}^2)} = \frac{-D(p)}{2 \, m^2} \, .
\end{align}

In obtaining \er{eqn: glue_diag}, we again used the parametrisation \er{eqn: gauge_tensor} for gluon components $D_{21} ~ \text{and} ~ D_{12}$, with $p = (m,m)$.~From 
the above result it quickly follows that $-p^2 \cdot G(p) = -2 \,m^2 \cdot G(p) = D(p)$.~With the help of \er{eqn: gauge_tensor}, this argument can be generalised to 
diagonal momenta $p$ of arbitrary dimension.~The behaviour $-p^2 \cdot G(p) \rightarrow D(p)$ can also be seen in Figure \ref{fig: 2d_glue}\,d), as one approaches the
rightmost point $p = (\pi, \pi)$.~Along the lattice diagonal, it thus holds that the form factors $E(p)$ and $F(p)$ are ill-defined, whereas the off-diagonal dressing 
$G(p)$ is proportional to the coefficient function $D(p)$.~This all implies that the continuum tensor description of the gluon propagator should suffice, which is confirmed 
numerically in Figure \ref{fig: 2d_glue}\,b).

We may now tackle the question on why the approximate equalities like $ F(p) \approx G(p) $ hold in the infrared region.~For this we will take a look at a specific IR momentum 
point, namely the kinematic choice $v = (1,1)$.~In a sense, this vector is doubly exceptional.~First, it is an example of diagonal kinematics, meaning that the relation $- 
v^2 \cdot G(v) = D(v)$ must hold exactly, as shown in \er{eqn: glue_diag}.~Second, $v$ is kinematically close to the on-axis point $ p = (0,1)$, for which one has the exact 
relations $ - p^2 \cdot F(p) = E(p)= D(p) $, as exemplified in \er{eqn: e_axis_one} and \er{eqn: e_axis_two}.~Putting these two tendencies together, one sees that for any 
points $k$ in the vicinity of $v = (1,1)$, the approximate equalities $ - k^2 \cdot F(k) \approx E(k) \approx -k^2 \cdot G(k)$ should hold, indicating the recovery of the
propagators continuum form.~Note that the coarseness of the lattice plays a central role here.~On very coarse lattices, with only a few momentum points in each direction, 
the diagonal vector $v= (1,1)$ is ``far away'' from the on-axis one $p = (0,1)$, and there is no reason to expect that the above relations should hold even approximately at 
the ``infrared'' energies. 

In most of the above arguments, the representation \er{eqn: gauge_tensor} played a crucial role:~without it, it is hard to imagine how the results of Figure \ref{fig:
2d_glue} could be explained analytically.~Nonetheless, at least some of the observations made here should hold regardless of \er{eqn: gauge_tensor}.~For instance, the 
applicability of the continuum basis along the lattice diagonal should follow solely from the fact that the description \er{eqn: lattice_glue} is redundant, if $p = (m,m)
$.~Also, the approximate equalities $ F(p) \approx G(p)$ ought to be true in the infrared region, without any reference to \er{eqn: gauge_tensor}, since one expects that 
the lattice tensor decomposition reduces to the continuum form at low momenta.~It would thus be interesting to see how some of these results hold up for second-rank lattice
tensors, whose basis elements are not determined (at least not fully) by gauge-fixing.~At the moment, we are not aware of any correlators which would constitute suitable 
candidates for such an investigation. 

To conclude this section, we want to comment on how the tensor representation \er{eqn: lattice_glue} may be used to test some of the continuum extrapolation methods.~We
know that in the continuum, the exact relations like $ - p^2 \cdot F(p) = E(p)$ ought to hold for arbitrary momentum $p$, and not just in the infrared.~It could thus be 
potentially useful to check if on a lattice, certain extrapolation methods can bring about the expected continuum behaviour(s) even at relatively high values of $p^2$.~This
would constitute one of the most direct possible tests of how successful some of these methods actually are, at least for the case of gluon two-point function.~In fact, if
one wanted \textit{only} to test such approaches, then there would be no need to consider actual Monte Carlo simulations, since it should be enough to look at (say) the gluon 
propagator of lattice perturbation theory.~This would make the corresponding calculations numerically far cheaper, and there would be virtually no restrictions on lattice 
sizes and the amount of data one can collect, to perform the said extrapolations with a desired accuracy.~We postpone such endeavours for future studies. 

\subsection{Gluon propagator in three dimensions}

\begin{figure}[!t]
\begin{center}
\graph[width = 0.40\tew]{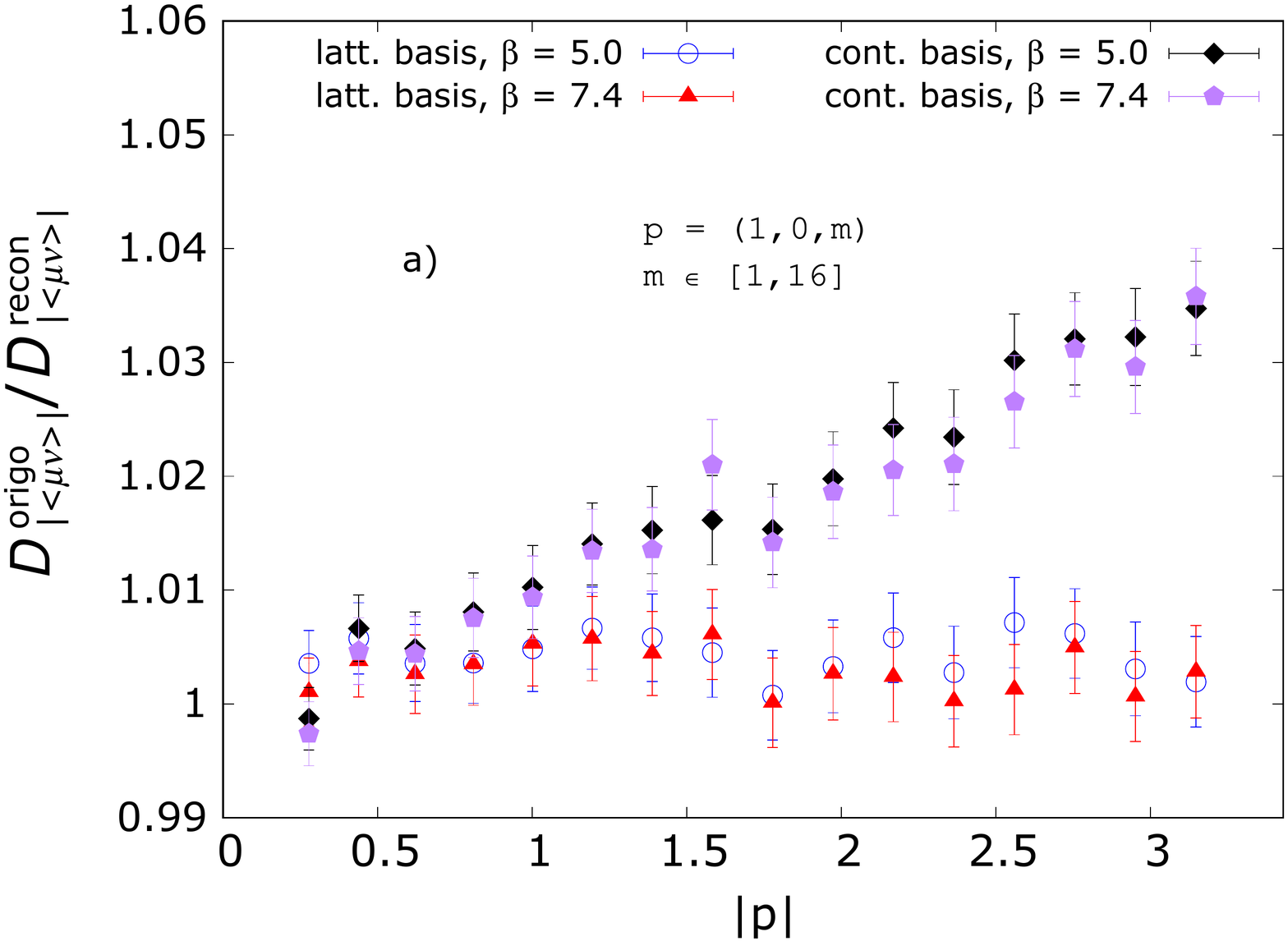}\graph[width = 0.40\tew]{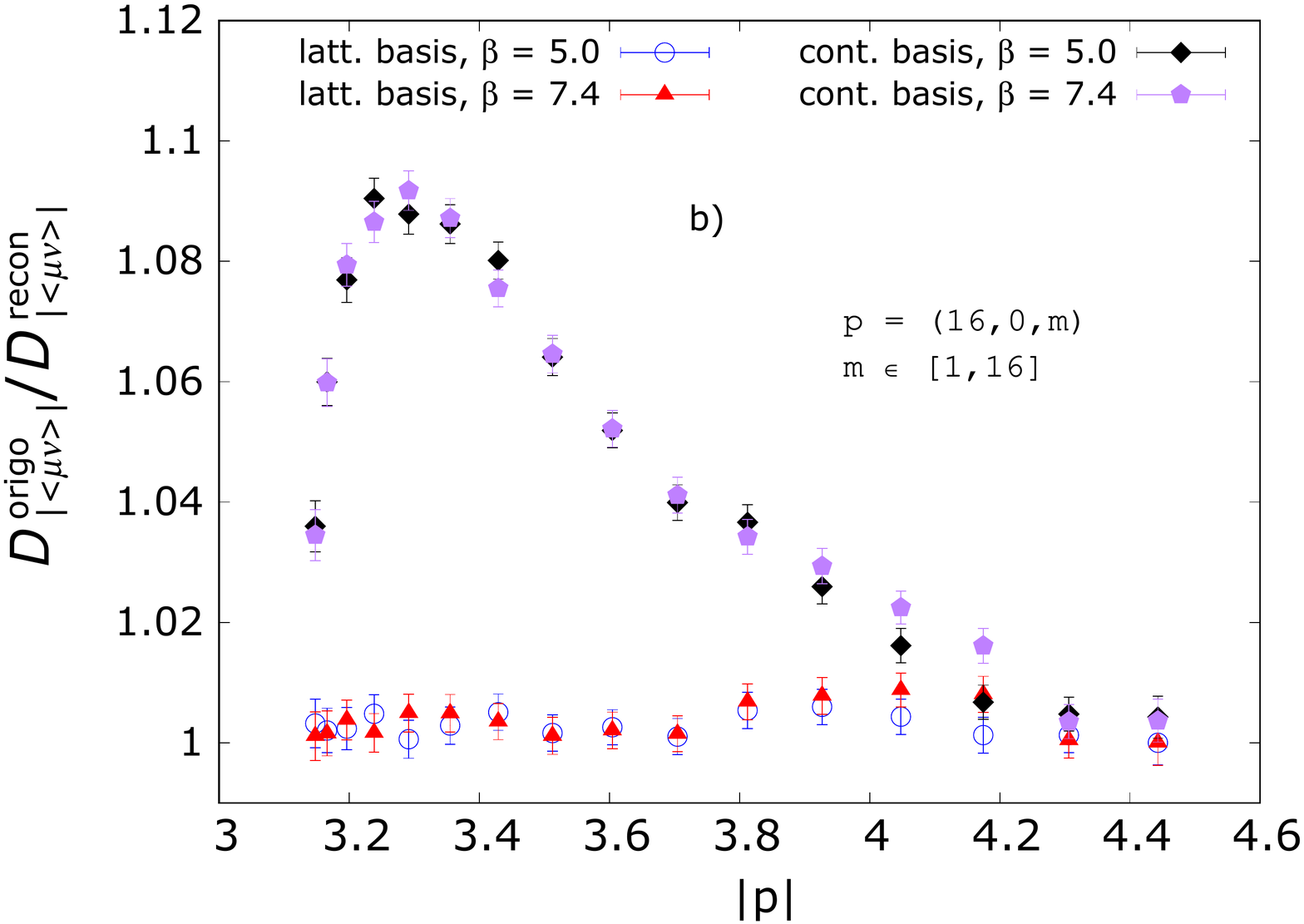} 
\graph[width = 0.40\tew]{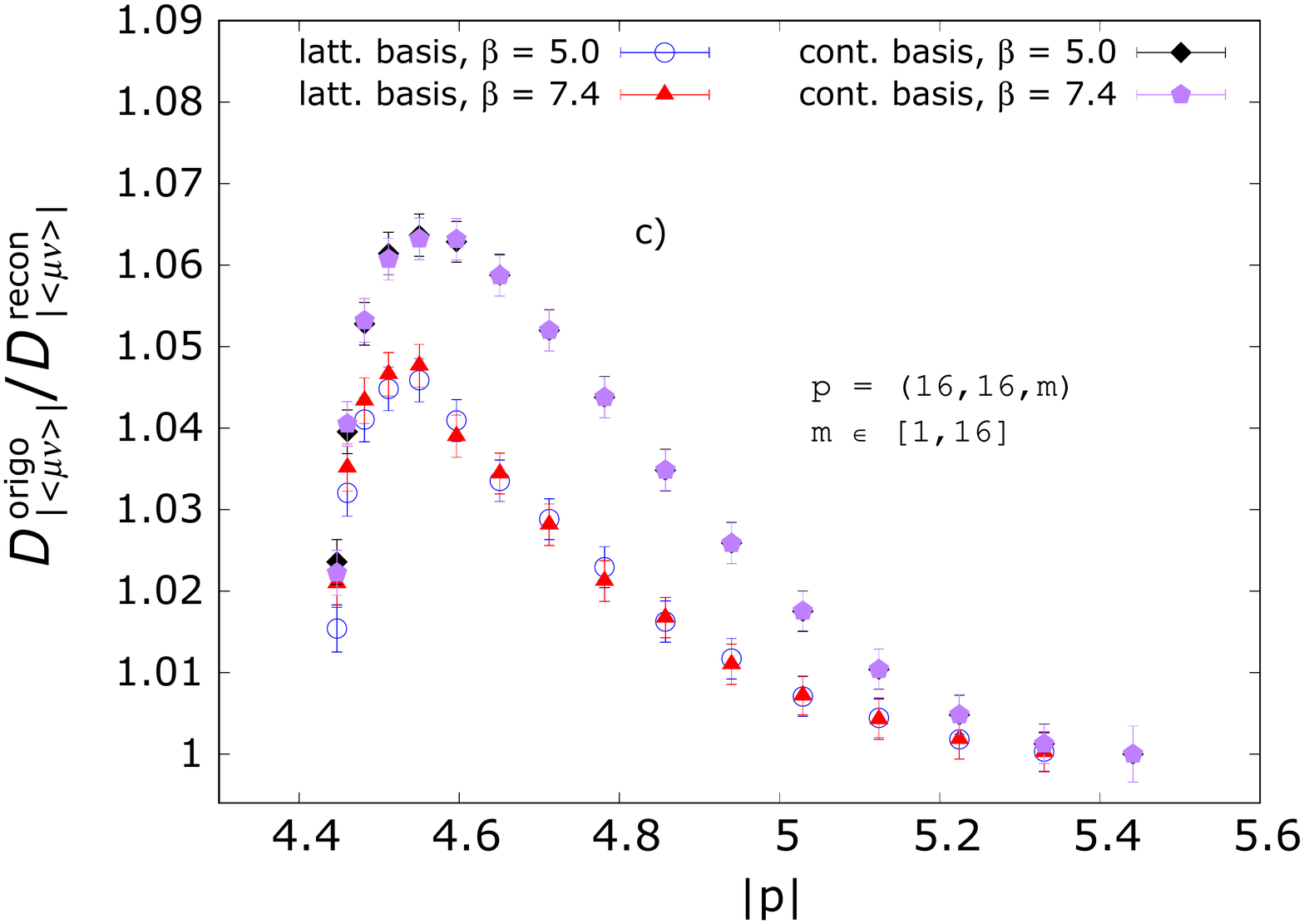}\graph[width = 0.40\tew]{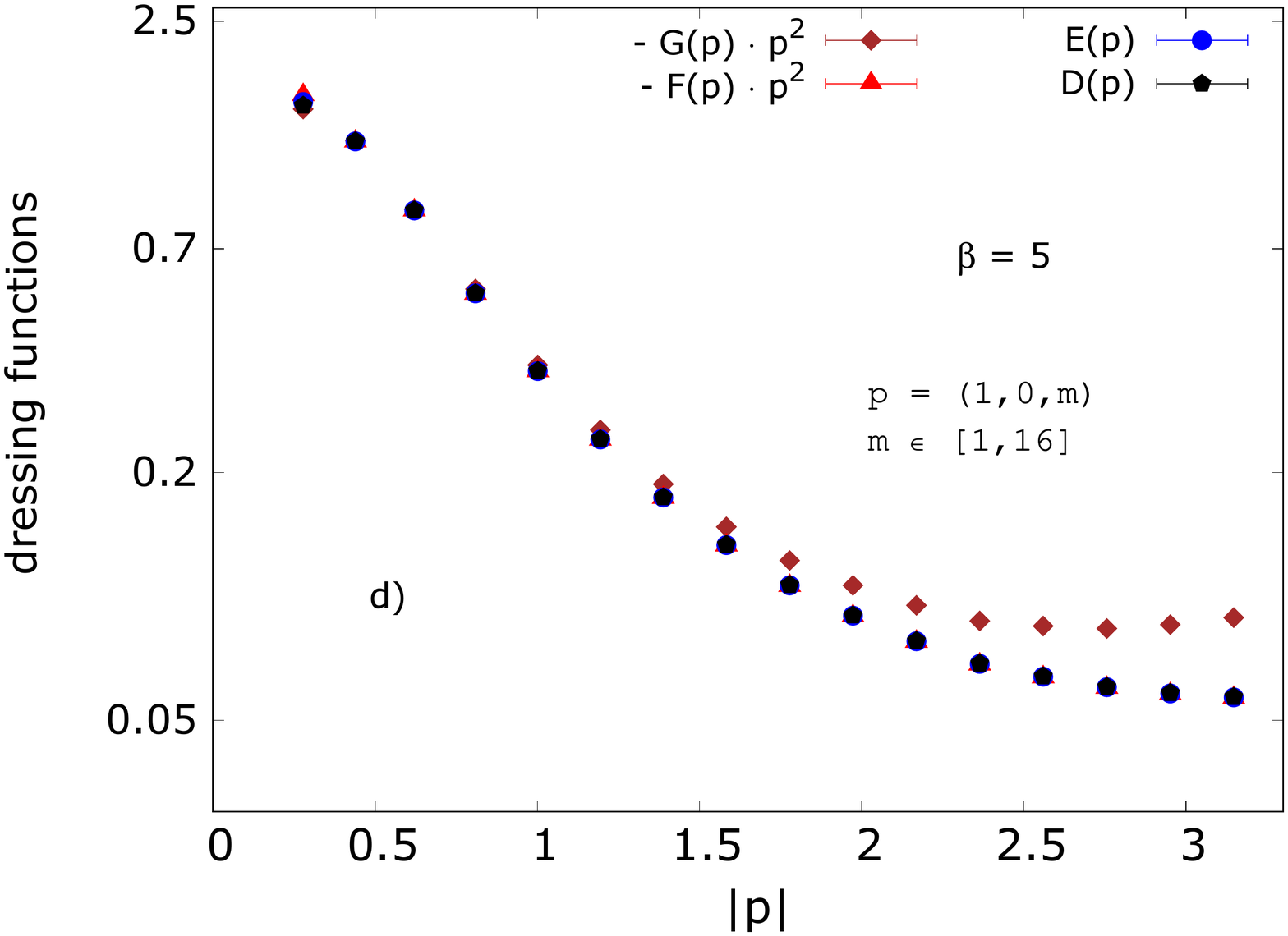}
\graph[width = 0.40\tew]{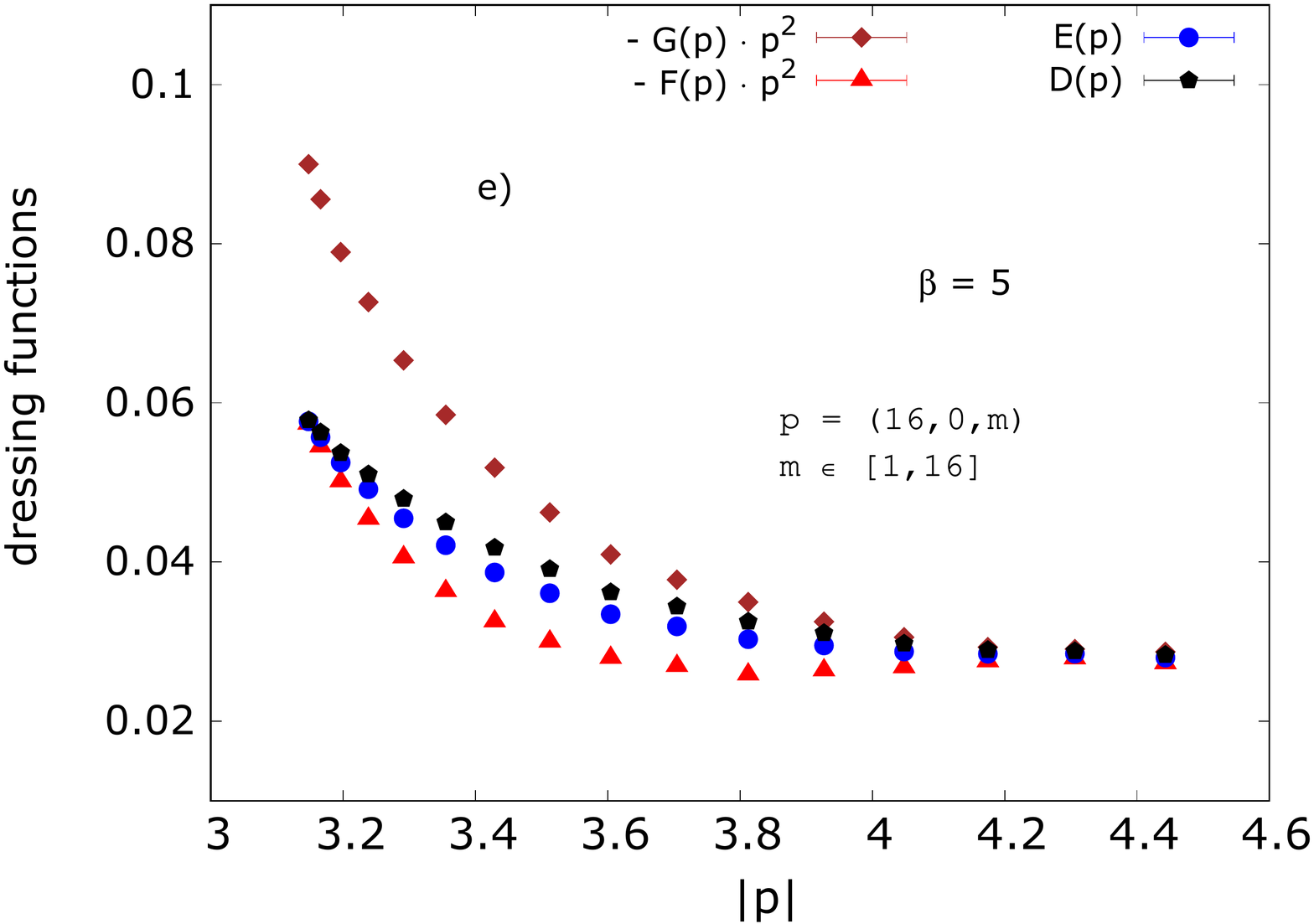}\graph[width = 0.40\tew]{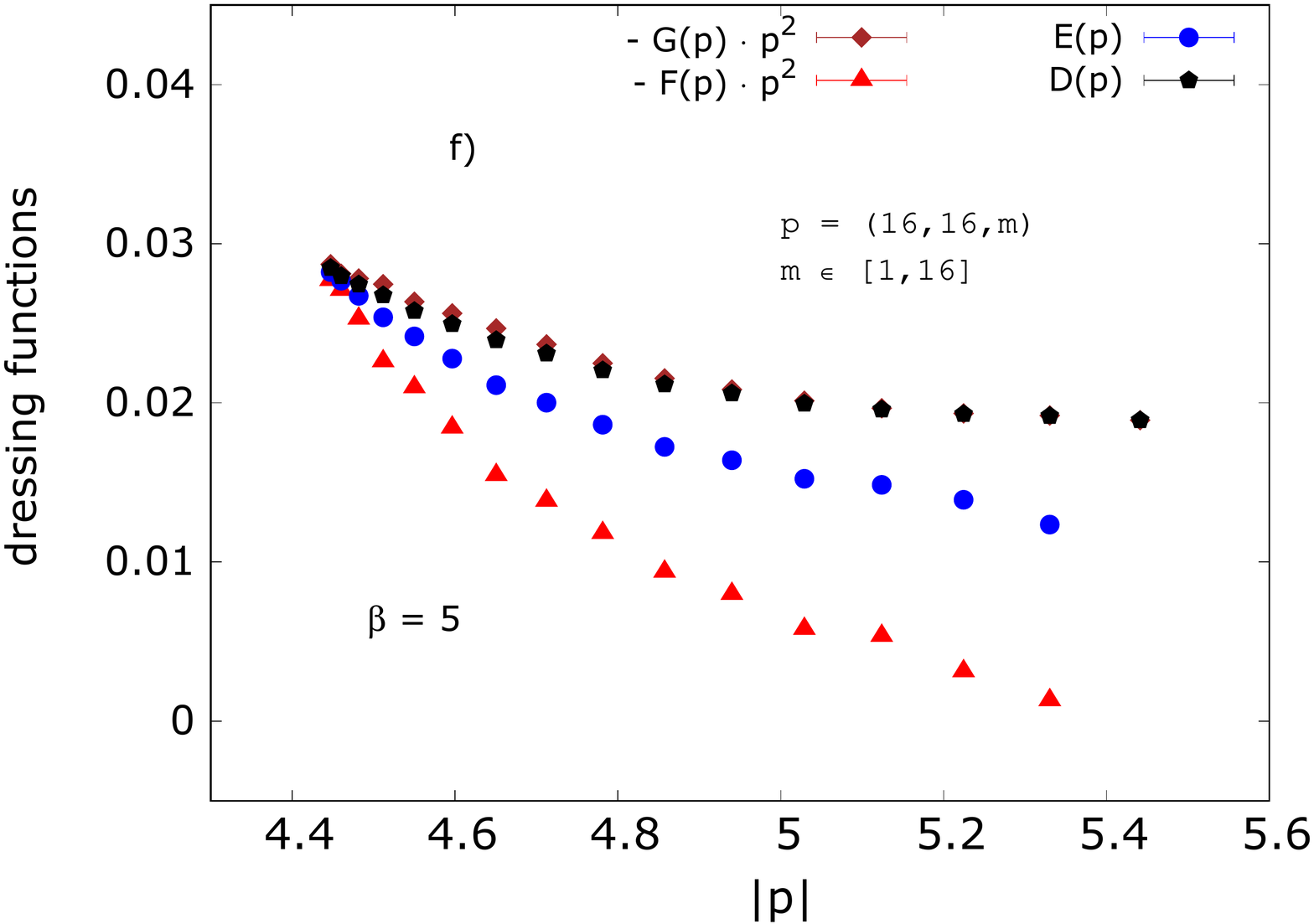}
\caption{Plots a) to c):~Results of vertex reconstruction on a $32^3$ lattice, according to continuum [\,equations \er{eqn: cont_glue} and \er{eqn: 
cont_project_glue}\,]  and lattice [\,equations \er{eqn: lattice_glue} and \er{eqn: latt_dress_glue}\,] decompositions.~Plots d) to f):~Data for form
factors of \er{eqn: dressing_d} and \er{eqn: latt_dress_glue}.~Results are plotted as functions of $|p| = \sqrt{p^2}$ (in lattice units), with momenta
defined in terms of vector $n_\mu$ in \er{eqn: fourier_glue}.~Note the logarithmic $y$ scale in plot d), see text for comments.~$\beta$ is the bare 
coupling of \er{eqn: wilson_action}.} 
\label{fig: 3d_big_glue}
\end{center}
\end{figure}

Some of our results for the gluon propagator on a three-dimensional lattice are given in Figure \ref{fig: 3d_big_glue}.~Concretely, in plots a) through c) we provide
the data regarding the propagator reconstruction at certain kinematic points, using the tensor bases of \er{eqn: cont_glue} and \er{eqn: lattice_glue}.~In graphs d) 
though f), one can find the results for the dressing functions of \er{eqn: dressing_d} and \er{eqn: latt_dress_glue}.~We note that the reconstruction results are given 
for two values of the parameter $\beta$ of \er{eqn: wilson_action}, whereas for correlator dressings only one gauge coupling value is considered, to prevent the plots
from getting too cluttered.~Apart from a noisier signal in the case of three dimensions, there are quite a few similarities with the two-dimensional situation.~For instance,
for the near-axis momentum $p = (1, 0, m) ~\, [\,\text{with} ~ m \in (1,16)\,] $, one can see the same general tendencies as for the corresponding vector $p =(1,m)$ in
two dimensions, both in terms of correlator reconstruction [\,compare graphs \ref{fig: 2d_glue}\,a) and \ref{fig: 3d_big_glue}\,a)\,] and the corresponding form factors 
[\,compare plots \ref{fig: 2d_glue}\,c) and \ref{fig: 3d_big_glue}\,d)\,].~Note that in Figure \ref{fig: 3d_big_glue}\,d), we use a logarithmic scale for the $y$ axis, 
as otherwise it would be very hard to make out the details at higher momentum values.

The above similarities notwithstanding, the case $d = 3$ also features some substantial differences, compared to the two-dimensional scenario.~Arguably the most obvious
one is the fact that the basis \er{eqn: lattice_glue} does not perform so well, with respect to the propagator reconstruction, as in two dimensions.~In particular, in graph
\ref{fig: 3d_big_glue}\,c) one can see that for certain momentum points of the form $p = (\pi, \pi, m)$, the reconstructed correlator deviates appreciably from the original 
one.~This is not surprising, as we argued at the end of section \ref{sec: gluon_lattice} that the representation \er{eqn: lattice_glue} is not complete for $d \geq 3$, and 
that additional structures of the kind \er{eqn: gluon_higher} ought to be added to the tensor basis for the gluon two-point function.

Another interesting feature of the three-dimensional propagator, which does not have a proper counterpart in two dimensions, is the recovery of the correlators continuum 
form at non-zero lattice momenta $p = (m, 0, m)$ (or any component permutations thereof).~To be more precise, all of the dressing functions in \er{eqn: latt_dress_glue}
are well-defined at such kinematics, and they are all proportional to the form factor $D(p)$, even at high values of $m$.~As an example, using the vector $p = (m,0,m)$ in
the definitions of $E(p)$ and $G(p)$ gives   
\begin{align}\label{eqn: special_mom_3d}
&E(p) = \frac{ 2 \, m^4\cdot(D_{11} + D_{22} + D_{33}) - 2 \, m^4 \, (D_{11} + D_{33})}{6 \, m^4 - 4 \, m^4} = \frac{ 2\, m^4 \, D_{22}}{ 2 \, m^4} = D(p) \, , \nonumber
\\[0.25cm] &G(p) = \frac{ m^2\cdot(D_{13} + D_{31}) }{4 \, m^4 - 2 \, m^4} = \frac{ - D(p)}{2\, m^2} \, .
\end{align}

In obtaining the final results in \er{eqn: special_mom_3d}, we again made use of the representation \er{eqn: gauge_tensor}, for momentum $p = (m , 0, m)$.~In the same
way, one may show that $ - p^2 \cdot F(p) = D(p) $ holds for the aforementioned vectors $p$.~Thus, the two-point function obtains its continuum tensor form.~These results
are confirmed numerically in plots \ref{fig: 3d_big_glue}\,b) and \ref{fig: 3d_big_glue}\,e), as the kinematic point $p = (\pi, 0, \pi)$ is approached from the left.~As already 
discussed in the previous section, all of these outcomes ultimately stem from the parametrisation \er{eqn: gauge_tensor}, but it would be interesting to see if they also remain 
true for second-rank lattice correlators whose tensor bases are not determined completely by gauge-fixing. 

It should also be pointed out that the results of \er{eqn: special_mom_3d} are not altered in any way if the decomposition \er{eqn: lattice_glue} is augmented by additional
tensor structures like \er{eqn: gluon_higher}, for momenta of the kind $p = (m,0,m)$.~This is because, for the said kinematics, all of the tensor elements with higher mass 
dimension are proportional to the continuum momentum factor $ p_\mu \, p_\nu $.~As an example, for the leading-order correction of \er{eqn: gluon_lead} it holds that  
\begin{align}\label{eqn: gluon_reduced}
\tau^\text{\,lead}_{\mu\nu}(p) \, = \, p_\nu \, p_\mu \, (p_\mu^2 \: + \: p_\nu^2) \, = m^2 \cdot p_\nu \, p_\mu  \, , \qquad \: \: \mu \: , \nu  = 1 \ldots 3 \, ,
\end{align} 

\begin{figure}[!t]
\begin{center}
\graph[width = 0.44\tew]{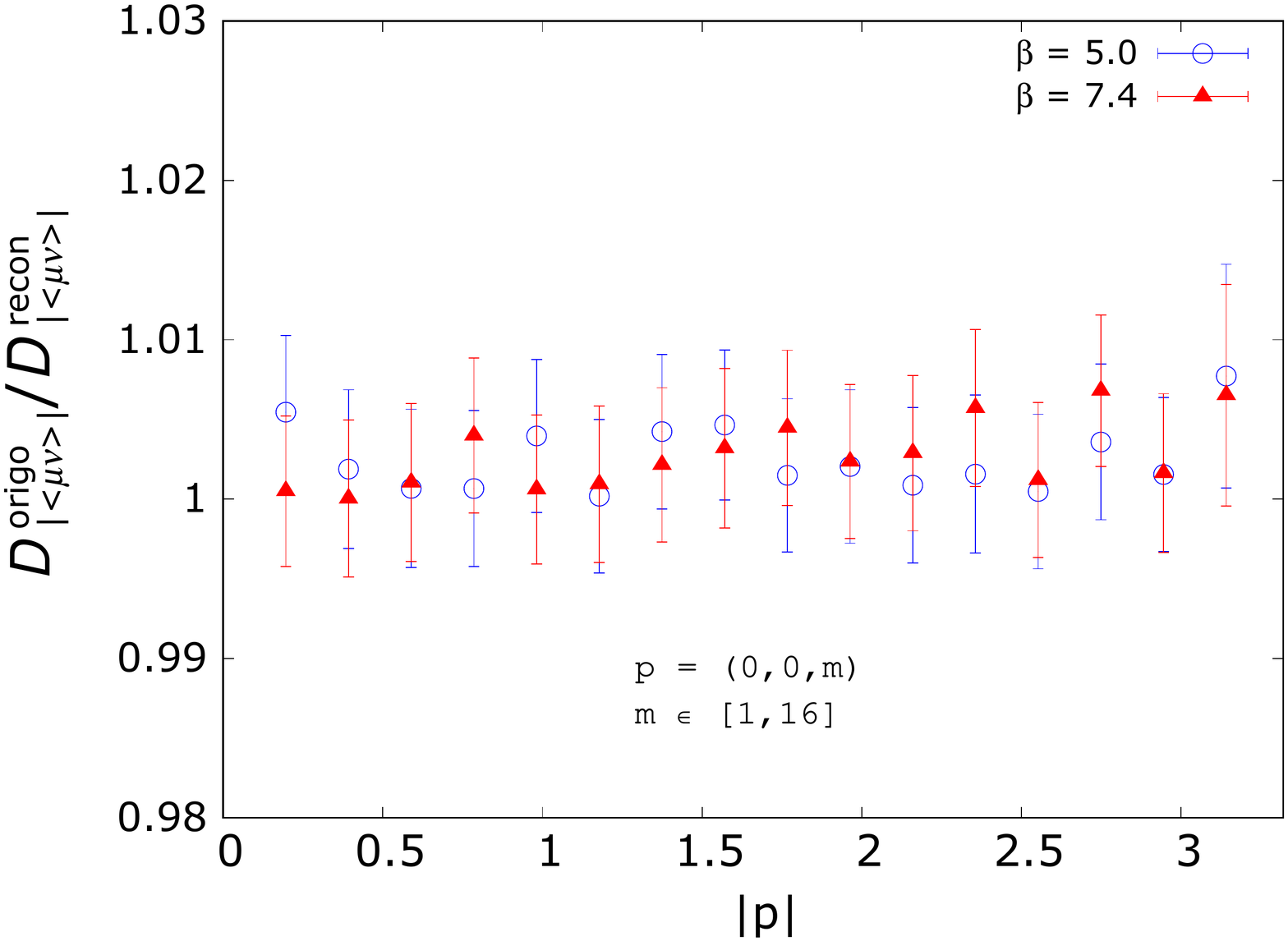}
\caption{Results for on-axis vertex reconstruction on a 32$^3$ lattice, according to the continuum [\,equations \er{eqn: cont_glue} and \er{eqn: cont_project_glue}\,] 
tensor decomposition.~Results are given as a function of $|p| = \sqrt{p^2}$, in lattice units.~$\beta$ is the bare lattice coupling of \er{eqn: wilson_action}. } 
\label{fig: on_axis_glue}
\end{center}
\end{figure}

\noindent
for the kinematic choice $p = (m, 0, m)$ (or any permutations theoreof).~The same remark holds for all of the structures akin to \er{eqn: gluon_higher}:~for appropriate
momentum $p$, they are all proportional to $p_\mu\,p_\nu$, and can thus be excluded from the propagators tensor description.~Besides the situation $ p= (m, 0, m)$, this 
argument also extends to on-axis configurations $p = (0,0,m)$, as well as the diagonal ones $p = (m,m,m)$.~For all of these kinematic points, the lattice propagator ought 
to be described fully by the continuum tensor representation.~Concerning the momenta like $p = (m,0,m)$, as well as the diagonal vectors, we already provided some 
numerical evidence for these claims, in Figure \ref{fig: 3d_big_glue}.~Up to now we have avoided looking at exact on-axis configurations, since the off-diagonal dressing
$G(p)$ is ill-defined at such points.~In Figure \ref{fig: on_axis_glue} we correct this ommision, by showing the numerical results which confirm that the on-axis gluon 
correlator is described exactly by the continuum tensor elements.

\begin{figure}[!t]
\begin{center}
\graph[width = 0.44\tew]{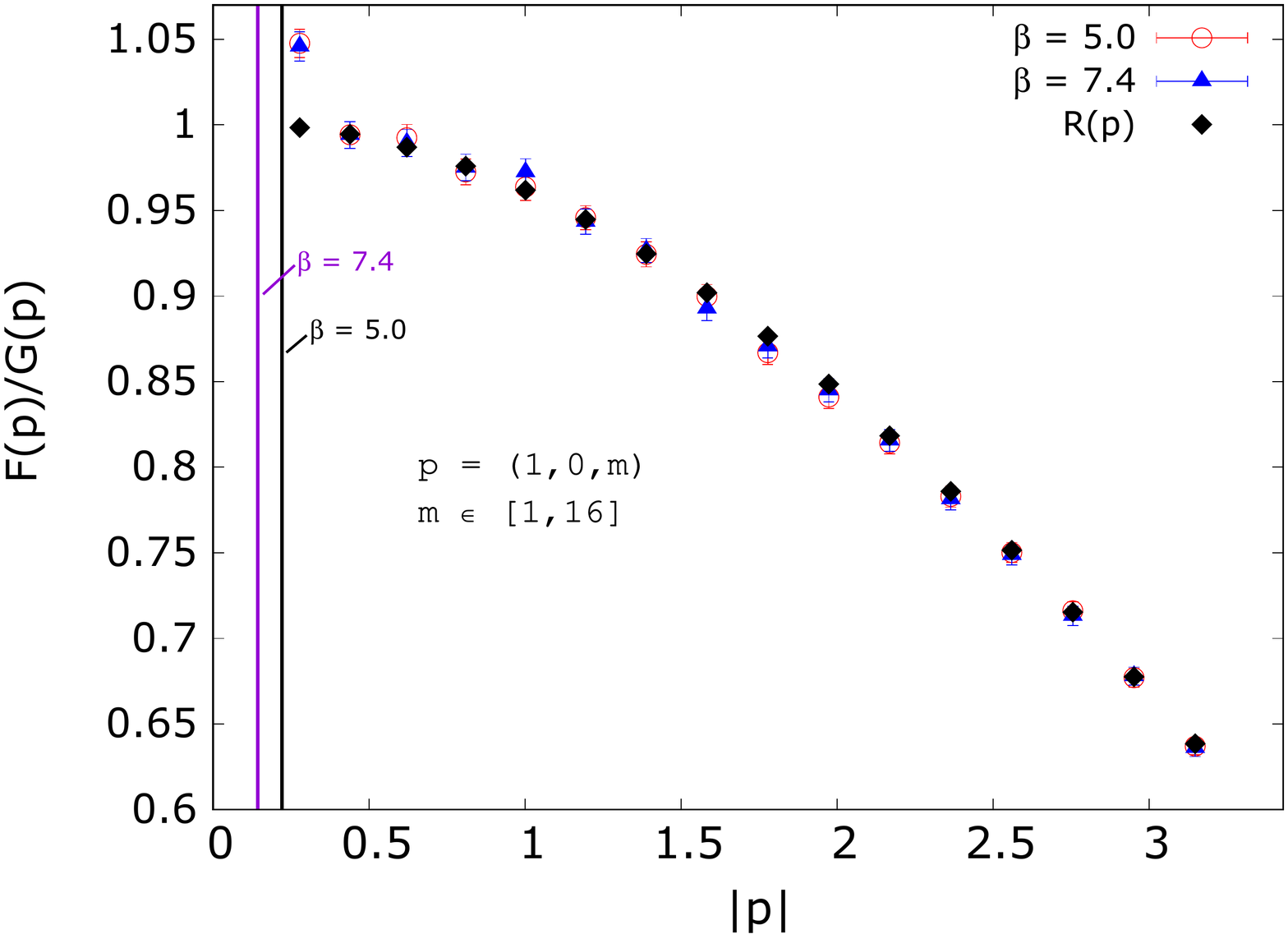}
\caption{Ratio of form factors $F(p)$ and $G(p)$ of \er{eqn: latt_dress_glue}, on a 32$^3$ lattice, as a function of $|p| = \sqrt{p^2}$ (in lattice units).~The values for 
the function $R(p) = |\hat{p}|/|p|$ are also provided, where $\hat{p} = 2 \, \sin{p/2}$.~The two vertical lines denote the physical momentum scale $|\hat{p}| = 0.316$ GeV,
for two $\beta$ couplings of \er{eqn: wilson_action} considered in our numerics.~The scale setting and significance of the physical momentum $\hat{p}^2 = 0.1$ GeV$^{\,2}$ 
are discussed in the text.} 
\label{fig: beta_glue}
\end{center}
\end{figure}

There is another interesting thing to be noted from the reconstruction results of Figures \ref{fig: 3d_big_glue} and \ref{fig: on_axis_glue}.~Namely, the gauge parameter 
$\beta$ of \er{eqn: wilson_action} seems to have little to no influence on the deviations between the original and the reconstructed propagator:~these discrepancies appear 
to depend almost exclusively on lattice kinematics.~This would also indicate that $\beta$ has no bearing on the rate at which the correlators continuum form is recovered,
as one goes deeper into the IR region.~To check this, we've taken a look at the ratio of form factors $F(p)$ and $G(p)$ from \er{eqn: lattice_glue}, at two different $\beta$
values, to see if the gauge coupling affects the way in which $F(p)/G(p)$ approaches unity at low momenta.~The results are shown in Figure \ref{fig: beta_glue}, and they 
support the notion that this ratio depends solely on kinematics, within statistical errors.~To further strengthen this argument, in the same plot we show the data for the 
function $R(p) = \sqrt{\hat{p}^2}/\sqrt{p^2}$, which can arguably be used as a measure of ``how fast'' the decomposition \er{eqn: gauge_tensor} reduces to the continuum 
propagator parametrisation, as the product $p^2$ decreases.~The fact that $R(p)$ describes most of the $F(p)/G(p)$ points with good accuracy shows that the latter ratio 
depends on kinematics alone.~Of course, the $\beta$-independence only holds when the results are given in lattice units, as the coupling controls the value of the lattice
spacing $a$ in physical units.

If one wished to improve the above situation, so that the ratio of functions $G(p)$ and $F(p)$ goes ``faster'' to unity at low(er) momentum, one would have a few
options to consider.~One possibility would be the use of continuum extrapolation methods, as was already discussed at the end of the previous section.~The other 
recourse is to modify the numerical gauge-fixing method, since it is ultimately this procedure, along with the `$p_\mu/2$' modification in \er{eqn: fourier_glue}, 
that brings about the tensor structure of \er{eqn: gauge_tensor}.~However, for numerical lattice simulations it is not yet known how to systematically improve the 
gauge-fixing algorithms, to a desired order in the lattice spacing $a$, even though some attempts in this direction have been made in the past \cite{Bonnet:1999bw}.
\!This could anyway be an interesting research topic for future studies.~Going back briefly to Figure \ref{fig: beta_glue}, one may also note a relatively large 
deviation between $F(p)$ and $G(p)$ at the lowest considered momentum,  $p = (1,0,1)$.~We are yet to check if this disagreement is a finite volume artifact, as the
basis \er{eqn: lattice_glue} does not take such effects into account.~Finally, in the Figure we also included two vertical lines, which mark the rough location of 
the physical momentum $|\hat{p}| \approx 0.312$ GeV, for the two considered $\beta$ values.~This physical scale ultimately has to do with the muon $g-2$ study 
of \cite{Aubin:2015rzx}, but since the full related discussion is lengthy and lies a bit outside our current main line of development, we will only provide the details
at the end of this section.   

As one of the last tests concerning the basis description \er{eqn: lattice_glue}, we want to show that the corresponding dressing functions are hypercubic invariants.~To do
so, we shall examine a collection  of relatively random momentum points $p$, which are not close to any of the special configurations like e.\,g.~on-axis or diagonal momenta.
\!Our goal is to show that averaging the functions over permutations and inversions of momentum components does not change their value, within statistical errors.~For this 
purpose, the form factors calculated at momenta $p = (p_1, p_2, p_3)$ [\,here denoted generically as $\mathcal{F}(p)$\,] will be compared with their appropriate permutation 
and parity averages, where for instance 
\begin{align}\label{eqn: permute_average}
\mathcal{F}^{\,\text{perms}} = \frac{1}{6} \cdot \left[\, \mathcal{F}(p_1, p_2, p_3) + \mathcal{F}(p_1, p_3, p_2) + \mathcal{F}(p_2, p_1, p_3) + \mathcal{F}(p_2, p_3, p_1) + 
\mathcal{F}(p_3, p_1, p_2)  + \mathcal{F}(p_3, p_2, p_1)\, \right] \, .
\end{align}

\begin{figure}[!t]
\begin{center}
\graph[width = 0.41\tew]{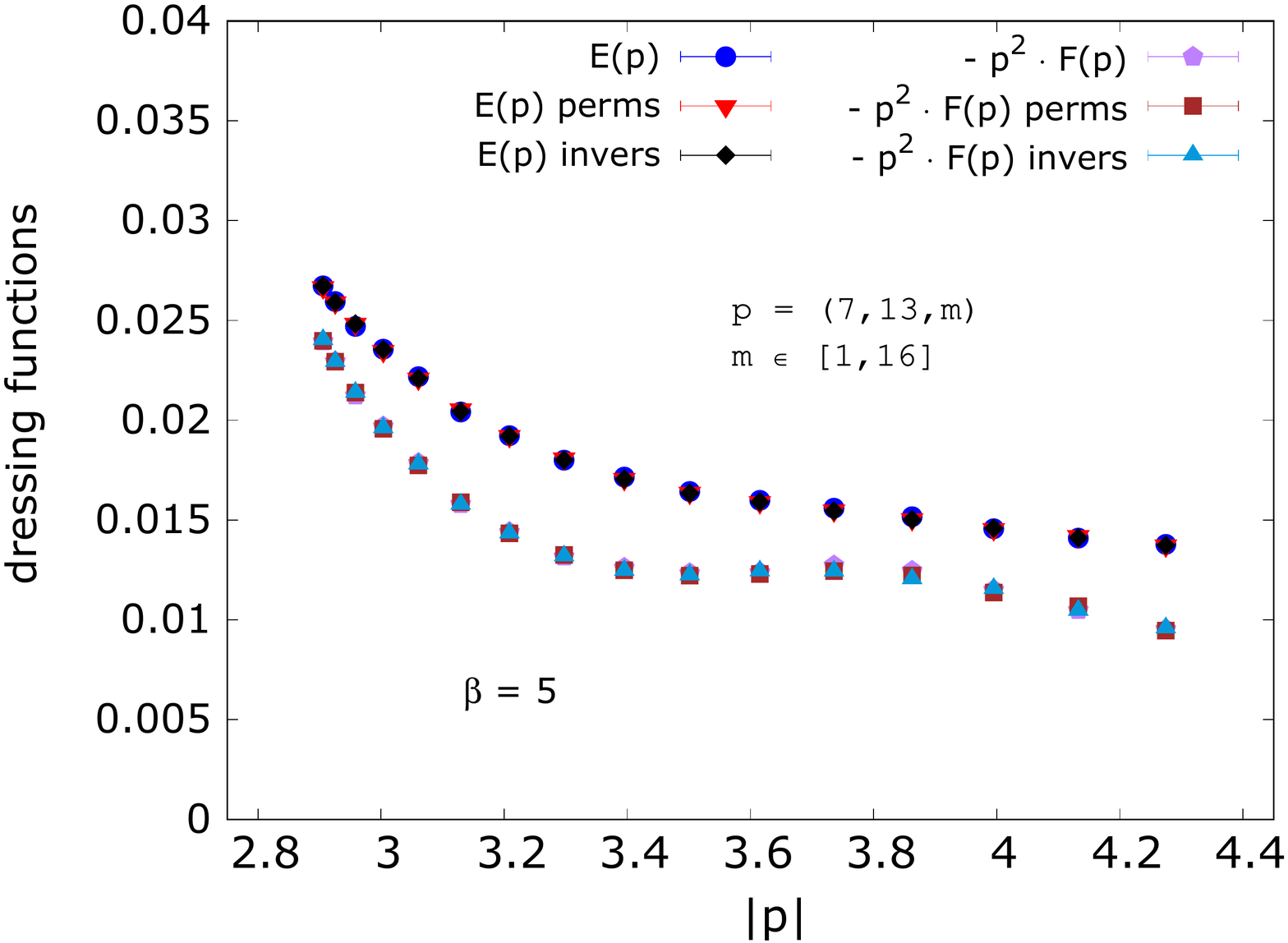}\graph[width = 0.41\tew]{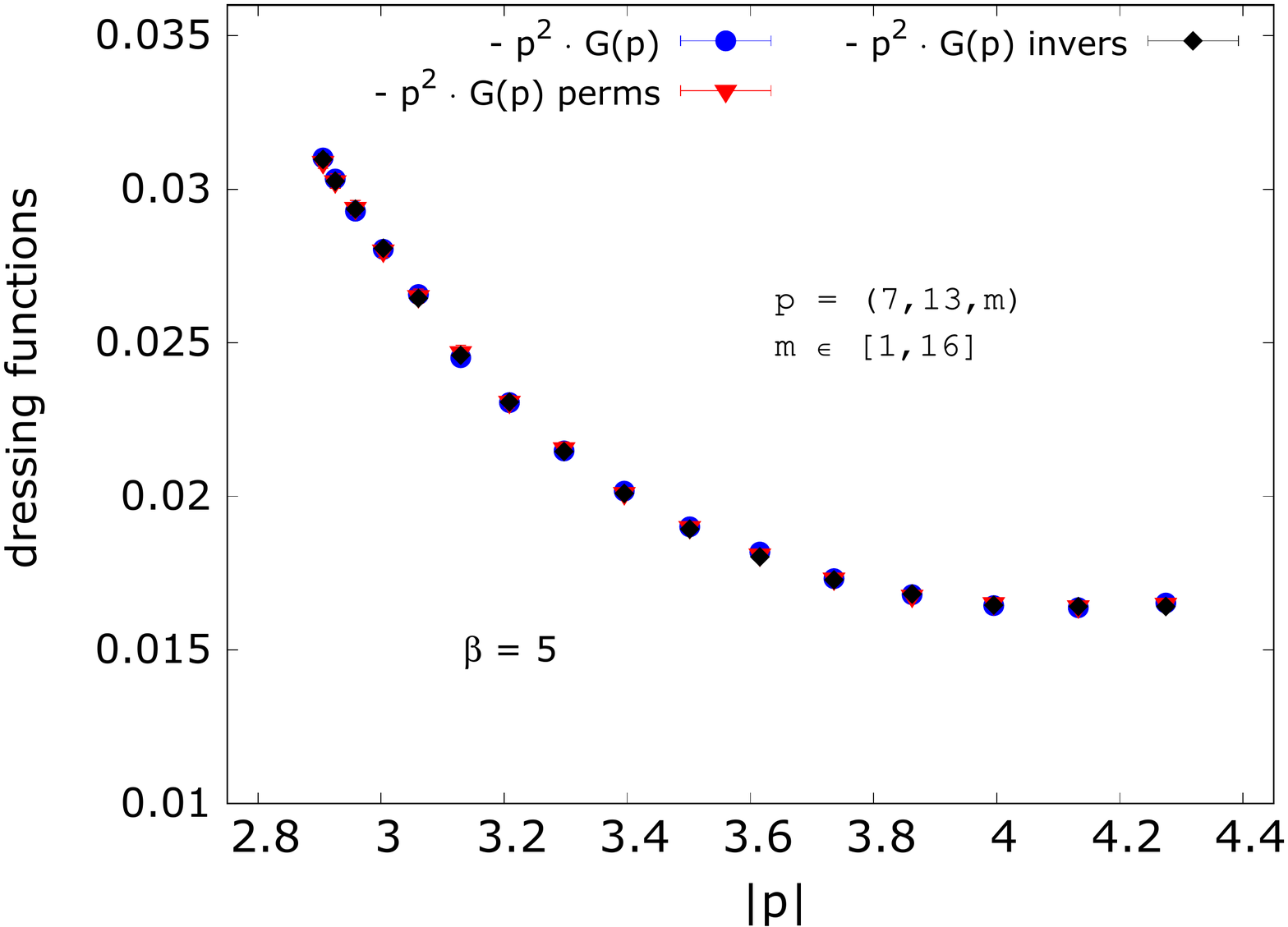}\\
\caption{\textit{Left}:~Test of permutation and inversion invariace of functions $E(p)$ and $F(p)$ of \er{eqn: latt_dress_glue}, on a $32^3$ lattice.~Labels ``$E(p)$'' and ``$
- p^2 \cdot F(p)$'' refer to results for a single momentum point $p$, while additional remarks ``perms'' and ``invers'' signify averages over permutations and inversions of $p$ 
components, see equation \er{eqn: permute_average} and accompanying text.\,\textit{Right}:~The same, for the dressing $G(p)$ of \er{eqn: latt_dress_glue}.~Data are provided as 
functions of $|p| = \sqrt{p^2}$, in lattice units.~$\beta$ is the gauge coupling of \er{eqn: wilson_action}.} 
\label{fig: aver_test_glue}
\end{center}
\end{figure}

In the same manner, the inversion average is obtained by going over all momenta of the form $ p^{\pm} = (\pm p_1, \pm p_2, \pm p_3)$, with all possible combinations of plus
and minus signs.~For both permutations and parity changes, we've checked that all of the functions which enter the sum like \er{eqn: permute_average} have the same sign, 
meaning that there can be no accidental cancellations during the averaging procedure.~The results are given in Figure \ref{fig: aver_test_glue}, and they indicate that the
form factors of \er{eqn: latt_dress_glue} are indeed invariant under hypercubic symmetry transformations.~This also (in)directly confirms that the gluon propagator itself 
constitutes a second-rank tensor with respect to these symmetry operations, a fact which is all but guaranteed by the tensor description \er{eqn: gauge_tensor}.~However, 
in Monte Carlo simulations, the validity of \er{eqn: gauge_tensor} depends crucially on the $p_\mu/2$ modification in the Fourier transform for the gluon potential, see
\er{eqn: fourier_glue}.~In the absence of this correction factor, the numerical propagator would in fact not transform as a second-rank tensor under inversions.~This issue 
is further discussed in Appendix \ref{eqn: inversions_append}.

\newpage

To conclude this part, we wish to briefly cover two more points.~First, in deriving the tensor decomposition \er{eqn: lattice_glue} we did not make any explicit reference to lattice
Landau gauge, and the basis itself should be applicable to virtually any covariant settings, wherein all of the coordinates are treated equally.~In our Monte Carlo simulations we 
chose to work in Landau gauge because it is the easiest one to implement, with numerical considerations of its (linear) covariant generalisations featuring some non-trivial complications,
see \cite{Giusti:1996kf,Cucchieri:2009kk,Bicudo:2015rma,Cucchieri:2018doy} for a related discussion.~Nonetheless, there should be no principal difficulties in using the basis \er{eqn: 
lattice_glue} and its modifications of the kind \er{eqn: gluon_higher} in any covariant calculations, once the numerical gauge-fixing part is done.~In the future we would thus like to
check how some of the more general conclusions of this section hold up in other covariant gauges. 

As a second point, we return to the Figure \ref{fig: beta_glue} and its vertical lines denoting the momentum scale $|\hat{p}| \approx 0.316$ GeV, for two $\beta$ couplings we considered 
in our simulations.~To convert the lattice spacing $a$ into physical units, we've set the string tension to $\sqrt{\sigma} = 0.44$ GeV and used the fit of equation (67) in \cite{
Teper:1998te}:~the fit requires the values of the 1$\times$1 Wilson loops, which are provided in Table \ref{tab: config_details}.~We marked the point(s) $|\hat{p}| \approx 0.316$  GeV
as significant since the momenta around or below the scale $p^2 \approx 0.1$ GeV$^{\,2}$ are presumably the ones for which the hadronic vacuum polarisation contributes the most to the 
anomalous magnetic moment of the muon \cite{Aubin:2015rzx, Golterman:2014ksa}.~Now, due to (lattice) vector-current conservation, the hadronic electromagnetic current $\Pi_{\mu\nu}(p)
$ will be described by equation \er{eqn: gauge_tensor}, up to higher-order scaling violations which can be ignored at low momenta \cite{Shintani:2010ph}.~This means that, up to certain 
effects which we shall discuss shortly, our results in Figure \ref{fig: beta_glue} also apply to $\Pi_{\mu\nu}(p)$ and subsequently to the vacuum polarisation $\Pi(p)$.~The main point 
here is that, at the relevant energy scale below 0.1 GeV$^{\,2}$, the discretisation artifacts seen in Fig.~\ref{fig: beta_glue} constitute a sub-percent effect with $F/G > 0.99
$:~we assume that the discrepancy at the lowest momentum is purely due to the finite volume.~To put these finite-spacing effects into perspective, in the muon $g-2$ study of \cite{
Aubin:2015rzx} the finite volume was estimated to incur a systematic uncertainty on the order of ten to fifteen percent.

Of course, our results in Figure \ref{fig: beta_glue} should not actually be directly applied/compared to \cite{Aubin:2015rzx}, because of different scale-setting procedures and 
completely different lattice setups (our symmetric three-dimensonal lattice versus the \textit{a}symmetric four-dimensional one of \cite{Aubin:2015rzx}).~Nonetheless, it is hard to
imagine that our conclusions on this matter could get modified drastically with more realistic comparisons, and it remains an almost absolute certainty that the discretisation effects
will be by far the sub-dominant source of systematic errors, in lattice studies of the anomalous muon magnetic moment.  

\subsection{Ghost-gluon vertex in three dimensions}\label{sec: ghost_glue}

As for the gluon two-point function, we start our discussion on the ghost-gluon vertex by specifying the corresponding numerical procedure.~On the lattice, the ghost-gluon 
Greens function can be obtained as the following Monte Carlo average \cite{Cucchieri:2004sq}: 
\begin{align}\label{eqn: green_ghost}
\Gamma_\mu^{\,abc}(p, q, k) = \frac{1}{V} \left\langle \left( M^{-1} \right)^{ab}(p,q) \, A_\mu^c (k) \right\rangle \, .
\end{align}           

In the above relation, $V$ stands for the lattice volume, $A_\mu^c (k)$ denotes the colour components of the gluon potential of \er{eqn: intro_latt_glue}, and $(M^{-1})^{ab}(p,
q)$ is the Fourier transform of the (inverse) Faddeev-Popov operator, i.\,e.
\begin{align}\label{eqn: faddeev_foury}
\left( M^{-1} \right)^{ab} (p,q) = \sum_{x, y} \, \e^{\, 2 \pi i \, ( p \cdot x \, + \, q\cdot y)} \: \left( M^{-1} \right)^{ab}(x, y) \, .
\end{align}

The Faddeev-Popov (FP) matrix itself is defined through its action on a scalar test function $\omega^b (x)$, where $b$ is a colour index and one has that \cite{Zwanziger:1993dh}
(in the following, the sums over $y$ and $b$ are implied):
\begin{align}\label{eqn: faddeev_define}
M^{ab}(x, y) \, \omega^b (y) = \delta_{xy} \, & \sum_{\mu} \, G_\mu^{\,ab} (y) \, [\, \omega^b(y) - \omega^b(y^+)\,] - G_\mu^{\,ab} (y^-) \, [\, \omega^b(y^-) - 
\omega^b(y)\,] \, + \nonumber \\ 
& \sum_c f^{abc} \, [\, A^b_\mu(y) \, \omega^c(y^+) - A^b_\mu(y^-) \, \omega^c(y^-) \,]  \, .
\end{align} 

In the above expression, $y^{\pm}$ stands for $y \pm e_\mu$, with $e_\mu$ being the unit vector in the $\mu$-th direction [\,the dummy index $\mu$ matches the one being
summed over in \er{eqn: faddeev_define}\,].~Also, the quantities $G_\mu^{\,ab} (y)$ used in the definition \er{eqn: faddeev_define} are equal to 
\begin{align}\label{eqn: g_faddeev_def}
G^{\,ab}_\mu (y) = \frac{1}{8} \, \text{Tr} \left( \{\sigma_a, \sigma_b\} \cdot \left[ \, U_\mu (y) \, + \, U_\mu^\dagger(y) \, \right]  \right) \: ,
\end{align} 

\noindent
where the curly brackets denote an anticommutator and $U_\mu (y)$ are the lattice links.~In writing down the relation \er{eqn: faddeev_define}, we've taken into account the fact
that we work in lattice Landau gauge, as otherwise there would be additional terms present.~To compute the Fourier transform of the inverse FP operator [\,i.\,e.~the quantity 
\er{eqn: faddeev_foury}\,], we used the so-called plane-wave source method \cite{Cucchieri:1997dx}.~The matrix inversion itself is performed via a preconditioned  conjugate 
gradient (CG) algorithm:~the preconditioning procedure is described in detail in \cite{Sternbeck:2005tk}.~At each iterative CG step we orthogonalise the prospective solution with 
respect to the constant subspace, since constant fields constitute zero modes of the FP matrix.~In the end, we use a total of 480 gauge field configurations for the evaluation of
the Greens functions \er{eqn: green_ghost}.~The vertex to be studied is extracted from $\Gamma_\mu^{\,abc}$ with a contraction 
\begin{align}\label{eqn: final_vertex}
\Gamma_\mu = \frac{1}{6} \, \sum_{abc} \, f^{\,abc} \, \Gamma^{\,abc}_\mu  \: .
\end{align}

The colour normalisation factor (1/6) stems from the $SU(N)$ identity $f^{\,ade} \, f^{\,bde} = N \, \delta^{\,ab}$, as applied to the particular case of $SU(2)$ gauge theory we
study here.~In lattice calculations, one is generally not really interested in correlators like \er{eqn: green_ghost}, but rather in the so-called amputated vertices, wherein
amputation includes (loosely speaking) dividing out the propagators pertaining to a function like \er{eqn: green_ghost} \cite{Parrinello:1994wd}.~Here, the procedure will be 
completely ignored, and we shall work directly with \er{eqn: final_vertex}.~We do this because amputation can potentially increase the overall statistical uncertainty [\,the 
amputated vertex inherits its errors from both \er{eqn: green_ghost} and the appropriate propagators\,], while changing none of the quantities we are mostly interested here.~In 
particular, it does not alter the tensor structure of the vertex, the symmetry properties of the dressing functions, nor the \textit{relative} values of vertex form factors, 
i.\,e.~the ``sizes'' of vertex dressings relative to one another.~Hence our focus on working directly with \er{eqn: final_vertex}.  
\begin{figure}[!t]
\begin{center}
\graph[width = 0.39\tew]{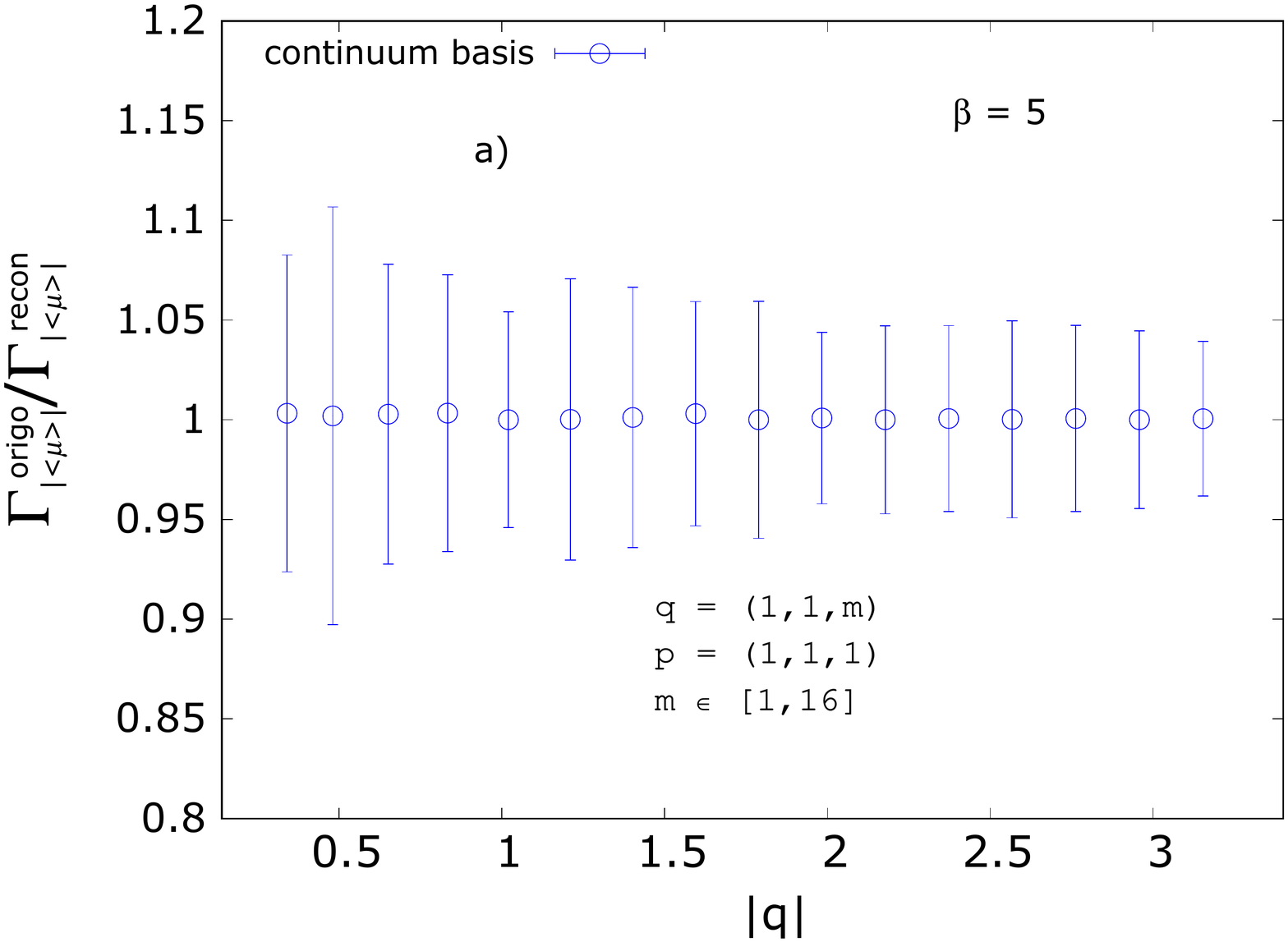}\graph[width = 0.39\tew]{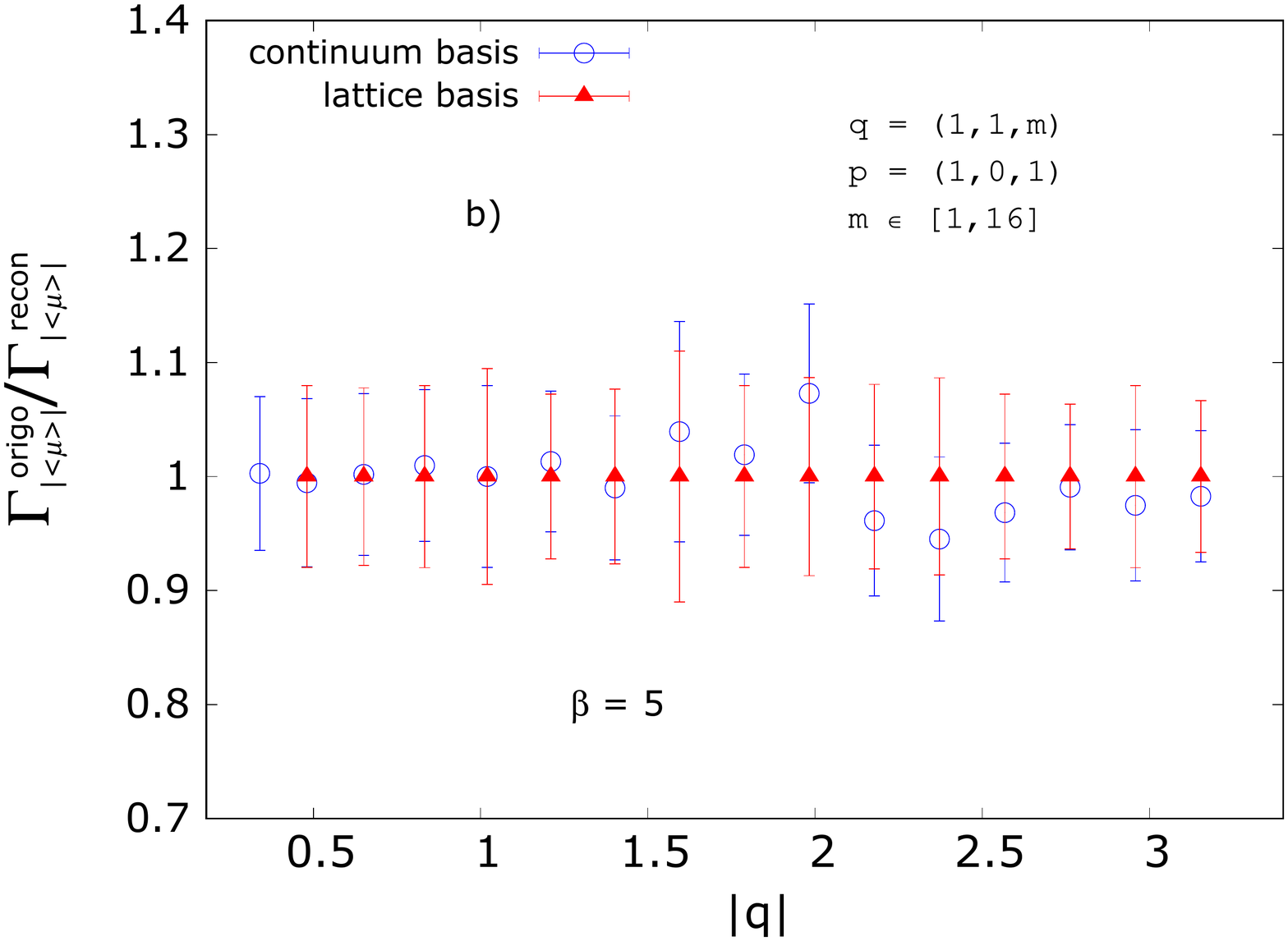}
\graph[width = 0.39\tew]{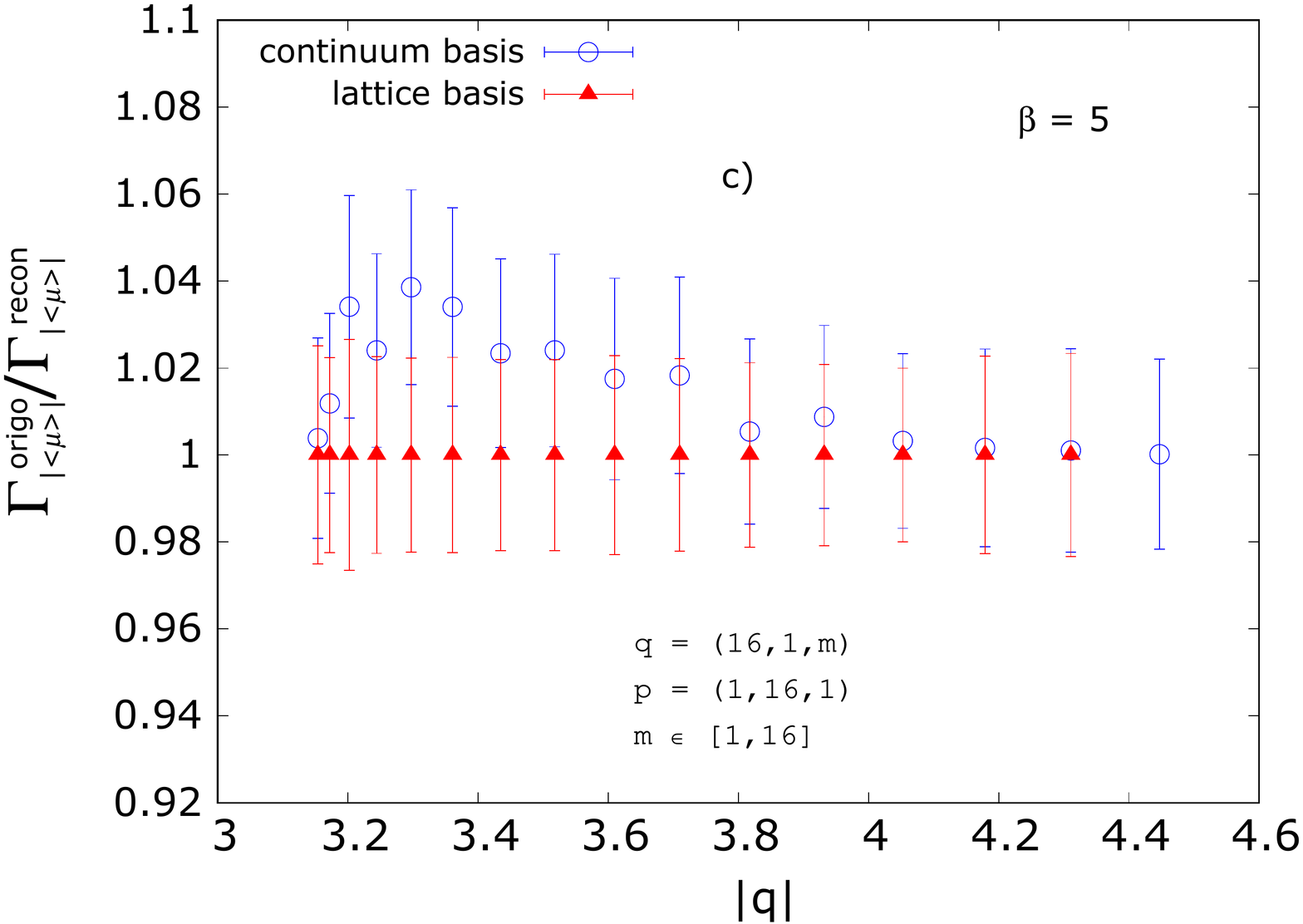}\graph[width = 0.39\tew]{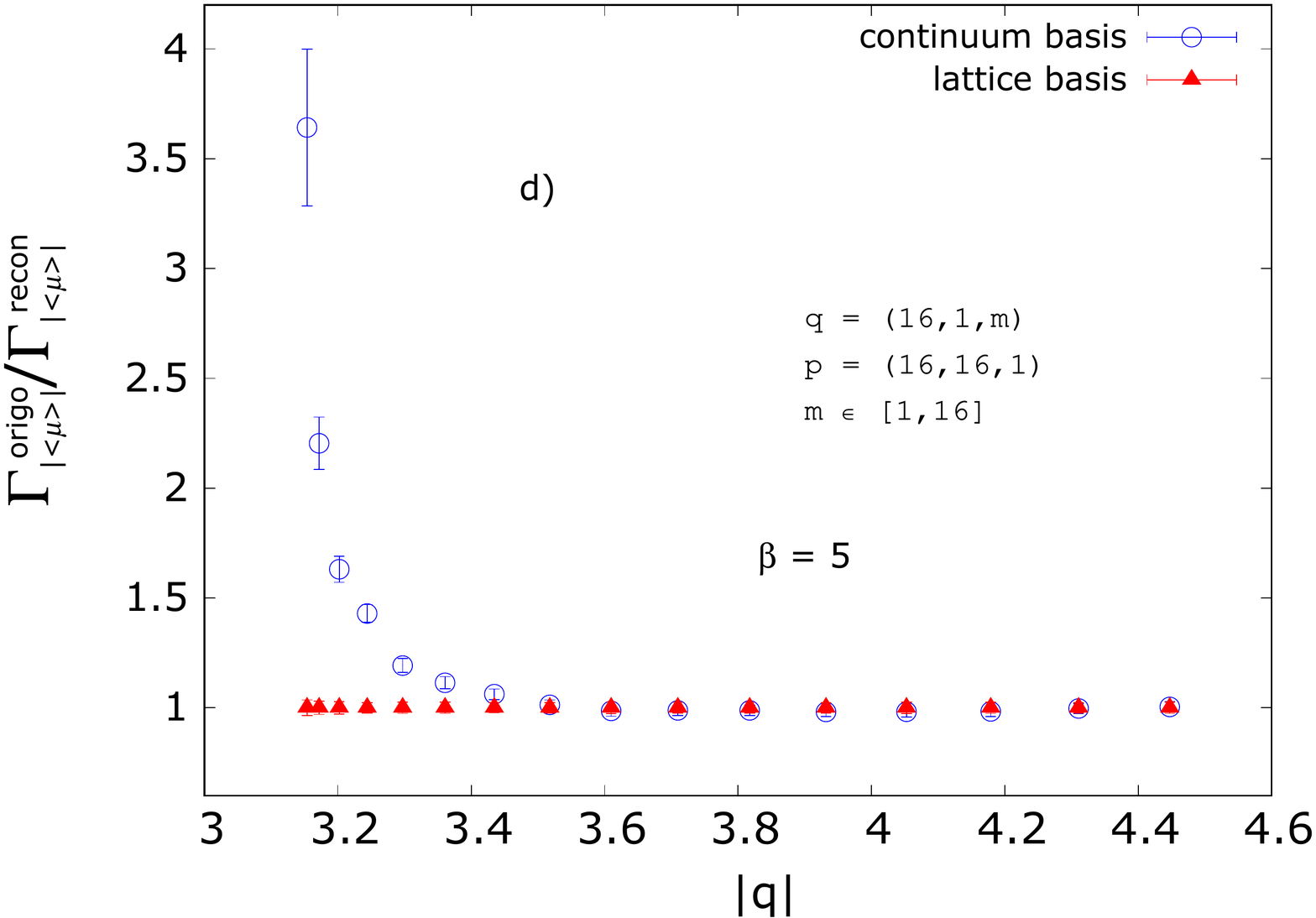}
\graph[width = 0.39\tew]{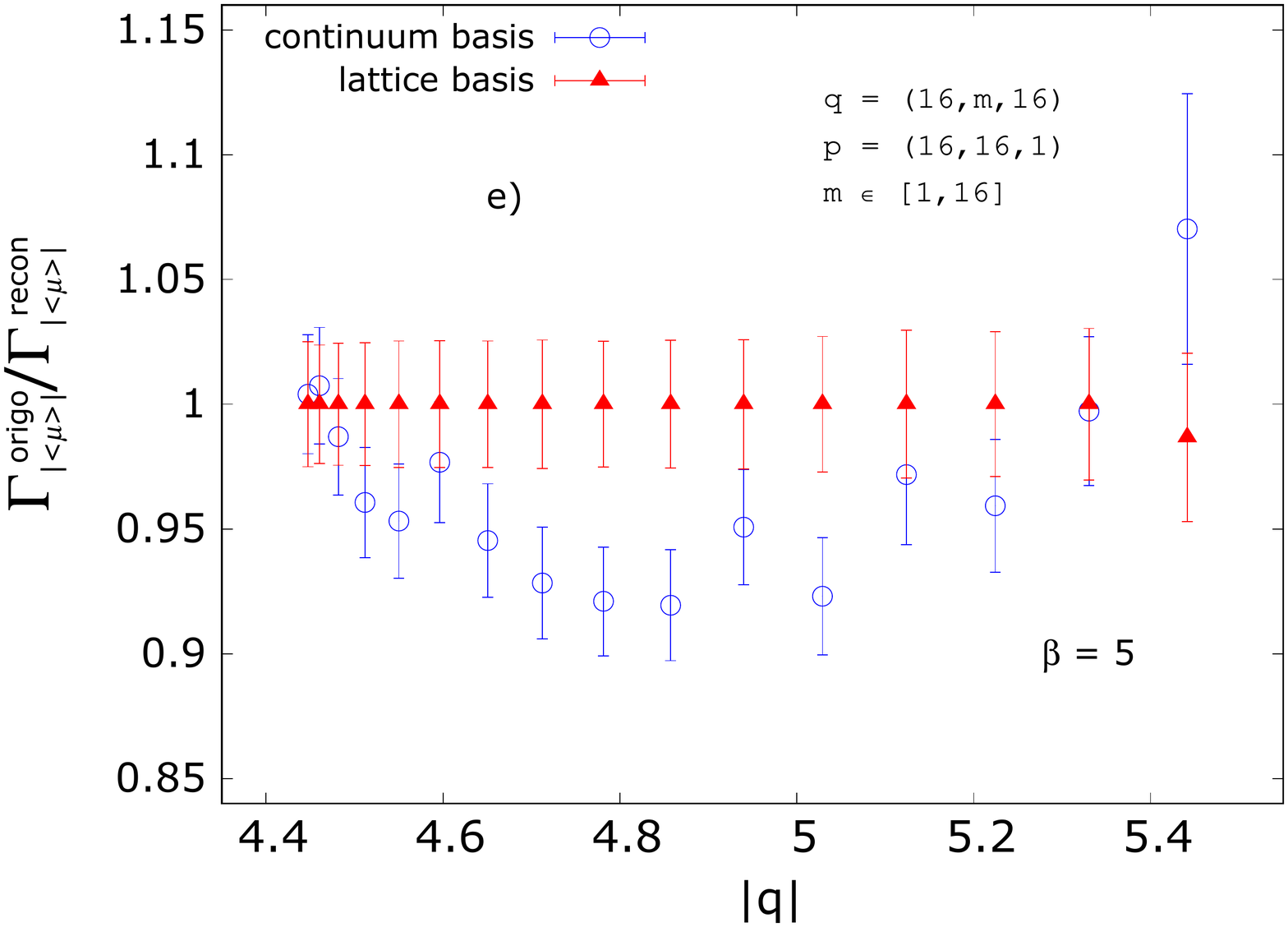}\graph[width = 0.39\tew]{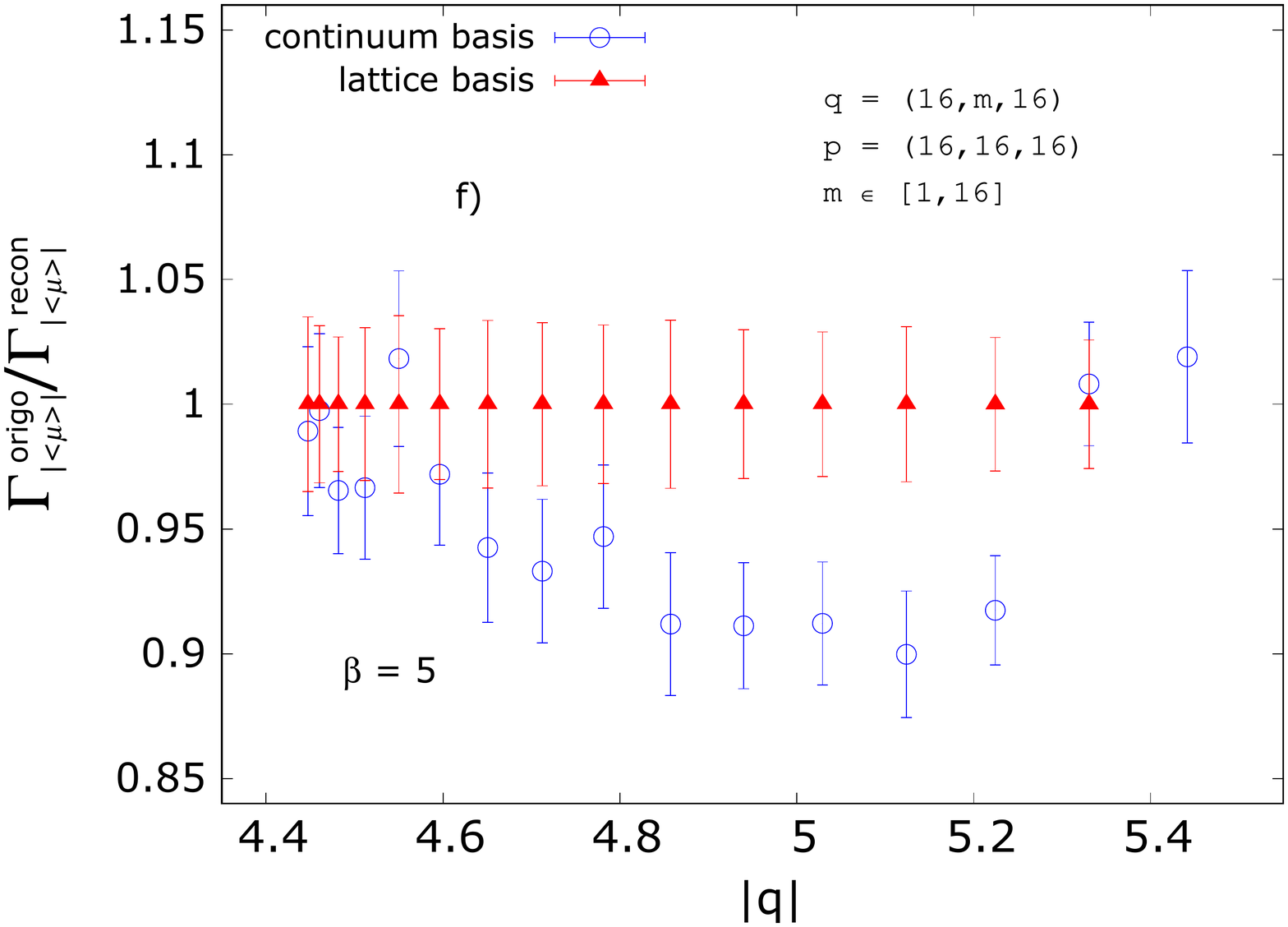}
\caption{Vertex reconstruction data for the correlator \er{eqn: final_vertex} on a 32$^3$ lattice, as functions of $|q| = \sqrt{q^2}$ (in lattice units).~We use the continuum 
[\,equations \er{eqn: ghost_cont} and \er{eqn: ghost_project}\,] and lattice-modified [\,equations \er{eqn: ghost_lattice}, \er{eqn: project_general}, \er{eqn: matrix_vertex} 
and \er{eqn: determinant}\,] tensor descriptions.~$\beta$ is the gauge coupling of \er{eqn: wilson_action}.} 
\label{fig: ghost_recon}
\end{center}
\end{figure}

In our vertex reconstruction tests, the correlator \er{eqn: final_vertex} will be described by two different tensor parametrisations.~One is the continuum basis, given by 
\begin{align}\label{eqn: ghost_cont}
\Gamma_\mu(q, k) = A(q,p) \, q_\mu + B(q,p) \, p_\mu \, ,
\end{align}

\noindent
with the appropriate projectors 
\begin{align}\label{eqn: ghost_project}
P_\mu^{\,A} = \frac{-p^2 \, q_\mu + q\cdot p \, p_\mu}{(q\cdot p)^2 - q^2 \, p^2} \, , \qquad \qquad \: P_\mu^{\,B} = \frac{ q\cdot p \, q_\mu -  \, q^2 \, p_\mu}{(q\cdot p)^2 - q^2 
\, p^2} \, .
\end{align}

The construction of the above projector functions is briefly discussed in Appendix \ref{sec: projectors}.~The other vertex decomposition to be studied here is (we will only 
consider a three-dimensional theory and hence three basis elements will suffice):
\begin{align}\label{eqn: ghost_lattice}
\Gamma_\mu(q, p) = E(q,p) \, q_\mu + F(q,p) \, p_\mu \, + G(q,p) \, q^{\,3}_\mu \, .
\end{align}

The explicit expressions for the projectors of \er{eqn: ghost_lattice} are also provided in Appendix \ref{sec: projectors}.~Reconstruction results for the correlator \er{eqn: 
final_vertex} are given in Figure \ref{fig: ghost_recon}, for the two above-mentioned tensor representations.~A brief glance at the corresponding data reveals a somewhat surprising 
fact:~namely, apart from a few ``critical points'' in Figure \ref{fig: ghost_recon}\,d), all the deviations pertaining to the continuum basis are within a twenty percent range, an 
arguably small discrepancy.~In fact, for most of the examined momentum points the continuum decomposition can be said to represent the vertex exactly, within somewhat large statistical 
uncertainties.~To the best of our knowledge, there is no \textit{a priori} reason that this should happen.~Unlike the case of the lattice gluon propagator, the tensor decomposition of
the ghost-gluon correlator is not determined (at least not completely) by gauge-fixing.~For the vertex, this means that there are no obvious constraints on possible deviations from 
the continuum tensor forms, and it is not clear why the function would show relative restraint in this regard.~It would be interesting to see if other lattice vector-valued quantities
like e.\,g.~the quark-gluon vertex, display similar tendencies (this would however be unexpected since fermions are usually significantly affected by finite spacing artifacts, see 
e.\,g.~\cite{August:2013jia}).

\begin{figure}[!t]
\begin{center}
\graph[width = 0.41\tew]{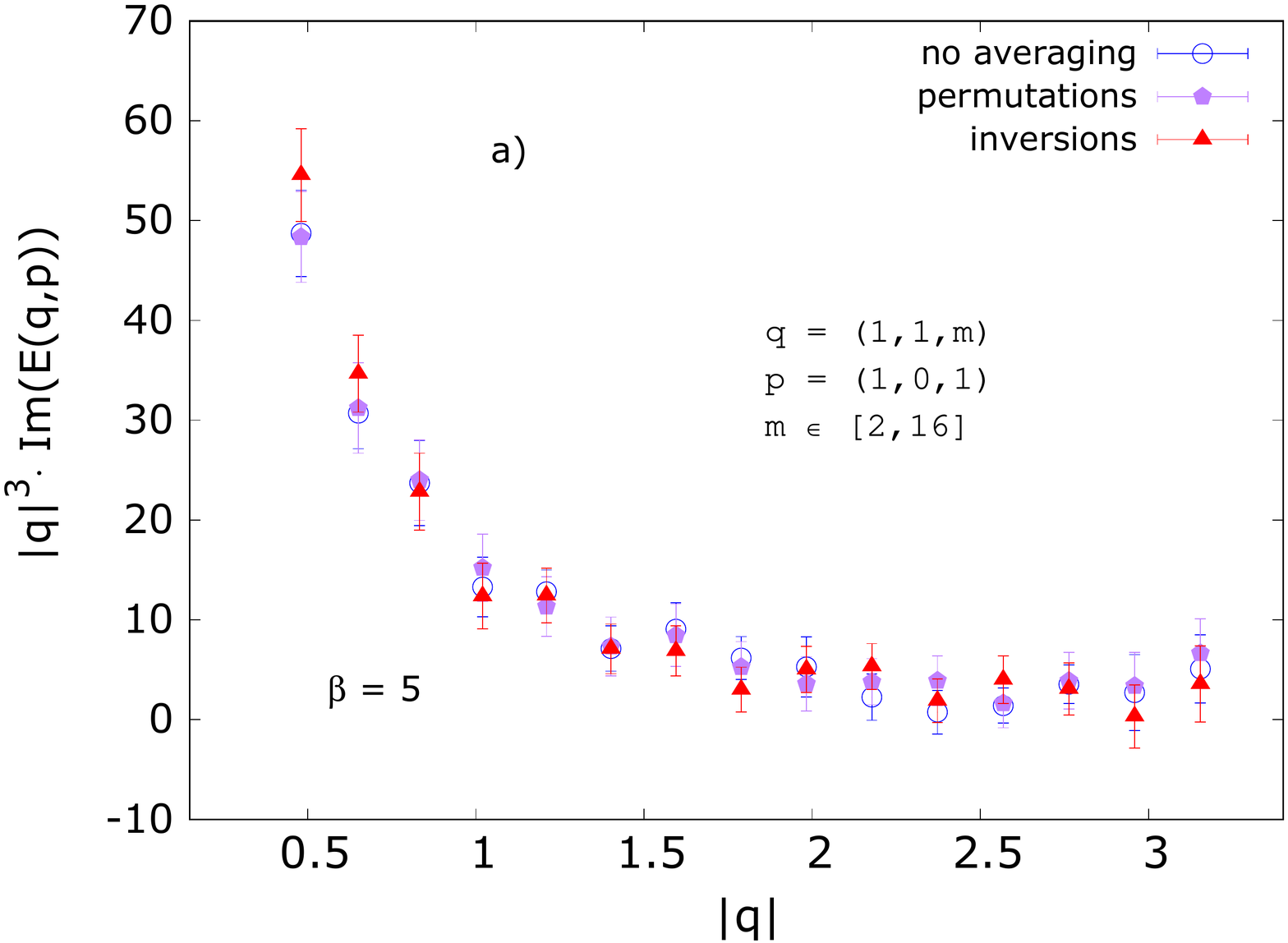}\graph[width = 0.41\tew]{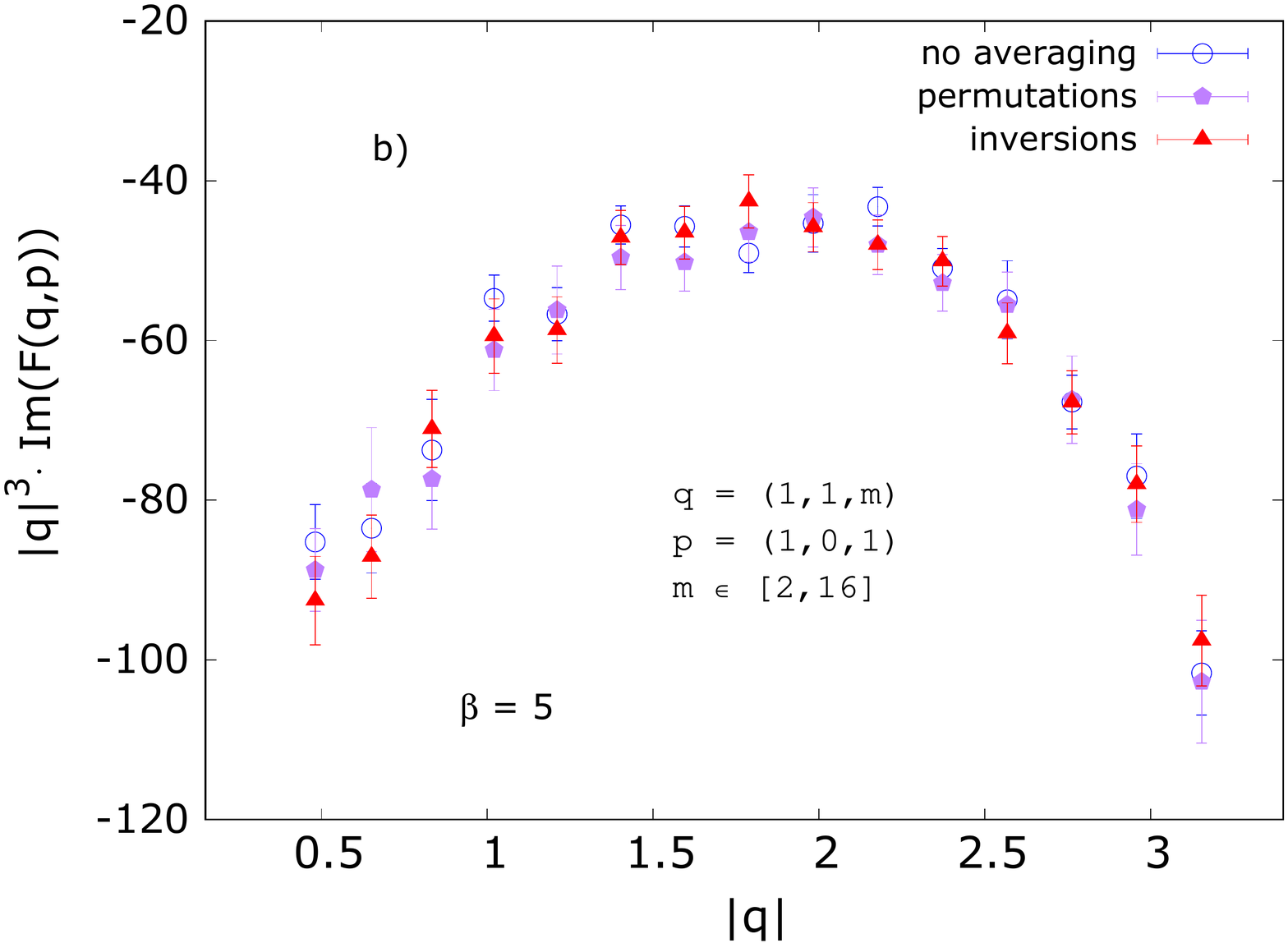}
\graph[width = 0.41\tew]{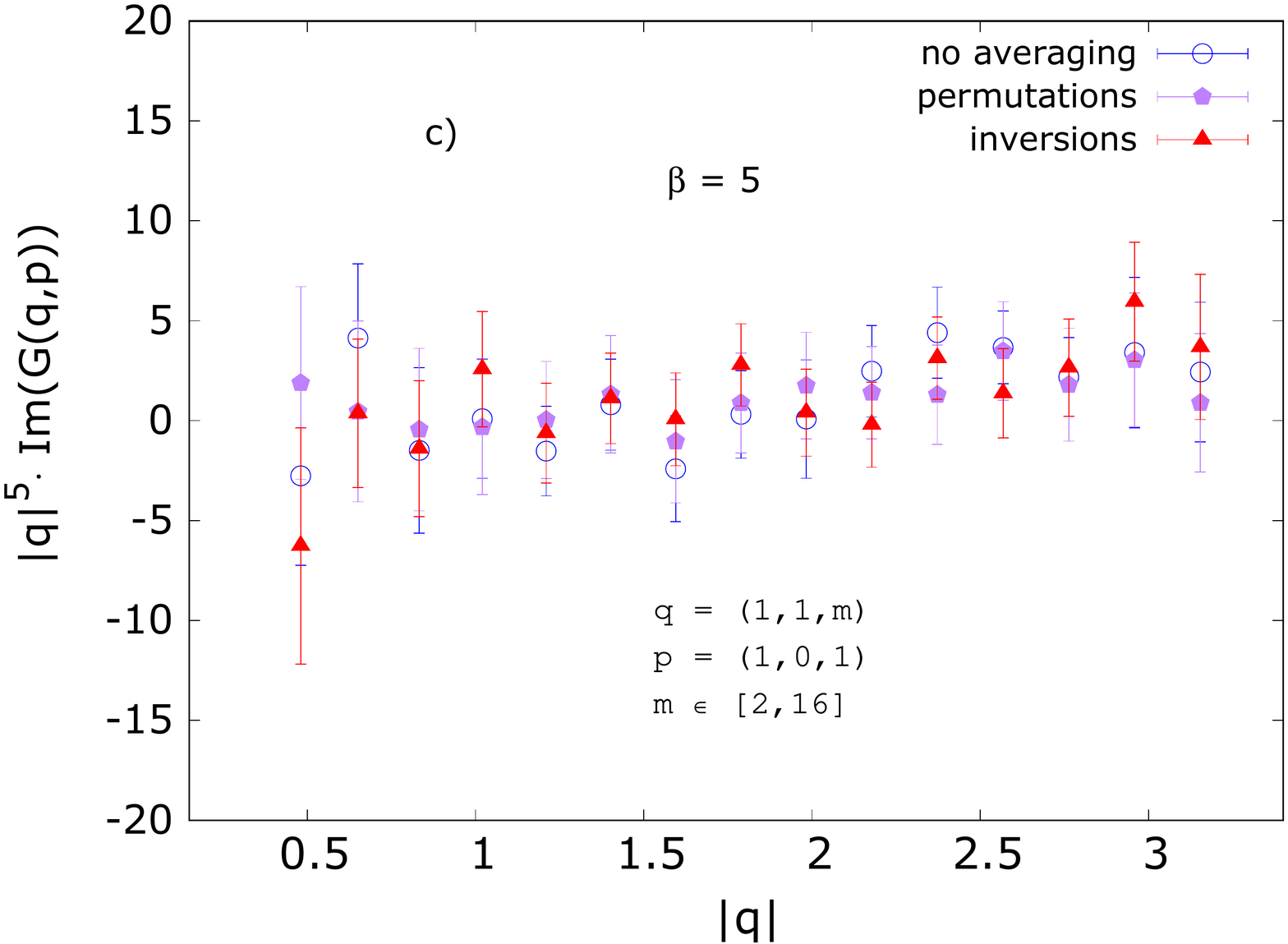}\graph[width = 0.41\tew]{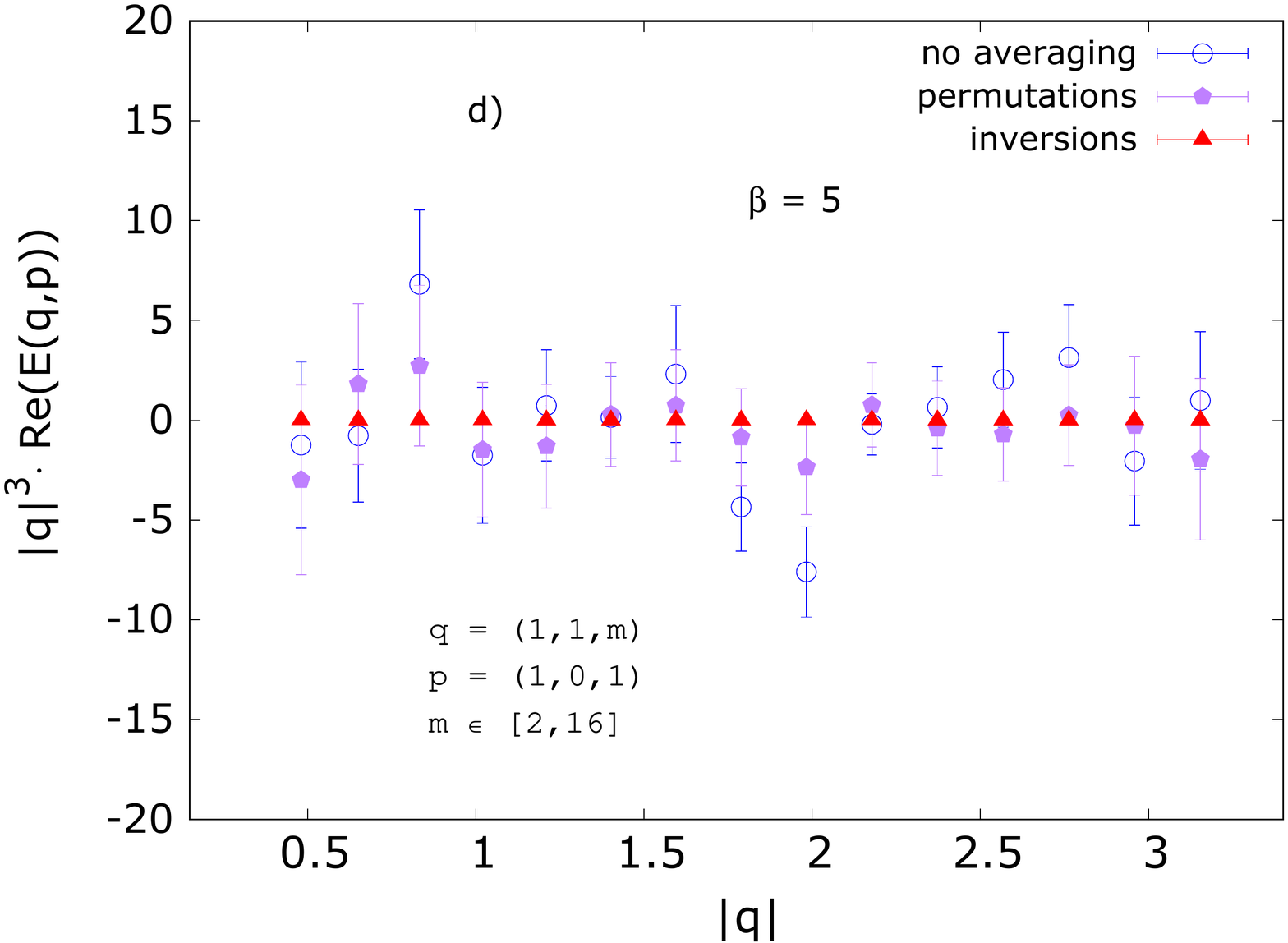}
\caption{Plots a) to c)\,:~Imaginary parts of lattice-modified vertex form factors [\,equations \er{eqn: ghost_lattice}, \er{eqn: project_general}, \er{eqn: matrix_vertex} and \er{eqn:
 determinant}\,] on a 32$^3$ lattice, as functions of $|q| = \sqrt{q^2}$ (in lattice units).~Labels ``permutations'' and ``inversions'' refer to results averaged (respectively) over all 
possible permutations and inversions of momentum components, as opposed to ``no averaging'' data obtained for a single kinematic set $(q,p)$.~Plot d)\,:~the real part of the vertex dressing 
$E(q,p)$ of \er{eqn: ghost_lattice}.~$\beta$ is the gauge coupling of \er{eqn: wilson_action}.} 
\label{fig: ghost_dress}
\end{center}
\end{figure}

In the absence of large lattice-induced modifications, it is also a bit challenging to precisely identify the special kinematic configurations.~In other words, from numerical results
it is hard to see where the continuum basis should describe the vertex exactly, due to linear dependencies among available tensor elements.~Arguably the most clear-cut examples of the
continuum description being sufficient are those in plots a),\,e) and f).~All the points in graph a) correspond to a situation with one diagonal vector [\,in this case $p = (1,\,1,\,
1)$\,] and the other one being almost on-axis [\,with $q = (1,\,1,\,m)$\,], which is a kinematic choice that was discussed in section \ref{sec: ghost_latt}.~In the same vein, the 
rightmost points in plots e) and f) feature a diagonal momentum $q$, with the other dynamic variable being equal or almost equal to the two special cases $p = (m,\,0,\,m)$ and $p = (m,
\,m,\,m)$:~hence the apparent applicability of the basis decomposition \er{eqn: ghost_cont}, see arguments of section \ref{sec: ghost_latt}.~Another set of interesting configurations
are the first and the last momentum points in Figure \ref{fig: ghost_recon}\,c).~We concentrate on the first one, where vector $p$ almost has the form $p \approx (\pi,\,0,\,0)$, and $q$ 
is close to the situation $q \approx (0,\,\pi,\,0)$.~One observes that with vectors $p = (\pi,\,0,\,0)$ and $q = (0,\,\pi,\,0)$, all of the mixed tensor structures of the kind $\tau^{\,
rs}_\mu = p_\mu^{\,r}\,q_\mu^{\,s}$, with appropriate non-zero integers $r$ and $s$ [\,see equation \er{eqn: taus_lattice_vertex} for details\,] will vanish, because $p_\mu \, q_\mu = 0$
(no summation implied), for all values of the index $\mu$.~Therefore only the lattice vectors $p_\mu^{\,2k + 1}$ and $q_\mu^{\,2k + 1}$ remain (with $k \in N_0$), which for the considered
kinematics are proportional to the continuum terms:~for an example, $p_\mu^{\,3}$ equals $\pi^2 \, p_\mu$ and so on.~This brings about the seeming near-completeness of the continuum 
description in the first momentum point in Figure \ref{fig: ghost_recon}\,c), and the same explanation holds for the rightmost kinematic choice in the plot.

The above interesting cases notwithstanding, most of the results in Figure \ref{fig: ghost_recon} are somewhat trivial, since they amount to a claim that a three-dimensional vector will
be described fully by a set of three linearly independent elements with an open vector index $\mu$.~However, what is not trivial is the claim that the form factors of the vertex basis 
\er{eqn: ghost_lattice} will be hypercubic invariants.~In Figure \ref{fig: ghost_dress} we provide the results of a hypercubic symmetry test, similar to the one we went over for the 
gluon propagator in Figure \ref{fig: aver_test_glue}.~Before we discuss the data points themselves, we need to clarify two things about the overall setup in the Figure.~First, instead of
the dressing functions of equation \er{eqn: ghost_lattice}, we plot the modified quantities $E', \, F'$ and $G'$, where $E'_{q,\,p} = |q|^3 \cdot E_{q,\,p}$, $F'_{q,\,p} = |q|^3 \cdot F_{
q,\,p}$, and $G'_{q,\,p} = |q|^5 \cdot G_{q,\,p}$\,:~here $|q|$ stands for $\sqrt{q}$.~This momentum-dependent alteration was done for presentation purposes, as without it the functions 
$F$ and $G$ would feature very different scales in the IR and UV energy regions, making it hard to distinguish any details at relatively high momentum $q$.~The form factors $F$ and $G$ 
seemingly diverge in the IR because we work with an un-amputated Greens function:~the amputated version should have the dressing functions which are far more flat at low momenta, see 
e.\,g.~\cite{Cucchieri:2004sq, Cucchieri:2006tf, Maas:2007uv, Cucchieri:2008qm}.~The second important thing concerning Figure \ref{fig: ghost_dress} is that we mostly examine the imaginary
parts of (generally complex-valued) vertex dressings.~This is because the corresponding real parts anyway vanish upon averaging over inversions of momentum components, as illustrated in 
plot \ref{fig: ghost_dress}\,d).~The reason that the real components are nullified upon parity-averaging is explained in Appendix \ref{sec: projectors}. 

\begin{figure}[!t]
\begin{center}
\graph[width = 0.39\tew]{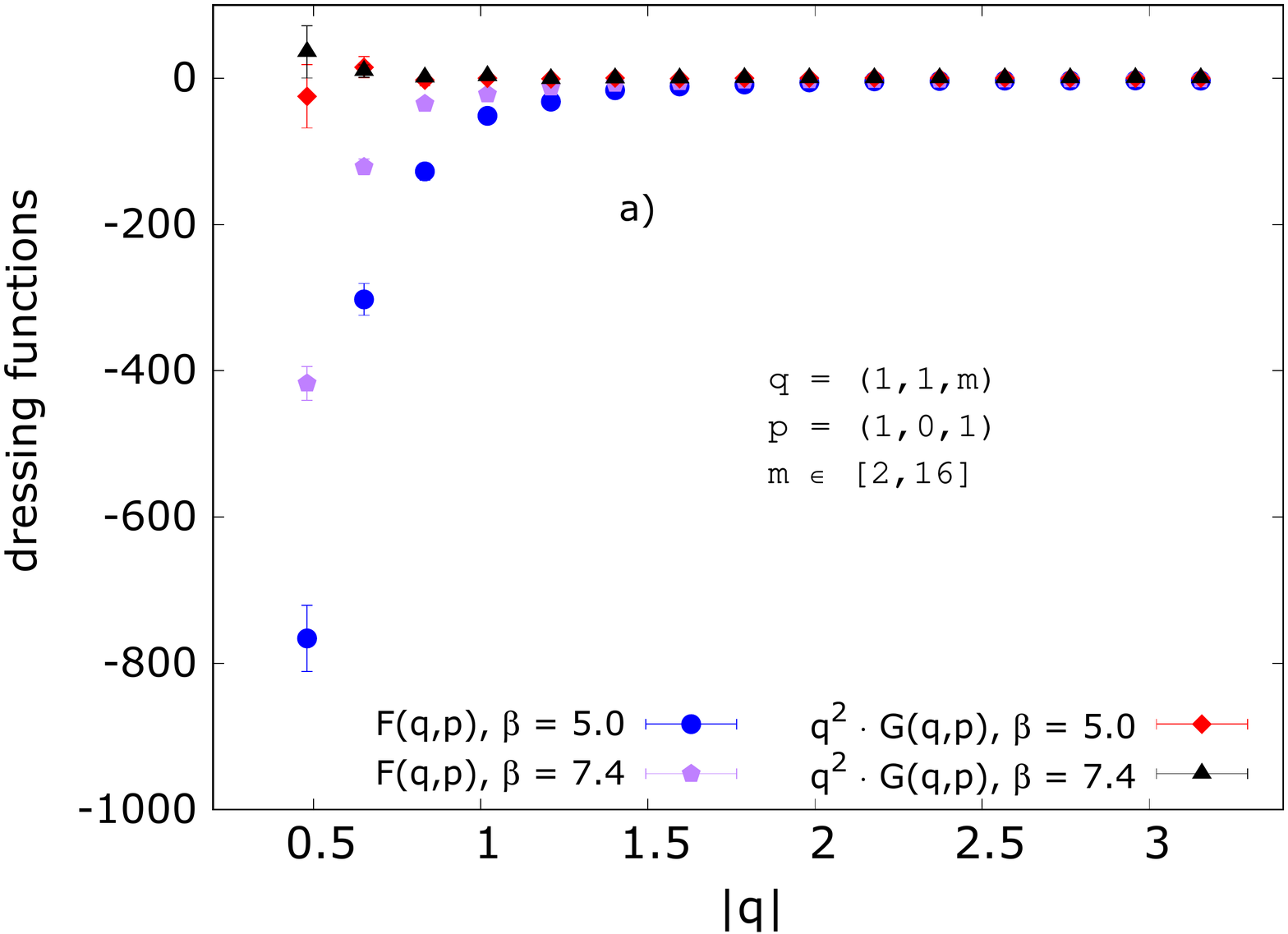}\graph[width = 0.39\tew]{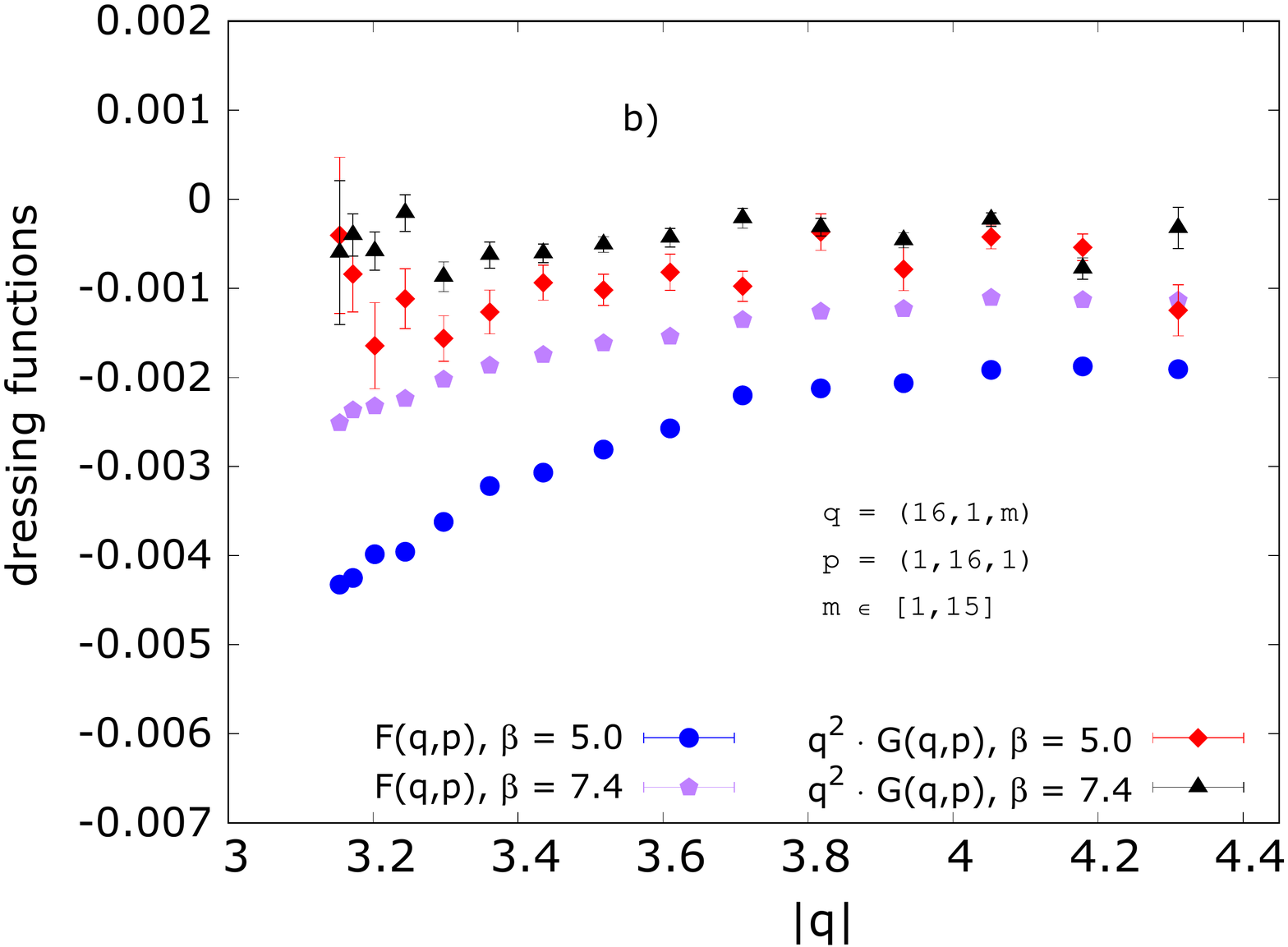}
\graph[width = 0.39\tew]{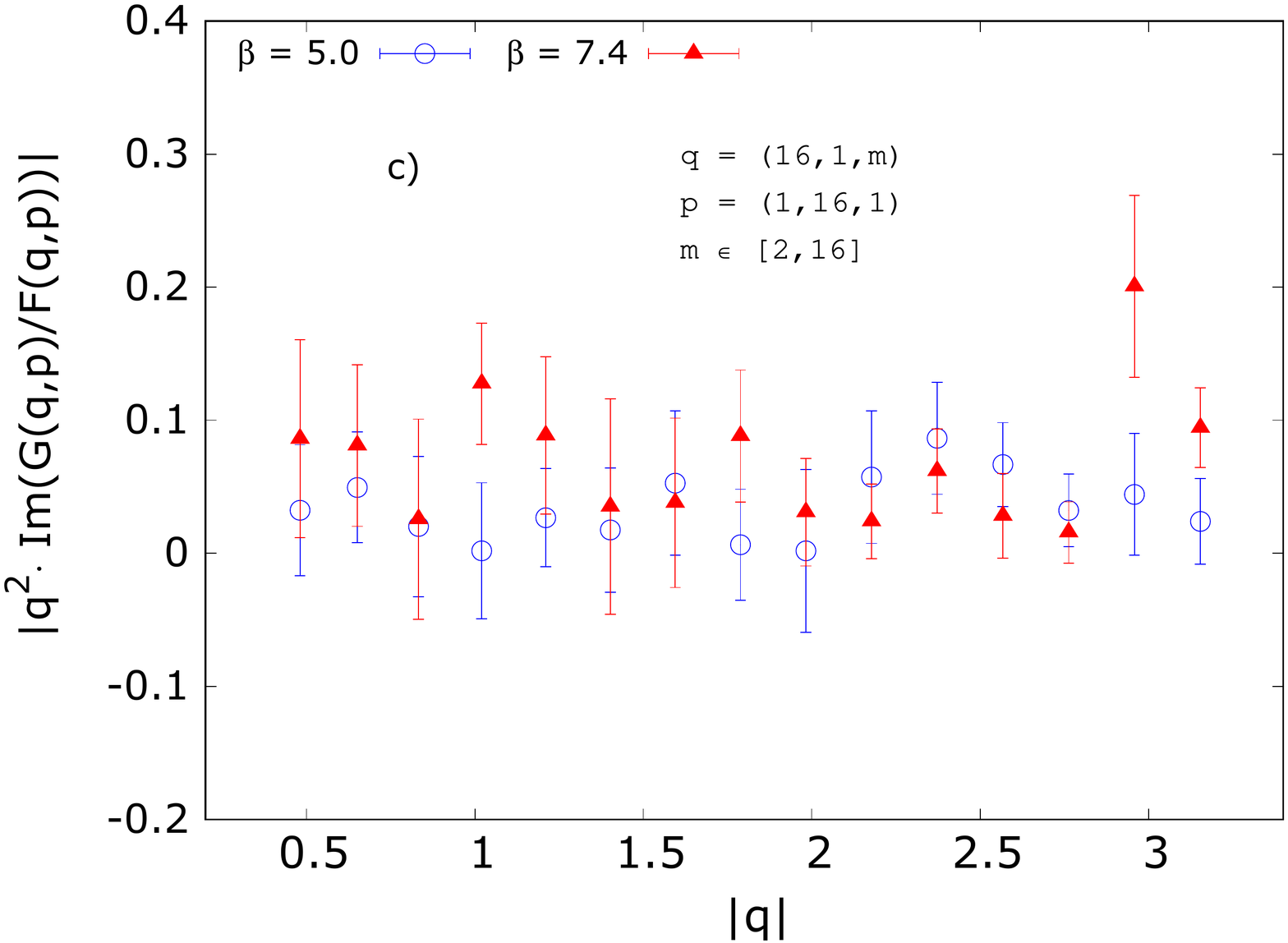}\graph[width = 0.39\tew]{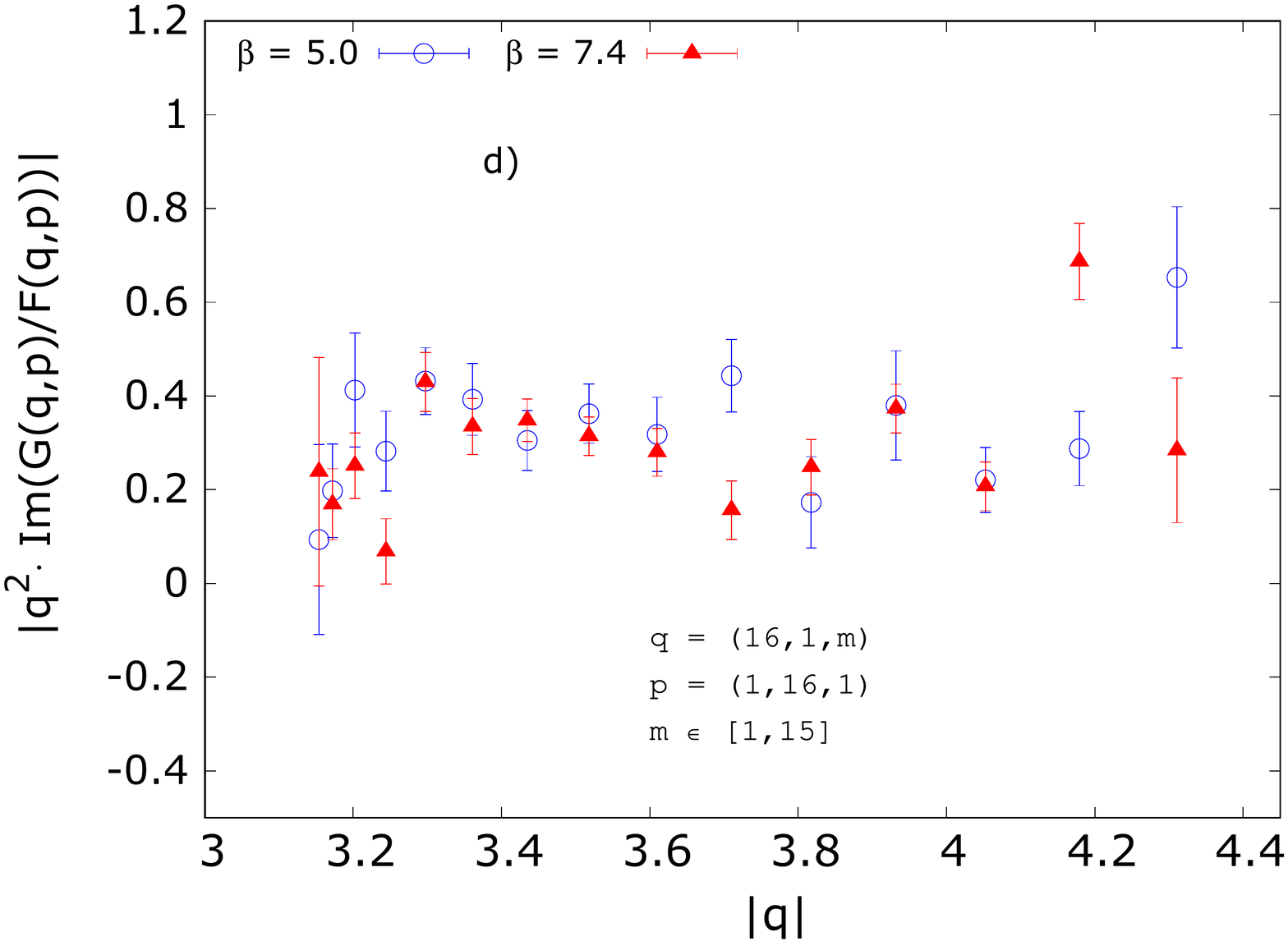}
\caption{Top panel:~Vertex form factors $F(q,p)$ and $q^2 \cdot G(q,p)$ [\,see equations \er{eqn: ghost_lattice}, \er{eqn: project_general}, \er{eqn: matrix_vertex} and \er{eqn: 
determinant}\,], as evaluated on a $32^3$ lattice.~Bottom panel:~absolute value of the ratio $q^2 \cdot G(q,p)/F(q,p)$, for the same kinematics as in the upper two plots.~All points are 
given as functions of $|q| = \sqrt{q^2}$, in lattice units.~$\beta$ is the gauge coupling of \er{eqn: wilson_action}.} 
\label{fig: ratio_vertex}
\end{center}
\end{figure} 

First and foremost, the plots from a) to c) in Figure \ref{fig: ghost_dress} confirm that (imaginary parts of) the dressing functions of \er{eqn: ghost_lattice} are invariant with respect 
to permutations and inversions of momentum components, albeit within rather ``generous'' error bars.~The precision of the hypercubic test can be improved with better statistics, but since
the evaluations of the lattice ghost-gluon correlator are computationally far more expensive than those of the gluon propagator (or three-gluon vertex), for now we've decided to stay with
a relatively modest sample of 480 gauge-fixed field configurations.~Going back to Fig.~\ref{fig: ghost_dress} itself, one may also note that the function $G'_{q,\,p} = |q|^5 \cdot G_{q,\,
p}$ (which has the same mass dimension as the dressings $E'$ and $F'$), is substantially suppressed in the IR region, compared to its continuum counterparts.~This is in accordance with the 
expectation that the vertex should be dominated by the continuum tensor structures, as one goes to lower values for both momenta $p$ and $q$.~However, what is arguably surprising about Figure
\ref{fig: ghost_dress}\,c) is that the values of $G'_{q,\,p}$ are consistent with zero (within big error bars) even when one of the components of $q$ is made to be relatively large.~This 
is probably more a sign of insufficient statistics, than an actual indication that $G'_{q,\,p}$ should be negligibly small in the `UV' region for momentum $q$.~Again, improved statistics 
could lead to more accurate conclusions.~In this regard it ought to be mentioned that extracting a good-quality signal for lattice operators with high(er) mass dimensions is a non-trivial
task even with comparatively large configuration samples:~as an example of this, one may consult Figure 12 in \cite{Vujinovic:2018nqc}.

Even though the signal for the form factor $G_{q,\,p}$ is not particularly good in our current setup, we decided to check if changing the gauge parameter $\beta$ has any appreciable
impact on the relative size of this function, compared to the continuum dressing $F$ (which seems to be the dominant contribution, for kinematics in Fig.~\ref{fig: ghost_dress}).~The
results are shown in Figure \ref{fig: ratio_vertex}, with two choices for the $\beta$ coupling and two kinematic configurations, one with small and the other with relatively large values 
for components of vectors $q$ and $p$.~In the Figure the functions $F_{q,p}$ and $q^2 \cdot G_{q,p}$, which have the same mass dimension, are compared both directly (upper panel) and as a
ratio $|q^2\cdot G/F|$ (lower panel), where $|.|$ denotes an absolute value.~As expected, the ratios $|q^2\cdot G/F|$ are considerably larger when both $q^2$ and $p^2$ have comparatively 
big values, signalling that the lattice corrections to continuum basis decompositions become more prominent in the UV energy region.~Nonetheless, even in the UV the dressing $q^2 \cdot
 G_{q,p}$ appears to be substantially smaller than $F_{q,p}$, which is a non-trivial result.~Concerning the lattice interaction parameter $\beta$, within statistical uncertainties it
has little to no impact on the relative sizes of the two form factors.~Again, this is probably more a consequence of modest statistics than a sign that the gauge coupling has no
influence on the ratios akin to $|q^2\cdot G/F|$.

Based on the data in Figures \ref{fig: ghost_recon} and \ref{fig: ratio_vertex}, it could be said that for most kinematic configurations on the lattice one may neglect the corrections 
to continuum tensor bases, if a quantitatively semi-accurate description of the vertex is desired.~More precisely, the continuum tensor decomposition should arguably be sufficient if 
one finds an uncertainty on the order of five to twenty percent tolerable in a given study.~Investigations where a more precise representation is desired ought to either consider only
special kinematic situations, or use the lattice-modified bases for the correlator $\Gamma_\mu$.~Besides improving the accuracy of the basis decomposition itself, an explicit evaluation
of the lattice-induced dressing functions like $G_{q,p}$ can be useful for testing the continuum extrapolation methods.~Namely, in the continuum a form factor like $G_{q,p}$ ought to 
vanish, and so one expects the ratios akin to those in the bottom panel of Fig.~\ref{fig: ratio_vertex} to go to zero as the continuum limit of the theory is approached.~Vanishing of
these ratios (within error bars) can be used as one of the indicators that the said extrapolations were successful.~On this note, we want to point out that these procedures for the 
ghost-gluon vertex (or indeed any functions beyond the propagators) are quite involved.~Since vertices generally feature multiple momentum variables, there are many more non-continuum 
hypercubic invariants than just those shown in \er{eqn: hyper_scalars}, see e.\,g.~equations \er{eqn: more_invariants} and \er{eqn: matrix_vertex} or section 5 in \cite{deSoto:2007ht}.
\!This means that, in order to eliminate all of the non-continuum scalars of a given mass dimension, many data points are needed to perform the extrapolations with reasonable precision.~In 
case of the ghost-gluon correlator the problem is further aggravated by the fact that the lattice Monte Carlo calculations of this function are numerically quite expensive, as mentioned 
before.~Attempts to improve the situation might constitute an interesting, albeit difficult, research topic for future studies. 

\section{Conclusions and outlook}\label{sec: conclude}

In this paper we have presented a way of deriving the tensor bases for lattice vertex functions, such that the corresponding form factors are invariant under the hypercubic symmetry 
transformations.~We've used the method to derive the most general possible (barring the finite volume artifacts) basis structures for lattice tensors of first and second rank, with 
up to two independent momenta in the former, and a single kinematic vector-like variable in the latter case.~The lowest-order non-continuum variants of these decompositions were 
applied to the ghost-gluon vertex and gluon propagator of lattice Monte Carlo simulations, resulting in a few interesting observations.~First, it was shown analytically and confirmed 
numerically that there exist special momentum configurations wherein the tensor structures of both correlators reduce to their continuum form.~For the gluon propagator $D_{\mu\nu}(p)$ 
in three dimensions, special kinematic situations correspond to the on-axis momentum $p = (m, 0, 0)$, the diagonal vector $p = (m,m,m)$, and the `in-between' points $p = (m,m,0)$:~any 
non-equivalent permutations of $p$ components are also allowed.~For the ghost-gluon vertex $\Gamma_\mu(q,\,p)$, all of the possible special combinations will not be given here, but we
merely state that one of these is the fully diagonal kinematic choice with $p = (m,m,m)$ and $q = (n,n,n)$.~The second notable result is that the rate at which the gluon propagator 
approaches its continuum form in the infrared is dictated solely by the numerical gauge-fixing algorithm:~it is however questionable if it is worthwhile to invest effort in improving 
this situation, since for momenta $\sqrt{p^2} \leq 1$ (in lattice units), the finite spacing incurs a quantitative effect below five percent, see Figure \ref{fig: beta_glue}.~We also
commented on how this reflects on the lattice investigations of the anomalous magnetic moment of the muon and argued that discretisation artifacts are negligible at the relevant energy 
scales.

As possible future applications of our framework, we've already discussed how it can be used to directly test some of the continuum extrapolation methods, but no actual results of this 
kind were provided.~We leave such endeavours for future investigations.~It also remains to be seen how the symmetry-based lattice modifications may affect some other correlators of 
interest, like the three-gluon or quark-gluon vertices.~While the three-gluon interaction kernel was briefly considered in section \ref{sec: gluon_lattice}, for now we've completely 
ignored the spinor fields and related $n$-point functions.~This is because we are not yet certain about all of the possible generalisations in this regard, when going from continuum to
discretised spacetimes:~for single-momentum functions we expect for hypercubic symmetry to allow for additional couplings apart from $\gamma \cdot p$, where $\gamma_\mu$ are the Euclidean
Dirac matrices and $p_\mu$ is the appropriate momentum vector.~This too constitutes an interesting research topic for the future, especially as one expects for finite spacing artifacts
to be more pronounced for fermions that bosons, see \cite{August:2013jia} as an example.~Finally, we are yet to check if the low-momentum discrepancy seen in Figure \ref{fig: beta_glue} 
for the gluon propagator is a finite volume effect:~if this turns out to be true, then perhaps our formalism may allow one to quantify such deviations as well, or indeed any (un)expected 
alterations with respect to the continuum tensor form, for lattice correlators of interest.    

\section*{ACKNOWLEDGMENTS} 

We gratefully acknowledge the support of the Austrian science fund FWF, under Schr\"odinger grant J3854-N36.~Parts of the numerical simulations were done on HPC clusters of the 
University of Graz.

\appendix

\section{Constructing the projectors for tensor basis elements}\label{sec: projectors}

\subsection{Basic principles}

In this Appendix, we will show how the projectors are constructed for various tensor decompositions employed in this paper.~We begin with a quick review of the 
basic ideas:~although this is standard textbook material (see e.\,g.~\cite{Hassani:1999sny}), we think that the description of the underlying procedure is useful 
since it will be used repeatedly in the following.~We start by assuming that one is working with a correlator $\Gamma$ whose tensor basis includes $N$ elements, or
explicitly
\begin{align}
\Gamma = \sum_{j = 1}^N \, \mathcal{F}^{\,j} \, \tau^{\,j} \, ,
\end{align} 

\noindent
where $\tau^{\,j}$ denotes the $j$-th tensor structure, and $\mathcal{F}^{\,j}$ is the corresponding form factor.~Whatever the quantum numbers of the vertex $\Gamma$ and 
its basis, the fundamental principle for obtaining the projectors for $\tau^{\,j}$ is always the same, and it amounts to a matrix inversion problem \cite{Hassani:1999sny}.
\!Namely, one starts by constructing the $N \times N$ matrix of products of basis elements, i.\,e.
\begin{align}\label{eqn: define_m}
M_{kj} = \tau^{\,k} \cdot \tau^{\,j} \, , \qquad k,\:j = 1, \ldots N \, ,
\end{align} 

\noindent
where the dot ($\cdot$) denotes whatever scalar product which is appropriate for the structures $\tau^{\,j}$.~The next step is to invert the matrix $M$, thus obtaining the 
operator $M^{-1}$.~From this, a projector for the $k$-th element $\tau^{\,k}$ follows as 
\begin{align}\label{eqn: project_general}
P^{\,k} = \sum_{j=1}^N \, M^{-1}_{kj} \, \tau^{\,j} \, , \qquad k = 1, \ldots N \, .
\end{align} 

From the above considerations, one can also easily see why no well-defined projectors can be constructed for redundant bases, i.\,e.~for tensor descriptions which feature linearly
dependent elements.~As an example, suppose that the structure (say) $\tau^{\,3}$ can be written as a linear combination of tensors $\tau^{\,1}$ and $\tau^{\,2}$, so that  
\begin{align}\label{eqn: linear_combo}
\tau^{\,3} = \mathcal{C}_1 \, \tau^{\,1} + \mathcal{C}_2 \, \tau^{\,2} \, , 
\end{align}

\noindent
with some non-vanishing constants $\mathcal{C}_1$ and $\mathcal{C}_2$.~In this situation, the third row of the matrix $M$ of \er{eqn: define_m} will be a linear combination
of its first and second rows.~However, it is well known that matrices which have at least one row or (respectively) column, that can be written as a linear combination of 
other rows, or (respectively) columns, are singular \cite{Hassani:1999sny}.~In other words, the said matrices have a vanishing determinant and cannot be inverted, thus making
the projectors of \er{eqn: project_general} ill-defined.~The argument straightforwardly generalises from the case \er{eqn: linear_combo} to whatever other linear combinations 
of basis structures one may think of.~With this, we may finally begin our projector constructions.

\subsection{Projectors for the gluon propagator}

We commence with a straightforward example of the continuum gluon propagator, with a tensor description 
\begin{align}\label{eqn: append_cont_glue}
D_{\mu\nu}(p) = A(p) \, \delta_{\mu\nu} + B(p) \, p_\mu \, p_\nu \, .
\end{align} 

As discussed above, to obtain the projectors for these basis elements one first computes the matrix $M$ akin to \er{eqn: define_m}, which in this case has the
form 
\begin{align}
M_{kj} = \tau^{\,k}_{\mu\nu} \cdot \tau^{\,j}_{\mu\nu} = \sum_{\mu = 1} ^d \sum_{\nu = 1}^d \tau^{\,k}_{\mu\nu} \, \tau^{\,j}_{\mu\nu} \, , \qquad k,\: j = 1, 2 \, , 
\end{align} 

\noindent
where $\tau_{\mu\nu}^1 = \delta_{\mu\nu}$ and $\tau_{\mu\nu}^2 = p_\mu \, p_\nu$.~The full matrix in $d$ dimensions is  
\begin{align}\label{eqn: matrix_cont}
M = \left[ \begin{array}{cc} d & \:\: p^2 \\[0.28cm] p^2 & \:\: \left(p^2\right)^2  \end{array} \right] \, , 
\end{align} 

\noindent
and its inverse is given by (for $p \neq 0$ and $d > 1$)\,:
\begin{align}
M^{-1} = \frac{\left(p^2\right)^{-2} }{(d - 1)} \cdot \left[ \begin{array}{cc} \left(p^2\right)^2 & \:\: - p^2 \\[0.27cm] - p^2 & \:\: d  \end{array} \right] \, . 
\end{align} 

According to the relation \er{eqn: project_general}, projectors for the basis elements $\tau_{\mu\nu}^1$ and $\tau_{\mu\nu}^2$ now follow as 
\begin{align}\label{eqn: proj_cont_glue}
& P^{\,1}_{\mu\nu} \: = \: P^{\,A}_{\mu\nu} \: = \: M^{-1}_{11} \, \tau_{\mu\nu}^1 \: + \: M^{-1}_{12} \, \tau_{\mu\nu}^2 \: = \: \frac{1}{d - 1} \left( \delta_{
\mu\nu} - \frac{p_\mu \, p_\nu}{p^2} \right) \, , \nonumber \\[0.17cm]
& P^{\,2}_{\mu\nu} \: = \: P^{\,B}_{\mu\nu} \: = \: M^{-1}_{21} \, \tau_{\mu\nu}^1 \: + \: M^{-1}_{22} \, \tau_{\mu\nu}^2 \: = \: \frac{1}{d - 1} \left( \frac{-
\delta_{\mu\nu}}{p^2} + \frac{d \, p_\mu\, p_\nu}{ \left(p^2\right)^2} \right) \, .
\end{align} 

We now want to apply a similar method to obtain the dressing functions of the lattice-modified basis for $D_{\mu\nu}(p)$, i.\,e.~the representation 
\begin{align}\label{eqn: append_latt_glue}         
&D_{\mu\mu} = E(p)\, \delta_{\mu\mu} \: + \: F(p) \, p_\mu^{\,2} \, , \qquad \quad \mu = 1, \ldots d \nonumber \\ 
&D_{\nu\mu} = G(p) \, p_\nu p_\mu \, , \qquad \qquad \qquad \quad \: \:  \mu, \, \, \nu = 1, \ldots d \, , \qquad \mu \neq \nu \, . 
\end{align} 

Diagonal and off-diagonal components of $D_{\mu\nu}$ transform independently of each other, under hypercubic symmetry transformations, and they will thus be treated separately 
in the following.~In particular, we will abandon the standard definition of a Lorentz-contraction of second rank tensors, i.\,e.  
\begin{align}       
\tau^{1}_{\mu\nu} \cdot \tau^{2}_{\mu\nu} = \sum_{\mu,\nu} \: \tau^{1}_{\mu\nu} \, \tau^{2}_{\mu\nu} \: ,  
\end{align} 

\noindent
in favour of its diagonal ($\ast$) and off-diagonal ($\star$) variants, as follows:
\begin{align}\label{eqn: new_products}      
&\tau^{1}_{\mu\nu} \ast \tau^{2}_{\mu\nu} = \sum_{\mu} \: \tau^{1}_{\mu\mu} \, \tau^{2}_{\mu\mu} \: , \nonumber \\ 
&\tau^{1}_{\mu\nu} \star \tau^{2}_{\mu\nu} = \sum_{\substack{\mu, \nu \\ \mu \neq \nu}} \: \tau^{1}_{\mu\nu} \, \tau^{2}_{\mu\nu} \: .
\end{align} 
   
Both of the above products will be hypercubic scalars (i.\,e.~they will be invariant under permutations of inversions of momentum components), provided that the momentum-space 
quantities $\tau^{1}_{\mu\nu}$ and $\tau^{2}_{\mu\nu}$ transform as tensors under the aforementioned symmetry operations.~To take the example of permutations, both contractions in 
\er{eqn: new_products} contain all possible combinations of tensor indices, all will thus be invariant under an arbitrary reshuffling of the said indices:~for instance, in a
three-dimensional theory the diagonal contraction is 
\begin{align}      
\tau^{1}_{\mu\nu} \ast \tau^{2}_{\mu\nu} \: = \: \tau^1_{11} \, \tau^2_{11} \, + \, \tau^1_{22} \, \tau^2_{22} \, + \, \tau^1_{33} \, \tau^2_{33} \, ,
\end{align} 

\noindent
and it remains unchanged under all elementary permutations $(1 \leftrightarrow 2, \, 1 \leftrightarrow 3, \: \text{and} \: 2 \leftrightarrow 3)$ and their combinations.~The same remark
holds for the off-diagonal contraction in \er{eqn: new_products}.~When it comes to parity changes, the invariance of the diagonal product is trivial because the diagonal components of
second-rank tensors do not change under such transformations, see equation \er{eqn: gluon_reflected_2d} as an example.~The off-diagonal terms do get modified under inversions [\,again,
see \er{eqn: gluon_reflected_2d} as an example\,], but since both $\tau^1_{\mu\nu}$ and $\tau^2_{\mu\nu}$ (with $\mu \neq \nu$) change in the same way, the overall off-diagonal product
will remain constant in this regard.~Now, let us go back to the gluon propagator, and in particular to its diagonal part $D_{\mu\mu}$.~From \er{eqn: append_latt_glue} one can see that
this function has two basis elements, $\delta_{\mu\mu}$ and $p_\mu^2$, and the corresponding projectors can be obtained with the same (essentially) procedure as for the continuum 
decomposition \er{eqn: append_cont_glue}.~The matrix of products of basis structures is 
\begin{align}
M^\text{\,diag}_{kj} = \tau^{\,k}_{\mu\nu} \ast \tau^{\,j}_{\mu\nu} = \sum_{\mu} \tau^{\,k}_{\mu\mu} \, \tau^{\,j}_{\mu\mu} \, , \qquad k,\: j = 1, 2 \, , 
\end{align}

\noindent
where $\tau^1_{\mu\nu} = \delta_{\mu\nu}$ and  $\tau^2_{\mu\nu} = p_\mu \, p_\nu$.~The explicit form of $M^\text{\,diag}$ is  
\begin{align}\label{eqn: diag_m}
M^\text{\,diag} = \left[ \begin{array}{cc} d & \quad \: p^2  \\[0.30cm] p^2 & \quad \: p^{\,[4]} \!\! \end{array} \right] \, , 
\end{align}

\noindent
with a hypercubic invariant $p^{\,[4]} = \sum_{\mu = 1}^{d} \, p_\mu^{\,4}$.~The inverse of the above operator is 
\begin{align}
M^{\text{\,diag}, \, -1 } = \frac{1}{d \, p^{\,[4]} - (p^2)^2} \cdot \left[ \begin{array}{cc}  p^{\,[4]} & \:\: - p^2 \\[0.27cm] - p^2 & \:\: d  \end{array} \right] \, . 
\end{align}

The projectors for the dressing functions $E(p)$ and $F(p)$ now follow as in equation \er{eqn: proj_cont_glue}, with an appropriate substitution $M^{-1} \rightarrow M^{
\text{\,diag},\, -1}$.~The form factors $E(p)$ and $F(p)$ themselves may be obtained as suitable (diagonal) contractions with the gluon propagator, i.\,e.
\begin{align}\label{eqn: diag_ffs}
&E(p) \: = P^{\,E}_{\mu\nu} \ast D_{\mu\nu} \: = \: \frac{ p^{\,[4]} \sum_{\mu} D_{\mu\mu} \:\: - \:\: p^2 \sum_{\mu} p_\mu^2 \, D_{\mu\mu} }{d \, p^{\,[4]} \, - \, (p^2)^2} \, , 
\nonumber \\[0.2cm] 
&F(p) \: = P^{\,F}_{\mu\nu} \ast D_{\mu\nu} \: = \: \frac{ - p^2 \sum_{\mu} D_{\mu\mu} \:\: + \:\: d \sum_{\mu} p_\mu^2 \, D_{\mu\mu} }{d \, p^{\,[4]} \, - \, (p^2)^2} \, ,
\end{align}

\noindent
wherein we used the fact that $\delta_{\mu\mu} = 1$.~Concerning the off-diagonal propagator $D_{\mu\nu} \, (\text{with}\: \mu \neq \nu)$ in \er{eqn: append_latt_glue}, it has only 
a single basis element, and the corresponding dressing function $G(p)$ can be immediately projected out as 
\begin{align}
G(p) \: = \: \frac{p_\mu \, p_\nu \star D_{\mu\nu}}{p_\mu \, p_\nu \star p_\mu \, p_\nu} \: =  \: \frac{\sum_{\substack{\mu, \nu \\ \mu \neq \nu}} p_\mu \, p_\nu \, D_{\mu\nu
}}{\sum_{\substack{\mu, \nu \\ \mu \neq \nu}} p_\mu^2 \, p_\nu^2} \, .
\end{align}

Now, for an arbitrary $d$-dimensional vector $p$ it holds that 
\begin{align}\label{eqn: mom_off_diag}
\sum_{\substack{\mu, \nu \\ \mu \neq \nu}} p_\mu^2 \, p_\nu^2 \: = \:  \left(p^2\right)^2 - p^{\,[4]} \, ,
\end{align}

\noindent 
so that $G(p)$ may also be written as    
\begin{align}\label{eqn: off_diag_ff}
G(p) \: = \: \frac{\sum_{\substack{\mu, \nu \\ \mu \neq \nu}} p_\mu \, p_\nu \, D_{\mu\nu}}{(p^2)^2 - p^{\,[4]}} \, , 
\end{align}

\noindent 
which is the form used in the main body of the paper.~The relation \er{eqn: mom_off_diag} will not be proven for a general dimension number $d$, but will only be demonstrated for 
a three-dimensional case:~from this the ingredients for a general proof can be easily deduced.~For any vector in three dimensions, $p = (p_1, p_2, p_3)$, one has that 
\begin{align}
\left(p^2\right)^2 \: = \: (p_1^2 + p^2_2 + p_3^2) \cdot (p_1^2 + p^2_2 + p_3^2) \: = \: p_1^4 \, + \, & p_2^4 \, + \, p_3^4 \, + \, p_1^2 \, p_2^2 \, + \, p_2^2 \, p_1^2 \, + \, 
p_1^2 \, p_3^2 \, + \, p_3^2 \, p_1^2 \, + \, p_2^2 \, p_3^2 \, + \, p_3^2 \, p_2^2 \: = \nonumber \\  
& p^{\,[4]} \, + \, \sum_{\substack{\mu, \nu \\ \mu \neq \nu}} p_\mu^2 \, p_\nu^2 \, ,
\end{align}

\noindent
from which a three-dimensional variant of \er{eqn: mom_off_diag} directly follows.~Now, in section \ref{sec: 2d_glue} we claimed that the form factors \er{eqn: diag_ffs} are ill-defined 
for diagonal momenta $p$.~We now want to show this explicitly.~In $d$ dimensions, a vector pointing along the lattice diagonal will have the form 
\begin{align}
p = \overbrace{(m, \, m, \, m, \, \ldots m)}^{d \, \text{terms}} \, , 
\end{align}  

\noindent 
and its relevant scalar invariants are $p^{\,2} = d \, m^2$ and $p^{\,[4]} = d \, m^4$.~Plugging these into \er{eqn: diag_ffs}, and using the fact that $p^{\,2}_\mu = m^2$ for all
values of the index $\mu$, one sees that both $F(p)$ and $E(p)$ evaluate to an ambiguous expression of the kind ``0/0''.~As mentioned before, this ambiguity has to do with the fact
that the diagonal basis elements $\delta_{\mu\mu}$ and $p_\mu^{\,2}$ are not linearly independent, for diagonal momentum vectors $p$, and the corresponding matrix $M^\text{\,diag}$ 
of \er{eqn: diag_m} is singular at such momentum points. 

At the end of section \ref{sec: gluon_lattice}, we also argued that the basis \er{eqn: append_latt_glue} is complete in two dimensions, since it exhausts all of the free parameters  
which may be present in a symmetric $2 \times 2$ matrix, see also equation \er{eqn: nd_def} and reference \cite{Morty:1962prc}.~Here, we want to show what this means ``in practice'',
by proving that the leading-order correction to the gluon tensor decomposition  [\,equation \er{eqn: gluon_higher}\,], can be described completely by the basis elements \er{eqn:
append_latt_glue}.~The second rank tensor of the kind \er{eqn: gluon_higher}, with the next-to-lowest mass dimension is (the lowest mass term is the continuum factor $p_\mu \, 
p_\nu$):
\begin{align}
\tau^\text{lead}_{\mu\nu}(p) \, = \, p_\mu \, p_\nu \, (p_\mu^2 \: + \: p_\nu^2) \, .
\end{align}

We first calculate the dressing functions of the basis \er{eqn: append_latt_glue}, pertaining to the above structure.~For the form factor $E(p)$ of the element $\tau^\text{lead}_{
\mu\nu}$, one gets (using the fact that $\tau^\text{lead}_{\mu\mu} = 2 \, p_\mu^4$)
\begin{align}
E(p) = \frac{ 2 \, \left(p^{\,[4]}\right)^2 - 2 \,p^2 \, p^{\,[6]}}{2 \, p^{\,[4]} - (p^2)^2} = - 2 \, p^2_1 \, p^2_2 \, , 
\end{align} 

\noindent
where $p_1$ and $p_2$ are the components of a two-dimensional vector $p = (p_1, p_2)$.~In the same vein, it follows that $F(p) = 2\, p^2 $ and $G(p) = p^2$.~The tensor quantity $
\tau^\text{lead}_{\mu\nu}$ can thus be represented as 
\begin{align}\label{eqn: mostra_higher}
&\tau_{\mu\mu}^\text{lead} \: = \: 2 \, p_\mu^4 = - \, 2 \, p^2_1 \, p^2_2 \, \delta_{\mu\mu} \, + \, 2 \, p^2 \, p_\mu^2 \: , \, \qquad \mu = 1, 2 \, , \nonumber \\ 
&\tau_{12}^\text{lead} \: = \: (p_1^2 + p_2^2) \, p_1 \, p_2 = p^2 \, p_1 \, p_2 \, ,   
\end{align} 

\noindent
and similarly for the other off-diagonal term $\tau^\text{lead}_{21}$.~One can easily check that the above relations are trivially true, i.\,e.~all the sides of the equation match 
each other.~Therefore, the quantity $\tau^\text{lead}_{\mu\nu}$ can indeed be described uniquely and completely by a decomposition of the kind \er{eqn: append_latt_glue}, in two 
dimensions.~Similar demonstrations can be done for any terms of the form \er{eqn: gluon_higher}, and the procedure can also be applied in higher-dimensional settings.~However,
even though the coefficients $E(p), \, F(p)$ and $G(p)$ would still be correct, when calculated in more than two dimensions, the equations like \er{eqn: mostra_higher} would no 
longer hold:~the left- and right-hand sides of such relations would not match.~This is because the representation \er{eqn: append_latt_glue} is incomplete, for $d \geq 3$, and it 
will not describe an arbitrary symmetric tensor $\tau_{\mu\nu}$ without loss of information.~One may consult \cite{Morty:1962prc} for further details.  

\subsection{Projectors for the ghost-gluon vertex} 

A continuum ghost-gluon vertex is described fully by two basis structures.~We shall take these to correspond to the antighost (denoted $q$ in this paper) and ghost ($p$) momentum,
so that 
\begin{align}
\Gamma_\mu(q,p) = A(q,p) \, q_\mu + B(q,p) \, p_\mu \, .   
\end{align}

Again, the method for obtaining the projector functions for the above elements is the same (in principle) as for the continuum gluon propagator \er{eqn: append_cont_glue}, or even
the diagonal lattice contribution $D_{\mu\mu}$ in \er{eqn: append_latt_glue}.~The appropriate matrix of products of basis tensors is (with $\tau^1_\mu = q_\mu$ and $\tau^2_\mu = 
p_\mu$)
\begin{align}
M^\text{\,vert} = \left[ \begin{array}{cc} q^2 & \quad  q\cdot p  \\[0.30cm] q\cdot p & \quad  p^2 \!\! \end{array} \right] \, , 
\end{align}

\noindent
and its inversion and subsequent application of \er{eqn: project_general} immediately gives the expressions of \er{eqn: ghost_project}.~We now proceed to the somewhat harder case of
a lattice-modified decomposition for the ghost-gluon correlator, where we choose
\begin{align}\label{eqn: append_ghost_latt}
\Gamma_\mu(q, p) = E(q,p) \, q_\mu + F(q,p) \, p_\mu \, + G(q,p) \, q^{\,3}_\mu \, .
\end{align}

For the above basis we will not write down explicitly the $3 \times 3$ operator $M$ akin to \er{eqn: define_m}.~However, we do want to point out that the said matrix will have 
some unusual entries, by standards of a continuum field theory.~By denoting the basis elements in numerical order as $\tau^1_\mu = q_\mu$, $\tau^2_\mu = p_\mu$ and $\tau^3_\mu
= q^{\,3}_\mu$, the matrix components (for example) $M_{23} = M_{32}$ and $M_{33}$ would be 
\begin{align}\label{eqn: more_invariants}
M_{23} \: = \: \sum_{\mu = 1}^3 \, p_\mu \, q^{\,3}_\mu \: = \: p^{(1} q^{3)} \:, \qquad \: \: M_{33} \: = \: \sum_{\mu = 1}^3 \, q^{\,6}_\mu = q^{\,[6]} \, ,  
\end{align}
 
\noindent
where we have introduced the notation 
\begin{align}
 p^{(2m + 1} \, q^{2 n + 1)} = \sum_{\mu = 1}^3 \,  p^{\,2 m + 1}_\mu \, q^{\,2 n + 1}_\mu \: , \qquad \: m, \, n \in N_0 \, ,  
\end{align}

\noindent
for hypercubic invariants consisting of two momentum vectors.~To avoid the possibility of confusion, until the end of this section the standard continuum scalar products will also be
written with hypercubic notation, so that (for example) $p^{\,2} = p^{\,[2]}$ and $p\cdot q = p^{(1} q^{1)}$.~Writing out the rows of $M^{-1}$ as row-vectors, one then has 
\begin{align}\label{eqn: matrix_vertex} 
&M^{-1}(1,:) \: = \: \left[\, \big( p^{(1} q^{3)} \big)^2 - p^{\,[2]} \, q^{\,[6]} \, , \qquad - p^{(1} q^{3)} \, q^{\,[4]} + p^{(1} q^{1)} \, q^{\,[6]} \, , \qquad - p^{(1}
q^{3)} \, p^{(1} q^{1)} + p^{\,[2]} \, q^{\,[4]}  \,\right] \, , \nonumber \\
&M^{-1}(2,:) \: = \: \Big[\,- p^{(1} q^{3)} \, q^{\,[4]} + p^{(1} q^{1)} \, q^{\,[6]} \, , \qquad \big( q^{\,[4]} \big)^2 - q^{\,[2]} \, q^{\,[6]} \, , \qquad - p^{(1} q^{3)} \, 
q^{\,[2]} - p^{(1} q^{1)} \, q^{\,[4]} \Big] \, ,  \\
&M^{-1}(3,:) \: = \: \left[\,- p^{(1} q^{3)} \, p^{(1} q^{1)} + p^{\,[2]} q^{\,[4]} \, , \qquad p^{(1} q^{3)} \, q^{\,[2]}  - p^{(1} q^{1)} \, q^{\,[4]} \, , \qquad 
( p^{(1} q^{1)})^2 - p^{\,[2]} \, q^{\,[2]} \right] \, . \nonumber
\end{align}

Additionally, all of the elements in $M^{-1}$ are to be divided out with the common prefactor 
\begin{align}\label{eqn: determinant}
(\text{det})^{-1} = \left( \big(p^{(1} q^{3)} \big)^2 \, q^{\,[2]} - 2\,p^{(1} q^{1)} \, p^{(1} q^{3)} \, q^{\,[4]} + p^{\,[2]} \, \big(q^{\,[4]}\big)^2 + \big(p^{(1} q^{1)}\big)^2
\,q^{\,[6]} - p^{\,[2]} \, q^{\,[2]} \, q^{\,[6]} \right)^{-1} \, ,
\end{align}

\noindent
which is the inverse of the determinant of $M$.~The projectors for \er{eqn: append_ghost_latt} are then obtained as linear combinations of basis elements themselves, according to 
the relation \er{eqn: project_general}, where $P^{\,1}_\mu = P^{\,E}_\mu$,\, $P^{\,2}_\mu = P^{\,F}_\mu$ and $P^{\,3}_\mu = P^{\,G}_\mu$.~Note that all of the momentum products 
which comprise the operator $M^{-1}$ are hypercubic scalars.~For instance, suppose that we flip the sign on a second momentum component, so that $q' = (q'_1, q'_2, q'_3) = (q_1,
-q_2, q_3)$, and similarly for the vector $p$.~Then the contraction (say) $p^{(1} q^{3)}$ would ``change'' as 
\begin{align}
p^{(1} q^{3)} \rightarrow (p^{(1} q^{3)})' \: = \: p'_{\,1} \, q'^{\,3}_1 + p'_{\,2} \, q'^{\,3}_2  + p'_{\,3} \, q'^{\,3}_3 \: = \: p_{\,1} \, q^{\,3}_1 + (-p_2 ) (-q^{\,3}_2) + 
p_{\,3} \, q^{\,3}_3 = p^{(1} q^{3)} \, . 
\end{align}

Thus, the said product is actually not modified in any non-trivial way under the considered inversion change.~In the same manner, all of the entries in $M^{-1}$ remain constant under
arbitrary permutations and parity transformations of momentum components.~The projectors for the basis \er{eqn: append_ghost_latt} therefore inherit the vector-like transformation 
properties from the basis elements themselves.

In section \ref{sec: ghost_glue} we claimed that the inversion averages of vertex dressing functions are purely imaginary.~In other words, if the form factors in \er{eqn: append_ghost_latt}
(which are in general complex numbers) were to be averaged over all momenta of the kind $q^{\pm} = (q_1^\pm, q_2^\pm, q_3^\pm)$, with all possible combinations of plus and minus signs, (and
similarly for momentum $p$), the end result will neccessarily have no real parts.~Here we will discuss why this happens.~For this demonstration we will use the identity 
\begin{align}\label{eqn: vertex_cc}
\Gamma_\mu(-q, -p) = \Gamma^\mathcal{\,C}_\mu(q,p) \, ,  
\end{align}

\noindent
for the numerical ghost-gluon vertex, where $\mathcal{C}$ denotes complex conjugation.~The above relation is proven in Appendix \ref{eqn: inversions_append}.~Under a full inversion, 
wherein \textit{all} momentum components change sign, all of the projectors for the basis \er{eqn: append_ghost_latt} transform as 
\begin{align}\label{eqn: inver_proj}
P_\mu(-q,-p) = - P_\mu(q,p) \, . 
\end{align}

The above rule follows from vector-like features of basis elements \er{eqn: append_ghost_latt}, under a full parity change.~Now suppose that we average the form factor (say) $E(q,p)$ 
over two sets of momenta, $(q,p)$ and $(-q,-p)$, so that 
\begin{align}\label{eqn: e_aver}
&E(q,p) \, + \,  E(-q,-p) \: = \: P_\mu^{\,E} (q,p) \cdot \Gamma_\mu(q,p) \, + \, P_\mu^{\,E} (-q,-p) \cdot \Gamma_\mu(-q,-p) \: \stackrel{!}{=} P_\mu^{\,E} (q,p) \cdot \Gamma_\mu(q,
p) \: - \nonumber \\ & P_\mu^{\,E} (q,p) \cdot \Gamma^\mathcal{\,C}_\mu(q,p) \: = \: P_\mu^{\,E} (q,p) \cdot \big(\Gamma_\mu(q,p) - \Gamma^{\, \mathcal{C}}_\mu(q,p)\big) \: = \: 2\, 
P_\mu^{\,E} (q,p) \cdot \text{Im}\left(\Gamma_\mu(q,p)\right) \, ,
\end{align}

\noindent
where ``Im'' denotes the imaginary part of the vertex.~In the ``!'' step above, we used both \er{eqn: vertex_cc} and \er{eqn: inver_proj}.~Note that in the above derivation we left out 
an averaging prefactor of 1/2, for reasons of brevity.~Equation \er{eqn: e_aver} should make it clear that the function $E(q,p)$, and indeed any of the vertex dressings in \er{eqn:
append_ghost_latt}, will be purely imaginary, when averaged over momentum sets $(q,p)$ and $(-q,-p)$.~Regarding the full parity averages, done over all possible inversions of momentum 
components, the same general conclusion applies.~To take the example of a three-dimensional vector $p = (p_1, p_2, p_3)$, the complete inversion average would include the following 
momenta:
\begin{align}\label{eqn: parity_aver}
& p^i = (p_1, p_2, p_3) \, , \quad p^{ii} = (-p_1, p_2, p_3) \, , \quad p^{iii} = (p_1, - p_2, p_3) \, , \quad p^{iv} = (p_1,  p_2, - p_3) \, , \\ 
& p^{v} = (-p_1, -p_2, p_3) \, , \quad p^{vi} = (-p_1, p_2, -p_3) \, , \quad p^{vii} = (p_1, - p_2, -p_3) \, , \quad p^{iv} = (-p_1,  -p_2, - p_3) \, . \nonumber
\end{align}

One may now observe that the above vectors can be organised into four pairs of ``parity opposites'', i.\,e.~pairs of momenta which differ from each other by a full inversion.~Concretely,  
momentum $p^{ii}$ of \er{eqn: parity_aver} is the parity opposite of $p^{vii}$,  $p^{iii}$ is opposite to $p^{vi}$, etc.~For each of these inversion-opposite pairs, one can separately 
do the same analysis as in \er{eqn: e_aver}, to conclude that the parity average of vertex form factors will contain no real components.~Note that this argument does not guarantee that
the remaining imaginary part of correlator dressings is an actual hypercubic invariant.~We are not entirely sure if such an invariance can even be proven analytically, starting from the
numerical ghost-gluon vertex itself.~Nonetheless, in Figure \ref{fig: ghost_dress} we subjected the coefficient functions of \er{eqn: append_ghost_latt} to a ``hypercubic invariance test'' 
and they seem to have passed it, within statistical uncertainties.   

\section{Numerical gluon propagator and ghost-gluon vertex under inversions}\label{eqn: inversions_append}

Throughout this paper we have assumed that lattice correlators of interest transform as tensors of appropriate rank, under hypercubic symmetry operators.~The said assertion might seem
trivial, but in numerical Monte Carlo (MC) simulations it is not always obvious that it should actually hold.~Put differently, for lattice vertex functions defined in terms of the link
variables $U_\mu(x)$, it is not always clear that they should obey the expected alteration rules with respect to hypercubic matrices.~In this Appendix we will look at the standard 
numerical lattice formulations for the gluon propagator and ghost-gluon vertex, and attempt to derive the corresponding transformation laws concerning the inversions of momentum 
components.~Permutations will not be considered here as in practice they boil down to ``mere'' relabelling of coordinate axes. 

In our upcoming proofs we will exploit some features specific to $SU(2)$ matrices.~In particular, our arguments will rely heavily on the fact that lattice links belonging to the $SU(
2)$ gauge group can be parametrised as $U_\mu = \mathbb{1} \, u_0 + i \, \vec{u} \cdot \vec{\sigma}$, with real coefficients $(u_0, \vec{u})$.~This does not neccessarily mean that our
final results do not apply to other $SU(N)$ gauge groups as well.~However, for such more general situations the proofs presented here will not hold, and alternative line of reasoning 
ought to be followed to deduce the transformation properties of lattice correlation functions. 

Let us start with an arguably easier case of the gluon two-point function.~By combining the definition of the gluon potential $A_\mu$ of \er{eqn: intro_latt_glue}, with the orthogonality
of Pauli matrices [\,$\text{Tr}(\sigma^a \, \sigma^b ) = 2 \, \delta^{ab}$, with $a, b = 1, \, 2, \, 3$\,], one quickly concludes that the colour components of $A_\mu$ are purely real,
in coordinate space.~In other words, the quantities 
\begin{align}\label{eqn: tralala}
A^b_\nu(x) \: = \: (1/2i) \cdot \text{Tr}\,[A_\nu(x) \, \sigma^b] \: = \: u^b_x , \qquad b = 1, \, 2, \, 3 \, , 
\end{align}

\noindent
have no imaginary components, since the coefficients $\vec{u}$ from the parametrisation of the lattice links $U_\mu(x)$ are real numbers.~The ``reality'' of $A^b_\nu(x)$, together with
the definition of the Fourier transform in \er{eqn: fourier_glue}, entails the momentum-space relation  
\begin{align}\label{eqn: a_conjug}
A^b_\nu(-p) \: = \: A^{b, \, \mathcal{C}}_\nu(p)
\end{align} 

\noindent
where $\mathcal{C}$ denotes complex conjugation.~Therefore, under a full parity transformation on momentum $p$, the imaginary part of $A^b_\nu(p)$ transforms as a vector, whereas the 
real part remains unaffected.~The non-vector nature of the real piece of $A^b_\nu(p)$ does not imply that this contribution ought to be simply ``thrown away'' in Monte Carlo simulations:
\!this term can still enter the evaluations of lattice correlators without affecting their correct symmetry features under parity changes.~As an example of this, one may take the diagonal
components of the gluon two-point function, defined in \er{eqn: d_propagator} (we assume that averaging over colour indices has been performed): 
\begin{align}
D_{\mu\mu} \: \sim \: A_\mu(p) \, A_\mu(-p) \: = \: A_\mu(p) \, A^{\,\mathcal{C}}_\mu(p) \: = \: |A_\mu(p)|^2 , \qquad \mu = 1, \ldots d \, ,
\end{align} 
 
\noindent
where $|z|^2 = a^2 + b^2$, for a complex number $z = a + i\,b$.~In evaluating the above expression we used the property \er{eqn: a_conjug}.~The diagonal propagator is thus guaranteed to be
purely real, symmetric (obviously) and invariant under arbitrary inversions of momentum components.~As for the off-diagonal terms, from the definition \er{eqn: d_propagator} alone one cannot
make any definitive statements regarding their real/imaginary components, symmetry under index swaps or correct properties with respect to parity.~Based on \er{eqn: d_propagator}, the only 
thing which can be claimed with certainty is that 
\begin{align}
D_{\mu\nu} \: = \: D^{\,\mathcal{C}}_{\nu\mu} , \qquad \mu , \, \nu = 1, \ldots d \, , \qquad \mu \neq \nu \, .
\end{align} 

Fortunately, in standard Monte Carlo calculations the off-diagonal contributions of the gluon two-point function are ``saved'' by gauge-fixing.~The usual (lattice) Landau gauge implementation, 
together with the $p_\mu/2$ correction in the Fourier transform \er{eqn: fourier_glue}, brings about the tensor structure given in \er{eqn: gauge_tensor}:~one may consult \cite{Alles:1996ka} 
for details.~Clearly, the equation \er{eqn: gauge_tensor} guarantees the absence of any imaginary pieces, as well as the symmetry under $\mu \leftrightarrow \nu$ and correct inversion features
for both the diagonal and off-diagonal propagator factors.~In the absence of the $p_\mu/2$ modification in \er{eqn: fourier_glue}, the resultant tensor representation would be 
\begin{align}\label{eqn: wrong_gluon}
D_{\mu\nu} \: = \: \left( \delta_{\mu\nu} - \frac{\bar{p}_\mu \, \bar{p}^{\,\mathcal{C}}_\nu}{\sum_\mu \, |\bar{p}_\mu|^2} \right) \: D(p) \, , 
\end{align}    

\noindent
where $\bar{p}_\mu = \cos p_\mu - 1 - i\, \sin p_\mu$  \cite{Alles:1996ka}.~The complex conjugation in the above expression comes from the minus sign in front of momentum $p$, in the 
$A_\nu$ part of \er{eqn: d_propagator}.~Strictly speaking, the same $\mathcal{C}$ operation also ought to be present in equation \er{eqn: gauge_tensor}, but in that case it would just 
eliminate the imaginary unit(s) $i$ which should be standing next to the lattice vector $\hat{p}_\mu = 2 \, \sin(p_\mu/2)$, see equation (15) in \cite{Alles:1996ka}.~In most Monte Carlo 
simulations featuring the structure \er{eqn: gauge_tensor}, the conjugation $\mathcal{C}$ is simply ignored and the imaginary pieces are removed by hand.~Now, it is straightforward to
check that the diagonal parts $D_{\mu\mu}$ in \er{eqn: wrong_gluon} will contain the momentum factors 
\begin{align}
\bar{p}_\mu \, \bar{p}^{\,\mathcal{C}}_\mu \: = \: 2 \, - \, 2 \, \cos p_\mu \, , \qquad \mu = 1, \ldots d \, .
\end{align}

The above functions are real and parity-invariant, since the cosine is an even function of its arguments.~On the other hand, the off-diagonal terms in \er{eqn: wrong_gluon} have both
real and imaginary components, they are not index-symmetric (i.\,e.~there is no symmetry under $\mu \leftrightarrow \nu$, if $\mu \neq \nu$), and they have no definitive properties 
regarding the parity transformations, as they feature sums of parity-even ($\cos p_\mu $) and parity-odd ($\sin p_\mu$) functions.~Some of these problems can be fixed by symmetrising 
the propagator \er{eqn: wrong_gluon} (so that $2 \, D^\text{\,symm}_{\mu\nu} = D_{\mu\nu} + D_{\nu\mu}$), which would result in a purely real function.~However, the non-diagonal terms 
of such a gluon would be invariant under arbitrary axis inversions, which is not supposed to happen, see e.\,g.~\er{eqn: gluon_reflected_2d}.~To the best of our knowledge, no modern 
lattice calculation uses the decomposition \er{eqn: wrong_gluon} (though see \cite{Mandula:1987rh} for an opposite example), but this function serves as an excellent example on how 
symmetry properties of (numerical) lattice vertices cannot be taken for granted.~In other words, one shouldn't simply assume that a given numerical function obeys the required 
tensor transformation laws:~whenever feasible, the presumed symmetry relations should be checked analytically and/or numerically, for Monte Carlo correlators of interest. 

This brings us to the numerical ghost-gluon vertex, defined through the (many) equations of \er{eqn: intro_latt_glue}, \er{eqn: fourier_glue}, \er{eqn: green_ghost}, \er{eqn: faddeev_foury},
\er{eqn: faddeev_define} and \er{eqn: g_faddeev_def}.~We wish to prove that the idendity \er{eqn: vertex_cc} holds for this correlator, which brings about the vanishing of the real part of 
the vertex form factors, upon parity-averaging.~Using the decomposition $U_\mu = \mathbb{1} \, u_0 + i \, \vec{u} \cdot \vec{\sigma}$ for the $SU(2)$ gauge links, along with the previously
mentioned orthogonality of the Pauli matrices $\sigma^a \, (a = 1, \, 2, \, 3)$, one may show that the quantities $G_\mu^{\,ab} (y)$ defined in \er{eqn: g_faddeev_def} ultimately evaluate to 
\begin{align}
G_\mu^{\,ab} (y) = \delta^{\, ab} \, u_0^{\, \mu, \, y} \, ,
\end{align}

\noindent
where $u^{\, \mu, \, y}_0$ is the real coefficient in front of the $\mathbb{1}$ term in $U_\mu(y)$.~Thus, the $G_\mu(y)$ contributions in \er{eqn: faddeev_define} are purely real, and so
are the coordinate-space potentials $A^b_\mu(y)$ [\,see equation \er{eqn: tralala}\,] and the structure constants $f^{\, abc}$ (we remind that $f^{\, abc} = \varepsilon^{\, abc}$, for an $
SU(2)$ group).~If there are any imaginary pieces in \er{eqn: faddeev_define}, they can only come from the function $\omega^{\,b} (y)$ alone.~This immediately entails the relation 
\begin{align}\label{eqn: omega_cc}
M^{ab}(x, y) \, \big( \omega^{\,b} (y) \big)^{\,\mathcal{C}} \: = \: \big( M^{ab}(x, y) \,  \omega^{\,b} (y) \big)^{\,\mathcal{C}}
\end{align}

\noindent
for the coordinate-space Faddeev-popov operator $M^{ab}(x, y)$.~In words, if the scalar test function $\omega^{\,b} (y)$ is complex-conjugated, the said conjugation can be applied to the
whole right-hand side of \er{eqn: faddeev_define}.~The equation \er{eqn: omega_cc} can be checked by splitting the test function into its real and imaginary parts and noting that, as there
are no imaginary terms in $M^{ab}_{x,\,y}$ apart from those coming from $\omega^{\,b} (y)$, the $\mathcal{C}$ operation on $\omega^{\,b} (y)$ carries over onto the whole FP matrix.~Now, in 
our lattice simulations one ``half'' of the Fourier transform in \er{eqn: faddeev_foury} is evaluated by inverting the operator $M^{ab}_{x,\,y}$ on a plane-wave source $\omega^{\,b} (y) = 
\delta^{\,bc} \, \e^{ \, 2 \, \pi i \, q \cdot y}$ (with $q \neq 0$) \cite{Cucchieri:1997dx}.~Symbolically, one may write 
\begin{align}
\left( M^{-1} \right)^{ab} (x,y) \, \delta^{\,bc} \, \e^{ \, 2 \, \pi i \, q \cdot y} \: \rightarrow \: \left( M^{-1} \right)^{ac} (x,q) \, . 
\end{align}

The above states in a symbolic way that inverting the FP operator with a plane-wave test function effectively replaces the coordinate variable(s) $y$ with the dependence on momentum $q$ (where 
$q \neq 0$), see \cite{Cucchieri:1997dx} for additional details.~The other half of the momentum-space transformation is done in the usual way, with 
\begin{align}\label{eqn: ajudar_cc}
\left( M^{-1} \right)^{ab} (p,q) \, \: = \: \sum_{x} \, \e^{\, 2 \, \pi i \, p \cdot x} \left( M^{-1} \right)^{ab} (x,q) \, . 
\end{align}

One may now observe that a full parity transformation on vector $q$ (so that $q \rightarrow q' = - q$) is equivalent to a complex conjugation of the source used for FP inversion, with $\e^{
\, 2 \, \pi i \, q \cdot y} \rightarrow \e^{- 2 \, \pi i \, q \cdot y}$.~By virtue of \er{eqn: omega_cc}, the said $\mathcal{C}$ transformation carries over onto the entire ``half-transformed''
operator, so that 
\begin{align}
\left( M^{-1} \right)^{ab} (x,-q) \: = \:  \left[ \left( M^{-1} \right)^{ab} (x,+q) \right]^{\,\mathcal{C}} \, .
\end{align}

Since the matrix $\left( M^{-1} \right)^{ab}_{x,\,q}$ is still real with respect to coordinates $x$ [\,we remind that all imaginary contributions come from $\omega^b (y)$\,],  by employing 
\er{eqn: ajudar_cc} one may see that the full inversion of momentum $p$ likewise translates to a complex conjugation of the whole operator $\left( M^{-1} \right)^{ab}_{p,\,q}$, or explicitly
\begin{align}\label{eqn: final_parity}
\left( M^{-1} \right)^{ab} (-p,-q) \: = \:  \left[ \left( M^{-1} \right)^{ab} (+p,+q) \right]^{\,\mathcal{C}} \, .
\end{align}

Finally, by combining the result \er{eqn: final_parity} with equations \er{eqn: green_ghost}, \er{eqn: final_vertex} and \er{eqn: a_conjug}, one obtains the transformation rule 
\begin{align}\label{eqn: again_cc}
\Gamma_\mu(-q, -p) = \Gamma^\mathcal{\,C}_\mu(q,p) \, ,  
\end{align}

\noindent
which is what we wanted to prove.~As shown at the end of the previous Appendix, the property \er{eqn: again_cc} guarantees that the parity averages of vertex form factors will be purely 
imaginary, which is also shown in Figure \ref{fig: ghost_dress}.~As also pointed out in the previous section, the relation \er{eqn: again_cc} does not ensure that the (imaginary part
of) the vertex transforms as a vector under partial inversions, wherein some (but not all\,!) momentum components get their signs flipped.~The latter quality is required for vertex form 
factors to be invariant under arbitrary parity changes of momentum components.~Still, results in Figure \ref{fig: ghost_dress} are suggestive of this invariance feature, admittedly within
sizeable error bars.~Unlike the expression \er{eqn: again_cc}, we are not sure if vertex transformation laws under arbitrary inversions can be derived starting from the definition \er{eqn: green_ghost}.~Also, it is not clear what happens to the above statements for symmetry groups other than $SU(2)$, as it is not obvious that relations like \er{eqn: omega_cc} would simply 
carry over:~for more general gauge groups it is possible, at least in principle, that the coordinate space FP operator $M^{ab}(x,y)$ will have imaginary contributions apart from those 
coming from the test function $\omega^{\,b} (y)$, which would make the equations akin to \er{eqn: omega_cc} invalid.~We leave a detailed consideration of these questions for future studies.  

\section{Symmetry rotations around lattice diagonals}\label{sec: diagonals}

In section \ref{sec: hyper_perms} we considered $\pi/2$ rotations around the coordinate axes, and their corresponding matrix representations.~We argued that these operators can always 
be broken down into combinations of inversions and pure permutations, and used this ``permutation + inversion'' viewpoint throughout the rest of the paper.~However, up to now we have 
avoided taking a detailed look at the symmetry rotations around diagonals of (hyper)cubes, and in this Appendix we would like to correct this ommision.~To be more precise, here we want
to explicitly demonstrate that the symmetry rotations around hypercube diagonals likewise fit the aforementioned ``permutation + inversion'' mold, and are thus in line with much of the 
analysis done in this work.

\begin{figure}[!t]
\begin{center}
\graph[width = 0.63\tew]{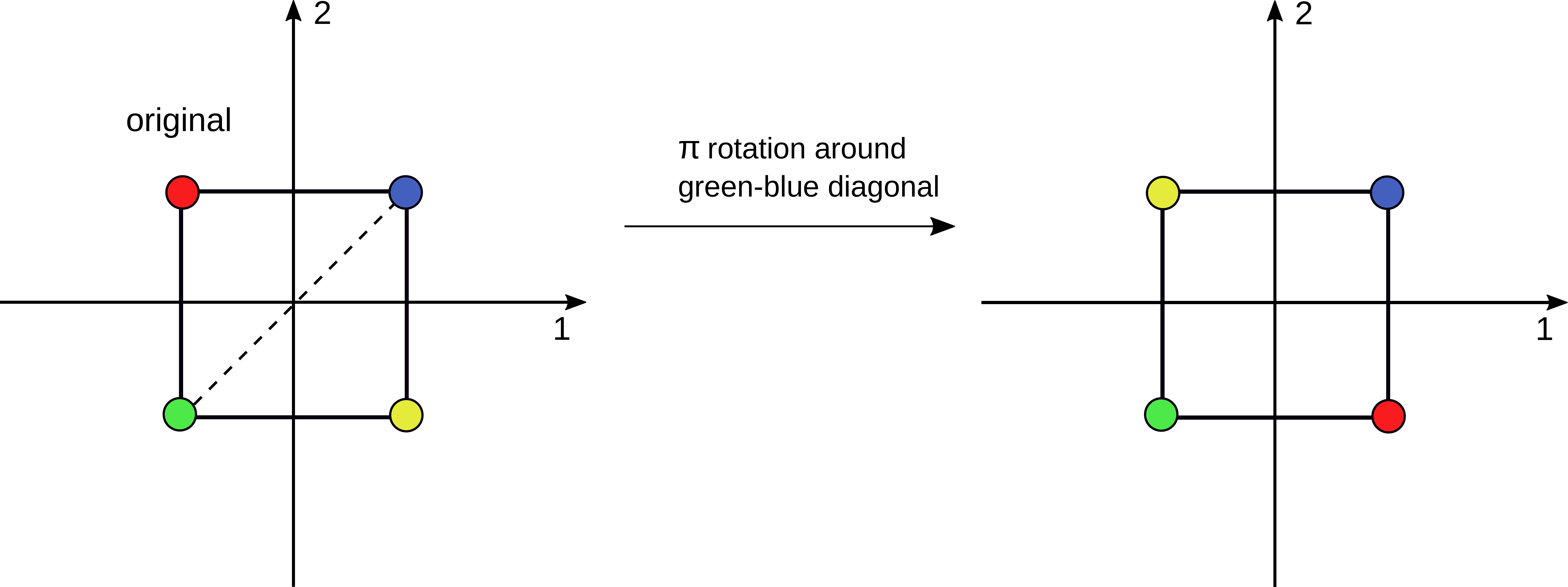} \\ 
\graph[width = 0.63\tew]{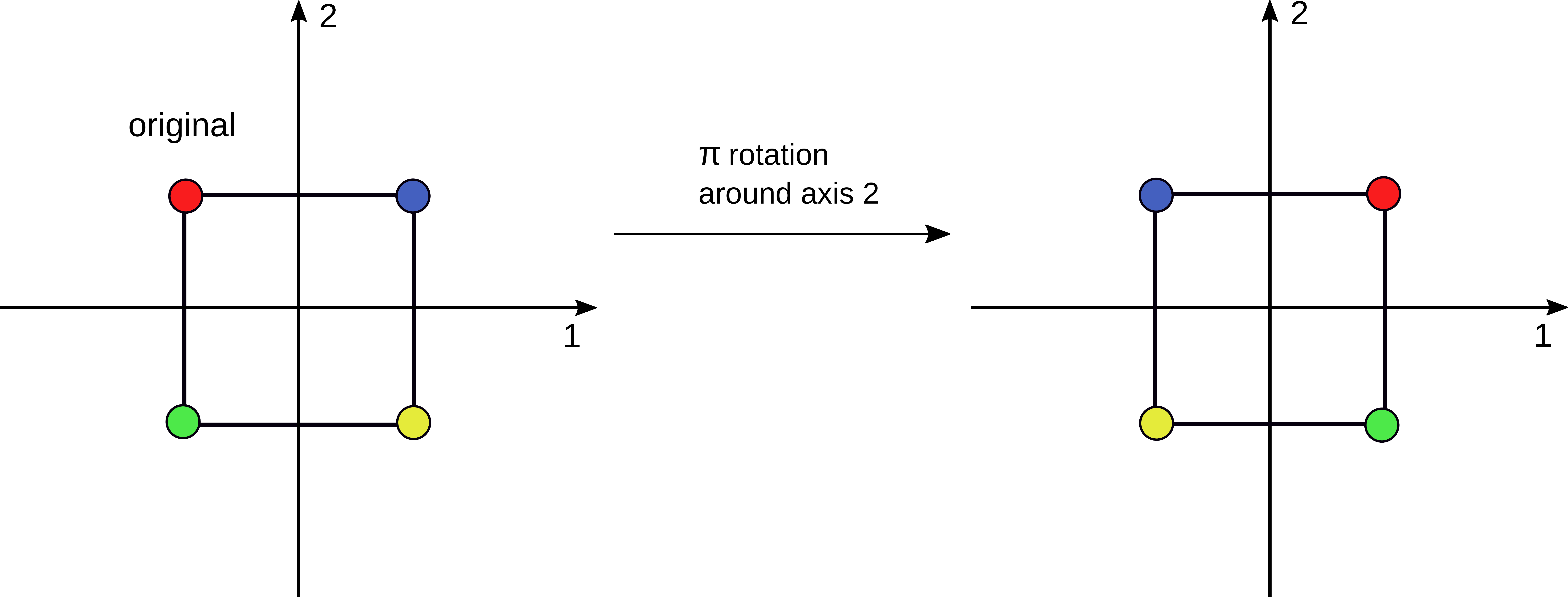} \\
\caption{\textit{Top}:~A $\pi$ degrees rotation of a square around one of its diagonals.~\textit{Bottom}:~A $\pi$ degrees rotation of a square around one of the coordinate 
axes.} 
\label{fig: square_off}
\end{center}
\end{figure}

As usual, we start the arguments with the simple case of two dimensions and the corresponding hypercube, i.\,e.~a square.~Any square is invariant under a rotation by $\pi$ degrees 
around any one of its two diagonals, as exemplified in the top panel of Figure \ref{fig: square_off}.~Note that the sign of the rotation (i.\,e.~clockwise versus counter-clockwise) 
does not matter here since the end result is exactly the same in both cases.~In Figure \ref{fig: square_off} we used colour-coding for the square corners, as in the absence of any 
labels there would be no difference between the original and the transformed object (of course, this is the whole point of a symmetry transformation).~Now the question is how does 
the operator which performs the action in the top plot of Figure \ref{fig: square_off} look like in matrix form.~In this simple situation, the correct answer can in fact be easily
guessed.~But since the same guessing method can hardly be generalised to higher dimensions, already here we will use a more rigorous approach.

One may start by answering an easier question, namely by finding the operator which flips the square by $\pi$ degrees around one of the coordinate axes.~An example with a rotation 
around axis 2 is given in the bottom panel of Figure \ref{fig: square_off}.~The corresponding operator is, in matrix form 
\begin{align}\label{eqn: flip_axis}
\rho_1 = \left[ \begin{array}{cc} -1 & \;\; 0 \\ 0 & \;\; 1  \end{array} \right] \, .
\end{align}
 
The fact that the matrix $\rho_1$ indeed effects the transformation depicted in the bottom graph of Figure \ref{fig: square_off} can be checked by writting down explicitly the 
coordinates for each of the square corners.~In the ``original'' position in Fig.~\ref{fig: square_off} the coordinates are 
\begin{align}\label{eqn: cut_corners}
&\text{red:}\: (-n,n) \, ,  \: \qquad  \: \hspace{0.47cm }\text{blue:}\: (n,n) \, , \nonumber \\
&\text{yellow:}\: (n,-n) \, , \: \qquad \: \text{green:}\: (-n,-n) \, ,
\end{align}

\noindent
with some number $n$.~From the above one can easily see that the operator $\rho_1$ in \er{eqn: flip_axis}, which changes the sign of the first component of all position vectors, will 
interchange the red corner with the blue one, as well as the green one with the yellow:~this is exactly the symmetry operation shown in the bottom panel of \ref{fig: square_off}.~We 
now want to turn $\rho_1$ into a matrix which flips the square by $\pi$ degrees around one of the diagonals.~This can be done in a standard way, via a similarity transformation \cite{
Morty:1962prc}:
\begin{align}\label{eqn: oper_diag}
L^{\,\pi, \, \text{diag}} \: = \: U \cdot \rho_1 \cdot U^{-1} \, ,
\end{align}

\noindent
where $U$ tranforms an unit vector pointing along the axis 2 into an unit vector pointing along one of the diagonals.~For concreteness, we choose the green-blue diagonal in Fig.~\ref{fig:
square_off} as the desired axis of rotation.~An unit vector pointing along this line (in the first quadrant) is of the form $e_\text{diag} = (\sqrt{2}/2, \, \sqrt{2}/2)$ and thus the 
operator $U$ can be obtained as a solution of the matrix-vector equation 
\begin{align}\label{eqn: uop_def}
U \cdot \left( \begin{array}{c} 0 \\ 1  \end{array} \right) \: = \: \left( \begin{array}{c} \sqrt{2}/2 \\ \sqrt{2}/2  \end{array} \right) \, .
\end{align}

The elements of $U$ are additionally constrained by the fact that $U$ needs to be an orthogonal transformation, so that $U\cdot U^{T} = \mathbb{1}$ (here $T$ denotes matrix transposition).
\!Combining the orthogonality conditions for $U$ with \er{eqn: uop_def} gives 
\begin{align}
U \: = \: \left[ \begin{array}{cc} \sqrt{2}/2 & \;\; \sqrt{2}/2 \\ - \sqrt{2}/2 & \;\; \sqrt{2}/2  \end{array} \right] \, , 
\end{align}

\noindent
so that the desired operator $L^{\,\pi, \, \text{diag}}$ of \er{eqn: oper_diag} is found to be
\begin{align}\label{eqn: explicit_diagonal}
L^{\,\pi, \, \text{diag}} \: = \: \left[ \begin{array}{cc} \sqrt{2}/2 & \;\; \sqrt{2}/2 \\ - \sqrt{2}/2 & \;\; \sqrt{2}/2  \end{array} \right] \cdot 
\left[ \begin{array}{cc} -1 & \;\; 0 \\ 0 & \;\; 1  \end{array} \right] \cdot \left[ \begin{array}{cc} \sqrt{2}/2 & \;\; -\sqrt{2}/2 \\  \sqrt{2}/2 & \;\; 
\sqrt{2}/2  \end{array} \right] \: = \: \left[ \begin{array}{cc} 0 & \;\; 1 \\ 1 & \;\; 0  \end{array} \right] \, .
\end{align}

$L^{\,\pi, \, \text{diag}}$ is obviously a pure permutation, which swaps the first and second components of two-dimensional vectors:~by applying this operation to all of the position
vectors in \er{eqn: cut_corners}, one sees that $L^{\,\pi, \, \text{diag}}$ will interchange the red and yellow square corners, while leaving the blue and green ones intact.~Therefore,
$L^{\,\pi, \, \text{diag}}$ is indeed a matrix representation of the symmetry rotation in the top panel of Figure \ref{fig: square_off}.~The operator which flips the square by $\pi$
degrees around the other diagonal (the red-yellow line in the top graph of \ref{fig: square_off}) is equal to $-L^{\,\pi, \, \text{diag}}$.~Thus, both symmetry operations match 
the ``permutations and inversions'' picture used in the main body of the paper.		

Essentially the same procedure which had led us to \er{eqn: explicit_diagonal} can be applied in any number of dimensions.~There are however some important differences compared to 
the two-dimensional case.~The first is the actual angle of rotation about the diagonals, which leaves a given (hyper)cube intact.~Whereas a two-dimensional square is invariant under
a rotation by $\pi$ degrees around the diagonal axes, the three-dimensional cube is left unaffected by $2\pi/3$ degrees rotations, whereas a four-dimensional hypercube will be 
indifferent to $\pi/2$ diagonal transformations \cite{Morty:1962prc}.~Thus, if one wanted to obtain an explicit matrix representation of a symmetry transformation about the diagonal
in (say) three dimensions, one would need to start with a $2\pi/3$ rotation around one of the axes, for instance axis number 3: 
\begin{align}\label{eqn: three_rotor}
L^{\,2\pi/3, \: 3} \: = \: \left[ \begin{array}{ccc} \:\:\:\: \cos(2\pi/3) & \;\; \sin(2\pi/3) & \;\; 0 \\ -\sin(2\pi/3) & \;\; \cos(2\pi/3) & \;\; 0 \\ 0 & \;\; 0 & \;\; 1  \end{array} 
\right] \: = \: \left[ \begin{array}{ccc} -1/2 & \;\; \sqrt{3}/2 & \;\; 0 \\ \sqrt{3}/2 & \;\; -1/2 & \;\; 0 \\ 0 & \;\; 0 & \;\; 1  \end{array} \right]
\, .
\end{align}

One other big difference with respect to the $d=2$ situation is the actual construction of the operator $U$ of \er{eqn: oper_diag}, which should transform a matrix like \er{eqn: 
three_rotor} into a symmetry rotation about one of the cube diagonals.~Namely, whereas equation \er{eqn: uop_def} is easily solved (coupled with orthogonality constraints for $U$), 
it's three-dimensional generalisation corresponding to the rotation \er{eqn: three_rotor} would be 
\begin{align}\label{eqn: uop_gener}
U \cdot \left( \begin{array}{c} 0 \\ 0 \\ 1 \end{array} \right) \: = \: \left( \begin{array}{c} \sqrt{3}/3 \\ \sqrt{3}/3 \\ \sqrt{3}/3  \end{array} \right) \, .
\end{align}

\noindent
which is very hard to solve analytically.~Fortunately, problems like \er{eqn: uop_gener} can always be broken down into multiple ``two-dimensional'' tasks.~Instead of solving the 
above equation directly, one may write the operator $U$ as a product of certain simpler matrices, where each matrix transforms a three-dimensional vector while keeping one of its 
components fixed.~In the case of \er{eqn: uop_gener}, it holds that 
\begin{align}\label{eqn: u_r1_r2}
U \: = \: (R^{\,2} \cdot R^{\,1})^{-1} \, , 
\end{align}

\noindent
where the matrices $R^{\,1}$ and $R^{\,2}$ are solutions of the equations 
\begin{align}\label{eqn: r_one_two}
R^{\,1} \cdot \left( \begin{array}{c} \sqrt{3}/3 \\ \sqrt{3}/3 \\ \sqrt{3}/3  \end{array} \right) \: = \: \left( \begin{array}{c} \sqrt{6}/3 \\ 0 \\ \sqrt{3}/3  \end{array} \right) \: , 
\qquad \text{and} \qquad 
R^{\,2} \cdot \left( \begin{array}{c} \sqrt{6}/3 \\ 0 \\ \sqrt{3}/3  \end{array} \right) \: = \: \left( \begin{array}{c} 0 \\ 0 \\ 1 \end{array} \right) \: .
\end{align}

Coupled with the orthogonality conditions for $R^{\,1}$ and $R^{\,2}$, each of the two equations in \er{eqn: r_one_two} are just as easy to handle as the two-dimensional problem 
\er{eqn: uop_def}.~Then, solving for $R^{\,1}$ and $R^{\,2}$, employing \er{eqn: u_r1_r2} and transforming the operator \er{eqn: three_rotor} according to \er{eqn: oper_diag} 
finally gives 
\begin{align}\label{eqn: finally!}
L^{\,2\pi/3, \: \text{diag}} \: = \: \left[ \begin{array}{ccc} 0 & \;\; 1 & \;\; 0 \\ 0 & \;\; 0 & \;\; 1 \\ 1 & \;\; 0 & \;\; 0  \end{array} \right] \: = \: \Pi^{13} \cdot \Pi^{23} \, .
\end{align}

As in a two-dimensional situation, the operator $L^{\,2\pi/3, \: \text{diag}}$ is a pure permutation, and in \er{eqn: finally!} we've written it as a product of two transpositions, 
wherein $\Pi^{\sigma\tau}$ exchanges the $\sigma$-th and $\tau$-th element of a vector with three components.~Likewise, all the other symmetry transformations around the diagonals 
of a three-dimensional cube are either pure permutations or combinations of permutations and inversions.~This kind of demonstration can straightforwardly be generalised to higher 
dimensions as well.~One may say that, regardless of the dimension number, symmetry rotations around (hyper)cube diagonals can always be decomposed into inversions and $\pi/2$ rotations 
around coordinate axes, since an elementary permutation (a transposition) is equvialent to a combination of a (coordinate) axial $\pi/2$ rotation and a parity operation, see \er{eqn: 
2d_rot_example} and \er{eqn: 3d_rot_example} as examples.~Thus, $\pi/2$ rotations around coordinate axes and inversions actually exhaust all of the hypercubic symmetry operations, in 
an arbitrary number of dimensions.

\section{Second-rank tensors under permutations}\label{sec: append_permutes}

In section \ref{sec: gluon_lattice} we provided some intuitive arguments for the claim that permutation operators cannot mix diagonal and off-diagonal components of tensors of second 
rank.~Here we wish to give a somewhat more formal treatment of this issue.~We will limit ourselves to considerations of transpositions, i.\,e.~transformations which exchange only two 
elements at a time (when acting on vector-valued quantities).~As already stated in \ref{sec: hyper_vector}, an analysis featuring transpositions alone is sufficient since any permutation
can be written as a composition of these elementary operators \cite{Hassani:1999sny}. 

For convenience, here we will repeat how a permutation $\Pi^{\sigma\tau}$, which exchanges the $\sigma$-th and $\tau$-th components of a vector (where both $\sigma$ and $\tau$ can
run from 1 to the dimension number $d$), can be represented in terms of matrices.~An operator $\Pi^{\sigma\tau}$ is equivalent to the identity element $\mathbb{1}$ whose $\sigma$-th 
and $\tau$-th rows have been permuted \cite{Hassani:1999sny}.~Therefore, when $\Pi^{\sigma\tau}$ is written out as a matrix, it will hold that
\begin{align}\label{eqn: delta_perm}
\Pi^{\sigma\tau}_{\mu\nu} \: = \: \delta_{\mu\nu} \, , \qquad \text{if} \:\: \mu \neq \sigma, \, \tau \, ,
\end{align} 

\noindent
whereas in the $\sigma$-th and $\tau$-th row of $\Pi^{\sigma\tau}$ the only non-vanishing elements are 
\begin{align}\label{eqn: sigma_tau_perm}
\Pi^{\sigma\tau}_{\sigma\tau} \: = \: \Pi^{\sigma\tau}_{\tau\sigma} \: = \: 1 \, .
\end{align} 

Now let us see what the above representation for $\Pi^{\sigma\tau}$ entails when it comes to second-rank tensors being transformed with transposition matrices.~The corresponding alteration 
rule is 
\begin{align}\label{eqn: second_append_law}
D''_{\mu\nu} \: = \: \sum_{\alpha = 1}^d \sum_{\beta = 1}^d \: \Pi^{\sigma\tau}_{\mu\alpha} \, \Pi^{\sigma\tau}_{\nu\beta} \, D_{\alpha\beta} \, .
\end{align} 

We start with the easiest possible example, where none of the external indices $\mu$ and $\nu$ in \er{eqn: second_append_law} matches the components $\sigma$ and $\tau$ that are being 
swapped.~In other words, we look at a situation where $\mu \neq \sigma, \, \tau$ and $\nu \neq \sigma, \, \tau$.~In this case, by employing \er{eqn: delta_perm} one quickly gets  
\begin{align}
D''_{\mu\nu} \: = \: \sum_{\alpha = 1}^d \sum_{\beta = 1}^d \: \delta_{\mu\alpha} \, \delta_{\nu\beta} \, D_{\alpha\beta} \: = \: D_{\mu\nu} \, .
\end{align}   

The above is a (trivial) statement that a tensor component $D_{\mu\nu}$ does not get altered, if its indices are not the ones being permuted by the operator $\Pi^{\sigma\tau}$.~As an 
example, terms of the kind $D_{14}$ and $D_{44}$ are indifferent to the action of a transformation like $\Pi^{23}$, as none of their indices are being changed by the said permutation.
\!Since in this trivial case a given component (either diagonal or off-diagonal) gets mapped onto itself, there can be no mixing between off-diagonal and diagonal tensor contributions.   

We now move on to the diagonal case, where (say) $\mu = \nu = \sigma$.~In this situation the general rule \er{eqn: second_append_law} reduces to 
\begin{align}\label{eqn: append_perm_one}
D''_{\sigma\sigma} \: = \: \sum_{\alpha = 1}^d \sum_{\beta = 1}^d \: \Pi^{\sigma\tau}_{\sigma\alpha} \, \Pi^{\sigma\tau}_{\sigma\beta} \, D_{\alpha\beta} \, \: = \: \Pi^{\sigma\tau}_{
\sigma\tau} \, \Pi^{\sigma\tau}_{\sigma\tau} \, D_{\tau\tau} \: = \: D_{\tau\tau} \, .
\end{align} 

In obtaining the above result we used the fact that $\Pi^{\sigma\tau}_{\sigma\tau} = 1$ is the only non-zero element in the $\sigma$-th row of $\Pi^{\sigma\tau}$.~In the same manner, a
consideration of the second (non-trivial) diagonal option, $\mu = \nu = \tau$ gives 
\begin{align}\label{eqn: append_perm_two}
D''_{\tau\tau} \: = \: D_{\sigma\sigma} \, .
\end{align} 

The important thing about relations \er{eqn: append_perm_one} and \er{eqn: append_perm_two} is that, whatever the actual numerical values of indices $\sigma$ and $\tau$, a diagonal tensor
term will always get mapped onto another diagonal element.~However, this still does not guarantee that an off-diagonal contribution cannot get mapped onto the diagonal subspace.~To check
for this possibility, we look at a case where (say) $\mu = \sigma$ and $\mu \neq \nu$ (obviously, this also means $\nu \neq \sigma$).~The appropriate transformation rule is 
\begin{align}\label{eqn: branching}
D''_{\sigma\nu} \: = \: \sum_{\beta = 1}^d \: \Pi_{\,\nu\beta}^{\sigma\tau} \, D_{\tau \beta} \, , \qquad \nu \neq \sigma \, ,
\end{align}  

\noindent
wherein we immediately used the property \er{eqn: sigma_tau_perm} to eliminate one of the sums.~There are now two possible options for the other external index $\nu$.~If it is equal 
to $\tau$, an application of \er{eqn: sigma_tau_perm} gives the final result
\begin{align}\label{eqn: off_diag_eg}
D''_{\sigma\tau} \: = \: D_{\tau \sigma} \, .
\end{align}

In obtaining the above expression we went in with the assumptions that $\mu = \sigma$, $\nu = \tau$ and $\mu \neq \nu$.~This obviously also entails that $\sigma \neq \tau$, and the 
equation \er{eqn: off_diag_eg} is thus an example of one off-diagonal term being transformed into another off-diagonal component.~Now going back to the more general case \er{eqn:
branching}, another possible option for $\nu$ is that $\nu \neq \tau$, for which the equation \er{eqn: delta_perm} quickly gives 
\begin{align}\label{eqn: final!}
D''_{\sigma\nu} \: = \: D_{\tau \nu} \, .
\end{align}   

In getting the result \er{eqn: final!}, the index $\nu$ was assumed to be different from both $\sigma$ and $\tau$, and therefore \er{eqn: final!} is a yet another example of an 
off-diagonal factor being changed into a different off-diagonal term.~Again, there is no mixing between diagonal and off-diagonal contributions involved here.~In deriving all of these 
``off-diagonal'' results, we started with the external index $\mu$ in \er{eqn: second_append_law} and equated it to $\sigma$.~One might just as well have started with the external number
$\nu$ and equated it to either $\sigma$ or $\tau$:~a consistent application of the representation \er{eqn: delta_perm} and \er{eqn: sigma_tau_perm} will always guarantee that an off-diagonal
tensor component gets transformed into some other part of the off-diagonal subspace.    

\vspace{0.81cm}  

\begin{table}[!h]
\begin{center}
\bgroup
\def\arraystretch{1.4}%
\begin{tabular}{||c|c|c|c|}
\hline\hline
$ V $   & $ \beta $  & $\langle W_{1,1}\rangle$ & $\omega$\\
\hline\hline
 $32^2$   & 8    & 0.81925(5)   &  1.93  \\  
\hline  
 $32^3$   & 5    & 0.786869(7)  &  1.96  \\     
\hline 
 $32^3$   & 7.4  & 0.859567(5)  &  1.95  \\
\hline\hline
\end{tabular}
\egroup
\caption{Some details concerning the gauge field configurations used in our simulations.~$\beta$ is the interaction parameter of the Wilson gauge action \er{eqn: wilson_action}.~$
\langle W_{1,1}\rangle$ is the expectation value of the $1 \times 1$ Wilson loop:~on a three-dimensional lattice this quantity is needed for the scale-setting fit of equation (67) 
in \cite{Teper:1998te}, with string tension $\sqrt{\sigma} = 0.44$ GeV.~$\omega$ is the parameter of the gauge fixing procedure employed in our numerics, the over-relaxation method 
\cite{Mandula:1990vs, Cucchieri:1995pn}.}
\label{tab: config_details}
\end{center}
\end{table}


\begin{thebibliography}{99} 

%\cite{Aoki:2016frl}
\bibitem{Aoki:2016frl}  
  S.~Aoki {\it et al.},
  %``Review of lattice results concerning low-energy particle physics,''
  Eur.\ Phys.\ J.\ C {\bf 77} (2017) no.2,  112
  doi:10.1140/epjc/s10052-016-4509-7
  [arXiv:1607.00299 [hep-lat]].
  %%CITATION = doi:10.1140/epjc/s10052-016-4509-7;%%
  %467 citations counted in INSPIRE as of 16 Apr 2019

%\cite{Eichmann:2016yit}
\bibitem{Eichmann:2016yit}
  G.~Eichmann, H.~Sanchis-Alepuz, R.~Williams, R.~Alkofer and C.~S.~Fischer,
  %``Baryons as relativistic three-quark bound states,''
  Prog.\ Part.\ Nucl.\ Phys.\  {\bf 91} (2016) 1
  doi:10.1016/j.ppnp.2016.07.001
  [arXiv:1606.09602 [hep-ph]].
  %%CITATION = doi:10.1016/j.ppnp.2016.07.001;%%
  %111 citations counted in INSPIRE as of 16 Apr 2019

%\cite{Gribov:1977wm}
\bibitem{Gribov:1977wm} 
  V.~N.~Gribov,
  %``Quantization of Nonabelian Gauge Theories,''
  Nucl.\ Phys.\ B {\bf 139} (1978) 1.
  doi:10.1016/0550-3213(78)90175-X
  %%CITATION = doi:10.1016/0550-3213(78)90175-X;%%
  %1589 citations counted in INSPIRE as of 23 Feb 2018

%\cite{Zwanziger:1993dh}    
\bibitem{Zwanziger:1993dh}
  D.~Zwanziger,
  %``Fundamental modular region, Boltzmann factor and area law in lattice gauge theory,''
  Nucl.\ Phys.\ B {\bf 412} (1994) 657.
  doi:10.1016/0550-3213(94)90396-4
  %%CITATION = doi:10.1016/0550-3213(94)90396-4;%%
  %267 citations counted in INSPIRE as of 23 Feb 2018

%\cite{Zwanziger:2001kw}
\bibitem{Zwanziger:2001kw}
  D.~Zwanziger,
  %``Nonperturbative Landau gauge and infrared critical exponents in QCD,''
  Phys.\ Rev.\ D {\bf 65} (2002) 094039
  doi:10.1103/PhysRevD.65.094039
  [hep-th/0109224].
  %%CITATION = doi:10.1103/PhysRevD.65.094039;%%
  %296 citations counted in INSPIRE as of 23 Feb 2018

%\cite{Zwanziger:2003cf}
\bibitem{Zwanziger:2003cf}
  D.~Zwanziger,
  %``Nonperturbative Faddeev-Popov formula and infrared limit of QCD,''
  Phys.\ Rev.\ D {\bf 69} (2004) 016002
  doi:10.1103/PhysRevD.69.016002
  [hep-ph/0303028].
  %%CITATION = doi:10.1103/PhysRevD.69.016002;%%
  %233 citations counted in INSPIRE as of 23 Feb 2018

%\cite{Kugo:1979gm}
\bibitem{Kugo:1979gm}
  T.~Kugo and I.~Ojima,
  %``Local Covariant Operator Formalism of Nonabelian Gauge Theories and Quark Confinement Problem,''
  Prog.\ Theor.\ Phys.\ Suppl.\  {\bf 66} (1979) 1.
  doi:10.1143/PTPS.66.1
  %%CITATION = doi:10.1143/PTPS.66.1;%%
  %947 citations counted in INSPIRE as of 23 Feb 2018

%\cite{Alkofer:2008tt}
\bibitem{Alkofer:2008tt}
  R.~Alkofer, C.~S.~Fischer, F.~J.~Llanes-Estrada and K.~Schwenzer,
  %``The Quark-gluon vertex in Landau gauge QCD: Its role in dynamical chiral symmetry breaking and quark confinement,''
  Annals Phys.\  {\bf 324} (2009) 106
  doi:10.1016/j.aop.2008.07.001
  [arXiv:0804.3042 [hep-ph]].
  %%CITATION = doi:10.1016/j.aop.2008.07.001;%%
  %177 citations counted in INSPIRE as of 16 Apr 2019

%\cite{Vujinovic:2014ioa}    
\bibitem{Vujinovic:2014ioa}           
  M.~Vujinovic and R.~Williams,
  %``Hadronic bound states in SU(2) from Dyson–Schwinger equations,''
  Eur.\ Phys.\ J.\ C {\bf 75} (2015) no.3,  100
  doi:10.1140/epjc/s10052-015-3324-x
  [arXiv:1411.7619 [hep-ph]].
  %%CITATION = doi:10.1140/epjc/s10052-015-3324-x;%%
  %8 citations counted in INSPIRE as of 23 Feb 2018

%\cite{Sanchis-Alepuz:2015qra}
\bibitem{Sanchis-Alepuz:2015qra}                                      
  H.~Sanchis-Alepuz and R.~Williams,
  %``Probing the quark–gluon interaction with hadrons,''
  Phys.\ Lett.\ B {\bf 749} (2015) 592
  doi:10.1016/j.physletb.2015.08.067
  [arXiv:1504.07776 [hep-ph]].
  %%CITATION = doi:10.1016/j.physletb.2015.08.067;%%
  %19 citations counted in INSPIRE as of 23 Feb 2018

%\cite{Williams:2015cvx}
\bibitem{Williams:2015cvx}
  R.~Williams, C.~S.~Fischer and W.~Heupel,
  %``Light mesons in QCD and unquenching effects from the 3PI effective action,''
  Phys.\ Rev.\ D {\bf 93} (2016) no.3,  034026
  doi:10.1103/PhysRevD.93.034026
  [arXiv:1512.00455 [hep-ph]].
  %%CITATION = doi:10.1103/PhysRevD.93.034026;%%
  %35 citations counted in INSPIRE as of 23 Feb 2018

%\cite{Binosi:2016rxz}
\bibitem{Binosi:2016rxz}
  D.~Binosi, L.~Chang, J.~Papavassiliou, S.~X.~Qin and C.~D.~Roberts,
  %``Symmetry preserving truncations of the gap and Bethe-Salpeter equations,''
  Phys.\ Rev.\ D {\bf 93} (2016) no.9,  096010
  doi:10.1103/PhysRevD.93.096010
  [arXiv:1601.05441 [nucl-th]].
  %%CITATION = doi:10.1103/PhysRevD.93.096010;%%
  %38 citations counted in INSPIRE as of 02 Jan 2019

%\cite{Eichmann:2016hgl}
\bibitem{Eichmann:2016hgl}
  G.~Eichmann, C.~S.~Fischer and H.~Sanchis-Alepuz,
  %``Light baryons and their excitations,''
  Phys.\ Rev.\ D {\bf 94} (2016) no.9,  094033
  doi:10.1103/PhysRevD.94.094033
  [arXiv:1607.05748 [hep-ph]].
  %%CITATION = doi:10.1103/PhysRevD.94.094033;%%
  %24 citations counted in INSPIRE as of 16 Apr 2019

%\cite{Sanchis-Alepuz:2017jjd}
\bibitem{Sanchis-Alepuz:2017jjd}
  H.~Sanchis-Alepuz and R.~Williams,
  %``Recent developments in bound-state calculations using the Dyson–Schwinger and Bethe–Salpeter equations,''
  Comput.\ Phys.\ Commun.\  {\bf 232} (2018) 1
  doi:10.1016/j.cpc.2018.05.020
  [arXiv:1710.04903 [hep-ph]].
  %%CITATION = doi:10.1016/j.cpc.2018.05.020;%%
  %10 citations counted in INSPIRE as of 16 Apr 2019

%\cite{Rodriguez-Quintero:2018wma}
\bibitem{Rodriguez-Quintero:2018wma}
  J.~Rodríguez-Quintero, D.~Binosi, C.~Mezrag, J.~Papavassiliou and C.~D.~Roberts,
  %``Process-independent effective coupling. From QCD Green's functions to phenomenology,''
  Few Body Syst.\  {\bf 59} (2018) no.6,  121
  doi:10.1007/s00601-018-1437-0
  [arXiv:1801.10164 [nucl-th]].
  %%CITATION = doi:10.1007/s00601-018-1437-0;%%
  %7 citations counted in INSPIRE as of 02 Jan 2019

%\cite{Vujinovic:2018nko}
\bibitem{Vujinovic:2018nko}
  M.~Vujinovic and R.~Alkofer,
  %``Low-energy spectrum of an SU(2) gauge theory with dynamical fermions,''
  Phys.\ Rev.\ D {\bf 98} (2018) no.9,  095030
  doi:10.1103/PhysRevD.98.095030
  [arXiv:1809.02650 [hep-ph]].
  %%CITATION = doi:10.1103/PhysRevD.98.095030;%%

%\cite{Eichmann:2019tjk}
\bibitem{Eichmann:2019tjk}
  G.~Eichmann, C.~S.~Fischer, E.~Weil and R.~Williams,
  %``Single pseudoscalar meson pole and pion box contributions to the anomalous magnetic moment of the muon,''
  arXiv:1903.10844 [hep-ph].
  %%CITATION = ARXIV:1903.10844;%%

%\cite{Schleifenbaum:2004id}
\bibitem{Schleifenbaum:2004id}                                    
  W.~Schleifenbaum, A.~Maas, J.~Wambach and R.~Alkofer,
  %``Infrared behaviour of the ghost-gluon vertex in Landau gauge Yang-Mills theory,''
  Phys.\ Rev.\ D {\bf 72} (2005) 014017
  doi:10.1103/PhysRevD.72.014017
  [hep-ph/0411052].
  %%CITATION = doi:10.1103/PhysRevD.72.014017;%%
  %106 citations counted in INSPIRE as of 06 Dec 2018

%\cite{Pawlowski:2005xe}
\bibitem{Pawlowski:2005xe}             
  J.~M.~Pawlowski,
  %``Aspects of the functional renormalisation group,''
  Annals Phys.\  {\bf 322} (2007) 2831
  doi:10.1016/j.aop.2007.01.007
  [hep-th/0512261].
  %%CITATION = doi:10.1016/j.aop.2007.01.007;%%
  %568 citations counted in INSPIRE as of 06 Dec 2018

%\cite{Kellermann:2008iw}
\bibitem{Kellermann:2008iw}  
  C.~Kellermann and C.~S.~Fischer,
  %``The Running coupling from the four-gluon vertex in Landau gauge Yang-Mills theory,''
  Phys.\ Rev.\ D {\bf 78} (2008) 025015
  doi:10.1103/PhysRevD.78.025015
  [arXiv:0801.2697 [hep-ph]].
  %%CITATION = doi:10.1103/PhysRevD.78.025015;%%
  %47 citations counted in INSPIRE as of 06 Dec 2018

%\cite{Alkofer:2008dt}
\bibitem{Alkofer:2008dt}
  R.~Alkofer, M.~Q.~Huber and K.~Schwenzer,
  %``Infrared Behavior of Three-Point Functions in Landau Gauge Yang-Mills Theory,''
  Eur.\ Phys.\ J.\ C {\bf 62} (2009) 761
  doi:10.1140/epjc/s10052-009-1066-3
  [arXiv:0812.4045 [hep-ph]].
  %%CITATION = doi:10.1140/epjc/s10052-009-1066-3;%%
  %53 citations counted in INSPIRE as of 06 Dec 2018


%\cite{Huber:2012zj}
\bibitem{Huber:2012zj}
  M.~Q.~Huber, A.~Maas and L.~von Smekal,
  %``Two- and three-point functions in two-dimensional Landau-gauge Yang-Mills theory: Continuum results,''
  JHEP {\bf 1211} (2012) 035
  doi:10.1007/JHEP11(2012)035
  [arXiv:1207.0222 [hep-th]].
  %%CITATION = doi:10.1007/JHEP11(2012)035;%%
  %40 citations counted in INSPIRE as of 06 Dec 2018

%\cite{Huber:2012kd}
\bibitem{Huber:2012kd}            
  M.~Q.~Huber and L.~von Smekal,
  %``On the influence of three-point functions on the propagators of Landau gauge Yang-Mills theory,''
  JHEP {\bf 1304} (2013) 149
  doi:10.1007/JHEP04(2013)149
  [arXiv:1211.6092 [hep-th]].
  %%CITATION = doi:10.1007/JHEP04(2013)149;%%
  %74 citations counted in INSPIRE as of 06 Dec 2018


%\cite{Aguilar:2013xqa}
\bibitem{Aguilar:2013xqa} 
  A.~C.~Aguilar, D.~Ibáñez and J.~Papavassiliou,
  %``Ghost propagator and ghost-gluon vertex from Schwinger-Dyson equations,''
  Phys.\ Rev.\ D {\bf 87} (2013) no.11,  114020
  doi:10.1103/PhysRevD.87.114020
  [arXiv:1303.3609 [hep-ph]].
  %%CITATION = doi:10.1103/PhysRevD.87.114020;%%
  %45 citations counted in INSPIRE as of 06 Dec 2018

%\cite{Aguilar:2013vaa}
\bibitem{Aguilar:2013vaa}  
  A.~C.~Aguilar, D.~Binosi, D.~Ibañez and J.~Papavassiliou,
  %``Effects of divergent ghost loops on the Green’s functions of QCD,''
  Phys.\ Rev.\ D {\bf 89} (2014) no.8,  085008
  doi:10.1103/PhysRevD.89.085008
  [arXiv:1312.1212 [hep-ph]].
  %%CITATION = doi:10.1103/PhysRevD.89.085008;%%
  %60 citations counted in INSPIRE as of 02 Jan 2019

%\cite{Pelaez:2013cpa}
\bibitem{Pelaez:2013cpa}                               
  M.~Pelaez, M.~Tissier and N.~Wschebor,
  %``Three-point correlation functions in Yang-Mills theory,''
  Phys.\ Rev.\ D {\bf 88} (2013) 125003
  doi:10.1103/PhysRevD.88.125003
  [arXiv:1310.2594 [hep-th]].
  %%CITATION = doi:10.1103/PhysRevD.88.125003;%%
  %73 citations counted in INSPIRE as of 06 Dec 2018

%\cite{Blum:2014gna}
\bibitem{Blum:2014gna}
  A.~Blum, M.~Q.~Huber, M.~Mitter and L.~von Smekal,
  %``Gluonic three-point correlations in pure Landau gauge QCD,''
  Phys.\ Rev.\ D {\bf 89} (2014) 061703
  doi:10.1103/PhysRevD.89.061703
  [arXiv:1401.0713 [hep-ph]].
  %%CITATION = doi:10.1103/PhysRevD.89.061703;%%
  %93 citations counted in INSPIRE as of 06 Dec 2018

%\cite{Eichmann:2014xya}
\bibitem{Eichmann:2014xya}
  G.~Eichmann, R.~Williams, R.~Alkofer and M.~Vujinovic,
  %``Three-gluon vertex in Landau gauge,''
  Phys.\ Rev.\ D {\bf 89} (2014) no.10,  105014
  doi:10.1103/PhysRevD.89.105014
  [arXiv:1402.1365 [hep-ph]].
  %%CITATION = doi:10.1103/PhysRevD.89.105014;%%
  %70 citations counted in INSPIRE as of 20 Feb 2018

%\cite{Cyrol:2014kca}
\bibitem{Cyrol:2014kca}
  A.~K.~Cyrol, M.~Q.~Huber and L.~von Smekal,
  %``A Dyson–Schwinger study of the four-gluon vertex,''
  Eur.\ Phys.\ J.\ C {\bf 75} (2015) 102
  doi:10.1140/epjc/s10052-015-3312-1
  [arXiv:1408.5409 [hep-ph]].
  %%CITATION = doi:10.1140/epjc/s10052-015-3312-1;%%
  %45 citations counted in INSPIRE as of 06 Dec 2018


%\cite{Binosi:2014kka}
\bibitem{Binosi:2014kka}
  D.~Binosi, D.~Ibañez and J.~Papavassiliou,
  %``Nonperturbative study of the four gluon vertex,''
  JHEP {\bf 1409} (2014) 059
  doi:10.1007/JHEP09(2014)059
  [arXiv:1407.3677 [hep-ph]].
  %%CITATION = doi:10.1007/JHEP09(2014)059;%%
  %31 citations counted in INSPIRE as of 06 Dec 2018


%\cite{Mitter:2014wpa}
\bibitem{Mitter:2014wpa}
  M.~Mitter, J.~M.~Pawlowski and N.~Strodthoff,
  %``Chiral symmetry breaking in continuum QCD,''
  Phys.\ Rev.\ D {\bf 91} (2015) 054035
  doi:10.1103/PhysRevD.91.054035
  [arXiv:1411.7978 [hep-ph]].
  %%CITATION = doi:10.1103/PhysRevD.91.054035;%%
  %79 citations counted in INSPIRE as of 10 May 2018


%\cite{Aguilar:2014lha}
\bibitem{Aguilar:2014lha}
  A.~C.~Aguilar, D.~Binosi, D.~Ibañez and J.~Papavassiliou,
  %``New method for determining the quark-gluon vertex,''
  Phys.\ Rev.\ D {\bf 90} (2014) no.6,  065027
  doi:10.1103/PhysRevD.90.065027
  [arXiv:1405.3506 [hep-ph]].
  %%CITATION = doi:10.1103/PhysRevD.90.065027;%%
  %54 citations counted in INSPIRE as of 06 Dec 2018 
 

%\cite{Pelaez:2015tba}
\bibitem{Pelaez:2015tba}
  M.~Peláez, M.~Tissier and N.~Wschebor,
  %``Quark-gluon vertex from the Landau gauge Curci-Ferrari model,''
  Phys.\ Rev.\ D {\bf 92} (2015) no.4,  045012
  doi:10.1103/PhysRevD.92.045012
  [arXiv:1504.05157 [hep-th]].
  %%CITATION = doi:10.1103/PhysRevD.92.045012;%%
  %14 citations counted in INSPIRE as of 06 Dec 2018


%\cite{Binosi:2016wcx}
\bibitem{Binosi:2016wcx}
  D.~Binosi, L.~Chang, J.~Papavassiliou, S.~X.~Qin and C.~D.~Roberts,
  %``Natural constraints on the gluon-quark vertex,''
  Phys.\ Rev.\ D {\bf 95} (2017) no.3,  031501
  doi:10.1103/PhysRevD.95.031501
  [arXiv:1609.02568 [nucl-th]].
  %%CITATION = doi:10.1103/PhysRevD.95.031501;%%
  %32 citations counted in INSPIRE as of 06 Dec 2018

%\cite{Cyrol:2016tym}
\bibitem{Cyrol:2016tym}
  A.~K.~Cyrol, L.~Fister, M.~Mitter, J.~M.~Pawlowski and N.~Strodthoff,
  %``Landau gauge Yang-Mills correlation functions,''
  Phys.\ Rev.\ D {\bf 94} (2016) no.5,  054005
  doi:10.1103/PhysRevD.94.054005
  [arXiv:1605.01856 [hep-ph]].
  %%CITATION = doi:10.1103/PhysRevD.94.054005;%%
  %57 citations counted in INSPIRE as of 06 Dec 2018
 

%\cite{Aguilar:2016lbe}
\bibitem{Aguilar:2016lbe}
  A.~C.~Aguilar, J.~C.~Cardona, M.~N.~Ferreira and J.~Papavassiliou,
  %``Non-Abelian Ball-Chiu vertex for arbitrary Euclidean momenta,''
  Phys.\ Rev.\ D {\bf 96} (2017) no.1,  014029
  doi:10.1103/PhysRevD.96.014029
  [arXiv:1610.06158 [hep-ph]].
  %%CITATION = doi:10.1103/PhysRevD.96.014029;%%
  %16 citations counted in INSPIRE as of 06 Dec 2018

%\cite{Cyrol:2017ewj}
\bibitem{Cyrol:2017ewj}
  A.~K.~Cyrol, M.~Mitter, J.~M.~Pawlowski and N.~Strodthoff,
  %``Nonperturbative quark, gluon, and meson correlators of unquenched QCD,''
  Phys.\ Rev.\ D {\bf 97} (2018) no.5,  054006
  doi:10.1103/PhysRevD.97.054006
  [arXiv:1706.06326 [hep-ph]].
  %%CITATION = doi:10.1103/PhysRevD.97.054006;%%
  %45 citations counted in INSPIRE as of 02 Jan 2019


%\cite{Huber:2017txg}
\bibitem{Huber:2017txg}
  M.~Q.~Huber,
  %``On non-primitively divergent vertices of Yang–Mills theory,''
  Eur.\ Phys.\ J.\ C {\bf 77} (2017) no.11,  733
  doi:10.1140/epjc/s10052-017-5310-y
  [arXiv:1709.05848 [hep-ph]].
  %%CITATION = doi:10.1140/epjc/s10052-017-5310-y;%%    
  %4 citations counted in INSPIRE as of 03 Dec 2018

%\cite{Oliveira:2018fkj}
\bibitem{Oliveira:2018fkj}
  O.~Oliveira, T.~Frederico, W.~de Paula and J.~P.~B.~C.~de Melo,
  %``Exploring the Quark-Gluon Vertex with Slavnov-Taylor Identities and Lattice Simulations,''
  Eur.\ Phys.\ J.\ C {\bf 78} (2018) no.7,  553
  doi:10.1140/epjc/s10052-018-6037-0
  [arXiv:1807.00675 [hep-ph]].
  %%CITATION = doi:10.1140/epjc/s10052-018-6037-0;%%
  %5 citations counted in INSPIRE as of 06 Dec 2018

%\cite{Corell:2018yil}
\bibitem{Corell:2018yil}
  L.~Corell, A.~K.~Cyrol, M.~Mitter, J.~M.~Pawlowski and N.~Strodthoff,
  %``Correlation functions of three-dimensional Yang-Mills theory from the FRG,''
  SciPost Phys.\  {\bf 5} (2018) 066
  doi:10.21468/SciPostPhys.5.6.066
  [arXiv:1803.10092 [hep-ph]].
  %%CITATION = doi:10.21468/SciPostPhys.5.6.066;%%
  %7 citations counted in INSPIRE as of 16 Apr 2019

%\cite{Aguilar:2018csq}
\bibitem{Aguilar:2018csq}
  A.~C.~Aguilar, M.~N.~Ferreira, C.~T.~Figueiredo and J.~Papavassiliou,
  %``Nonperturbative structure of the ghost-gluon kernel,''
  Phys.\ Rev.\ D {\bf 99} (2019) no.3,  034026
  doi:10.1103/PhysRevD.99.034026
  [arXiv:1811.08961 [hep-ph]].
  %%CITATION = doi:10.1103/PhysRevD.99.034026;%%
  %6 citations counted in INSPIRE as of 16 Apr 2019

%\cite{Parrinello:1994wd}
\bibitem{Parrinello:1994wd}
  C.~Parrinello,
  %``Exploratory study of the three gluon vertex on the lattice,''
  Phys.\ Rev.\ D {\bf 50} (1994) R4247
  doi:10.1103/PhysRevD.50.R4247
  [hep-lat/9405024].
  %%CITATION = doi:10.1103/PhysRevD.50.R4247;%%
  %30 citations counted in INSPIRE as of 08 May 2018

%\cite{Alles:1996ka}
\bibitem{Alles:1996ka}
  B.~Alles, D.~Henty, H.~Panagopoulos, C.~Parrinello, C.~Pittori and D.~G.~Richards,
  %%``$\alpha_s$ from the nonperturbatively renormalised lattice three gluon vertex,''
  Nucl.\ Phys.\ B {\bf 502} (1997) 325
  doi:10.1016/S0550-3213(97)00483-5
  [hep-lat/9605033].
  %%%%CITATION = doi:10.1016/S0550-3213(97)00483-5;%%%%
  %%101 citations counted in INSPIRE as of 06 Feb 2018

%\cite{Boucaud:1998bq}                                                         
\bibitem{Boucaud:1998bq}                                          
  P.~Boucaud, J.~P.~Leroy, J.~Micheli, O.~Pene and C.~Roiesnel,      
  %``Lattice calculation of alpha(s) in momentum scheme,''             
  JHEP {\bf 9810} (1998) 017
  doi:10.1088/1126-6708/1998/10/017
  [hep-ph/9810322].
  %%CITATION = doi:10.1088/1126-6708/1998/10/017;%%
  %91 citations counted in INSPIRE as of 19 Feb 2018

%\cite{Skullerud:2002ge}
\bibitem{Skullerud:2002ge}
  J.~Skullerud and A.~Kizilersu,
  %``Quark gluon vertex from lattice QCD,''
  JHEP {\bf 0209} (2002) 013
  doi:10.1088/1126-6708/2002/09/013
  [hep-ph/0205318].
  %%CITATION = doi:10.1088/1126-6708/2002/09/013;%%
  %73 citations counted in INSPIRE as of 23 Feb 2018

%\cite{Skullerud:2003qu}
\bibitem{Skullerud:2003qu}
  J.~I.~Skullerud, P.~O.~Bowman, A.~Kizilersu, D.~B.~Leinweber and A.~G.~Williams,
  %``Nonperturbative structure of the quark gluon vertex,''
  JHEP {\bf 0304} (2003) 047
  doi:10.1088/1126-6708/2003/04/047
  [hep-ph/0303176].
  %%CITATION = doi:10.1088/1126-6708/2003/04/047;%%
  %105 citations counted in INSPIRE as of 23 Feb 2018

%\cite{Cucchieri:2004sq}                                     
\bibitem{Cucchieri:2004sq}
  A.~Cucchieri, T.~Mendes and A.~Mihara,
  %``Numerical study of the ghost-gluon vertex in Landau gauge,''
  JHEP {\bf 0412} (2004) 012
  doi:10.1088/1126-6708/2004/12/012
  [hep-lat/0408034].
  %%CITATION = doi:10.1088/1126-6708/2004/12/012;%%
  %114 citations counted in INSPIRE as of 23 Feb 2018

%\cite{Cucchieri:2006tf}
\bibitem{Cucchieri:2006tf}  
  A.~Cucchieri, A.~Maas and T.~Mendes,
  %``Exploratory study of three-point Green's functions in Landau-gauge Yang-Mills theory,''
  Phys.\ Rev.\ D {\bf 74} (2006) 014503
  doi:10.1103/PhysRevD.74.014503
  [hep-lat/0605011].
  %%CITATION = doi:10.1103/PhysRevD.74.014503;%%
  %90 citations counted in INSPIRE as of 23 Feb 2018

%\cite{Ilgenfritz:2006he}
\bibitem{Ilgenfritz:2006he}
  E.-M.~Ilgenfritz, M.~Muller-Preussker, A.~Sternbeck, A.~Schiller and I.~L.~Bogolubsky,
  %``Landau gauge gluon and ghost propagators from lattice QCD,''
  Braz.\ J.\ Phys.\  {\bf 37} (2007) 193
  doi:10.1590/S0103-97332007000200006
  [hep-lat/0609043].
  %%CITATION = doi:10.1590/S0103-97332007000200006;%%
  %80 citations counted in INSPIRE as of 06 Dec 2018

%\cite{Maas:2007uv}
\bibitem{Maas:2007uv}          
  A.~Maas,
  %``Two and three-point Green's functions in two-dimensional Landau-gauge Yang-Mills theory,''
  Phys.\ Rev.\ D {\bf 75} (2007) 116004
  doi:10.1103/PhysRevD.75.116004
  [arXiv:0704.0722 [hep-lat]].
  %%CITATION = doi:10.1103/PhysRevD.75.116004;%%
  %113 citations counted in INSPIRE as of 07 May 2018

%\cite{Cucchieri:2008qm}
\bibitem{Cucchieri:2008qm}                    
  A.~Cucchieri, A.~Maas and T.~Mendes,
  %``Three-point vertices in Landau-gauge Yang-Mills theory,''   
  Phys.\ Rev.\ D {\bf 77} (2008) 094510
  doi:10.1103/PhysRevD.77.094510
  [arXiv:0803.1798 [hep-lat]].
  %%CITATION = doi:10.1103/PhysRevD.77.094510;%%
  %112 citations counted in INSPIRE as of 23 Feb 2018

%\cite{Maas:2011se}
\bibitem{Maas:2011se}
  A.~Maas,
  %``Describing gauge bosons at zero and finite temperature,''
  Phys.\ Rept.\  {\bf 524} (2013) 203
  doi:10.1016/j.physrep.2012.11.002
  [arXiv:1106.3942 [hep-ph]].
  %%CITATION = doi:10.1016/j.physrep.2012.11.002;%%
  %141 citations counted in INSPIRE as of 07 May 2018

%\cite{Sternbeck:2012mf}
\bibitem{Sternbeck:2012mf}
  A.~Sternbeck and M.~Müller-Preussker,
  %``Lattice evidence for the family of decoupling solutions of Landau gauge Yang-Mills theory,''
  Phys.\ Lett.\ B {\bf 726} (2013) 396
  doi:10.1016/j.physletb.2013.08.017
  [arXiv:1211.3057 [hep-lat]].
  %%CITATION = doi:10.1016/j.physletb.2013.08.017;%%
  %53 citations counted in INSPIRE as of 16 Apr 2019


%\cite{Boucaud:2013jwa}
\bibitem{Boucaud:2013jwa}
  P.~Boucaud, M.~Brinet, F.~De Soto, V.~Morenas, O.~Pène, K.~Petrov and J.~Rodriguez-Quintero,
  %``Three-gluon running coupling from lattice QCD at $N_f=2+1+1$: a consistency check of the OPE approach,''
  JHEP {\bf 1404} (2014) 086
  doi:10.1007/JHEP04(2014)086
  [arXiv:1310.4087 [hep-ph]].
  %%CITATION = doi:10.1007/JHEP04(2014)086;%%
  %8 citations counted in INSPIRE as of 23 Feb 2018


%\cite{Maas:2013aia}
\bibitem{Maas:2013aia}
  A.~Maas and T.~Mufti,
  %``Two- and three-point functions in Landau gauge Yang-Mills-Higgs theory,''
  JHEP {\bf 1404} (2014) 006
  doi:10.1007/JHEP04(2014)006
  [arXiv:1312.4873 [hep-lat]].
  %%CITATION = doi:10.1007/JHEP04(2014)006;%%
  %41 citations counted in INSPIRE as of 16 Apr 2019

%\cite{Duarte:2016jhj}
\bibitem{Duarte:2016jhj}
  A.~G.~Duarte, O.~Oliveira and P.~J.~Silva,
  %``Landau gauge gluon vertices from Lattice QCD,''
  PoS LATTICE {\bf 2016} (2016) 351
  doi:10.22323/1.256.0351
  [arXiv:1610.10096 [hep-lat]].
  %%CITATION = doi:10.22323/1.256.0351;%%

%\cite{Athenodorou:2016oyh}   
\bibitem{Athenodorou:2016oyh}
  A.~Athenodorou, D.~Binosi, P.~Boucaud, F.~De Soto, J.~Papavassiliou, J.~Rodriguez-Quintero and S.~Zafeiropoulos,
  %``On the zero crossing of the three-gluon vertex,''
  Phys.\ Lett.\ B {\bf 761} (2016) 444
  doi:10.1016/j.physletb.2016.08.065
  [arXiv:1607.01278 [hep-ph]].
  %%CITATION = doi:10.1016/j.physletb.2016.08.065;%%
  %23 citations counted in INSPIRE as of 23 Feb 2018


%\cite{Sternbeck:2016ltn}
\bibitem{Sternbeck:2016ltn}
  A.~Sternbeck,
  %``QCD propagators and vertices from lattice QCD (in memory of Michael Müller-Preußker),''
  EPJ Web Conf.\  {\bf 137} (2017) 01020
  doi:10.1051/epjconf/201713701020
  [arXiv:1612.06106 [hep-lat]].
  %%CITATION = doi:10.1051/epjconf/201713701020;%%
  %1 citations counted in INSPIRE as of 06 Dec 2018


%\cite{Boucaud:2017obn}            
\bibitem{Boucaud:2017obn}
  P.~Boucaud, F.~De Soto, J.~Rodriguez-Quintero and S.~Zafeiropoulos,
  %``Refining the detection of the zero crossing for the three-gluon vertex in symmetric and asymmetric momentum subtraction schemes,''
  Phys.\ Rev.\ D {\bf 95} (2017) no.11,  114503
  doi:10.1103/PhysRevD.95.114503
  [arXiv:1701.07390 [hep-lat]].
  %%CITATION = doi:10.1103/PhysRevD.95.114503;%%
  %13 citations counted in INSPIRE as of 23 Feb 2018

%\cite{Sternbeck:2017ntv}
\bibitem{Sternbeck:2017ntv}
  A.~Sternbeck, P.~H.~Balduf, A.~Kizilersu, O.~Oliveira, P.~J.~Silva, J.~I.~Skullerud and A.~G.~Williams,
  %``Triple-gluon and quark-gluon vertex from lattice QCD in Landau gauge,''
  PoS LATTICE {\bf 2016} (2017) 349
  doi:10.22323/1.256.0349
  [arXiv:1702.00612 [hep-lat]].
  %%CITATION = doi:10.22323/1.256.0349;%%
  %4 citations counted in INSPIRE as of 22 May 2018

%\cite{Vujinovic:2018nqc}
\bibitem{Vujinovic:2018nqc}
  M.~Vujinovic and T.~Mendes,
  %``Probing the tensor structure of lattice three-gluon vertex in Landau gauge,''
  Phys.\ Rev.\ D {\bf 99} (2019) no.3,  034501
  doi:10.1103/PhysRevD.99.034501
  [arXiv:1807.03673 [hep-lat]].
  %%CITATION = doi:10.1103/PhysRevD.99.034501;%%
  %1 citations counted in INSPIRE as of 26 Feb 2019

%\cite{Maas:2018ska}
\bibitem{Maas:2018ska}                      
  A.~Maas, S.~Raubitzek and P.~Törek,
  %``Exploratory study of the off-shell properties of the weak vector bosons,''
  arXiv:1811.03395 [hep-lat].
  %%CITATION = ARXIV:1811.03395;%%
  %2 citations counted in INSPIRE as of 16 Apr 2019

%\cite{Maas:2019tnm}
\bibitem{Maas:2019tnm}
  A.~Maas,
  %``The quenched SU(2) scalar-gluon vertex in minimal Landau gauge,''
  arXiv:1902.10568 [hep-lat].
  %%CITATION = ARXIV:1902.10568;%%


%\cite{Hassani:1999sny}
\bibitem{Hassani:1999sny}
 \textit{Mathematical Physics:\,A Modern Introduction to Its Foundations}, S.\,Hassani (Springer-Verlag New York, USA, 1999), 1003 p.

%\cite{Weyl:1939prc}
\bibitem{Weyl:1939prc}
 \textit{The classical groups:\,Their invariants and Representations}, H.\,Weyl (Princeton University Press, USA, 1939, 1946), 336 p.

%\cite{Morty:1962prc}
\bibitem{Morty:1962prc}
 \textit{Group Theory and Its Application to Physical Problems}, M.\,Hamermesh (Original print:~Addison-Wesley Publishing, USA, 1962,
Reprinted:~Dover Publications, USA, 1989), 509 p.

%\cite{Kawai:1980ja}
\bibitem{Kawai:1980ja}                        
  H.~Kawai, R.~Nakayama and K.~Seo,
  %``Comparison of the Lattice Lambda Parameter with the Continuum Lambda Parameter in Massless QCD,''
  Nucl.\ Phys.\ B {\bf 189} (1981) 40.
  doi:10.1016/0550-3213(81)90080-8
  %%CITATION = doi:10.1016/0550-3213(81)90080-8;%%
  %230 citations counted in INSPIRE as of 16 Apr 2019

%\cite{Weisz:1983bn}
\bibitem{Weisz:1983bn}
  P.~Weisz and R.~Wohlert,
  %%``Continuum Limit Improved Lattice Action for Pure Yang-Mills Theory. 2.,''
  Nucl.\ Phys.\ B {\bf 236} (1984) 397
   Erratum: [Nucl.\ Phys.\ B {\bf 247} (1984) 544].
  doi:10.1016/0550-3213(84)90563-7, 10.1016/0550-3213(84)90543-1
  %%%%CITATION = doi:10.1016/0550-3213(84)90563-7, 10.1016/0550-3213(84)90543-1;%%%%
  %%182 citations counted in INSPIRE as of 22 Jan 2018

%\cite{Aubin:2015rzx}
\bibitem{Aubin:2015rzx}
  C.~Aubin, T.~Blum, P.~Chau, M.~Golterman, S.~Peris and C.~Tu,
  %``Finite-volume effects in the muon anomalous magnetic moment on the lattice,''
  Phys.\ Rev.\ D {\bf 93} (2016) no.5,  054508
  doi:10.1103/PhysRevD.93.054508
  [arXiv:1512.07555 [hep-lat]].
  %%CITATION = doi:10.1103/PhysRevD.93.054508;%%
  %37 citations counted in INSPIRE as of 16 Apr 2019

%\cite{Wilson:1974sk}
\bibitem{Wilson:1974sk}
  K.~G.~Wilson,
  %``Confinement of Quarks,''
  Phys.\ Rev.\ D {\bf 10} (1974) 2445.
  doi:10.1103/PhysRevD.10.2445
  %%CITATION = doi:10.1103/PhysRevD.10.2445;%%
  %4801 citations counted in INSPIRE as of 22 Feb 2019

%\cite{Weisz:1982zw}
\bibitem{Weisz:1982zw} 
  P.~Weisz,
  %%``Continuum Limit Improved Lattice Action for Pure Yang-Mills Theory. 1.,''
  Nucl.\ Phys.\ B {\bf 212} (1983) 1.
  doi:10.1016/0550-3213(83)90595-3
  %%%%CITATION = doi:10.1016/0550-3213(83)90595-3;%%%%
  %%361 citations counted in INSPIRE as of 22 Jan 2018

%\cite{Symanzik:1983gh}
\bibitem{Symanzik:1983gh}
  K.~Symanzik,
  %%``Continuum Limit and Improved Action in Lattice Theories. 2. O(N) Nonlinear Sigma Model in Perturbation Theory,''
  Nucl.\ Phys.\ B {\bf 226} (1983) 205.
  doi:10.1016/0550-3213(83)90469-8
  %%%%CITATION = doi:10.1016/0550-3213(83)90469-8;%%%%
  %%493 citations counted in INSPIRE as of 22 Jan 2018


%\cite{Becirevic:1999uc}
\bibitem{Becirevic:1999uc}         
  D.~Becirevic, P.~Boucaud, J.~P.~Leroy, J.~Micheli, O.~Pene, J.~Rodriguez-Quintero and C.~Roiesnel,
  %``Asymptotic behavior of the gluon propagator from lattice QCD,''
  Phys.\ Rev.\ D {\bf 60} (1999) 094509
  doi:10.1103/PhysRevD.60.094509
  [hep-ph/9903364].
  %%CITATION = doi:10.1103/PhysRevD.60.094509;%%
  %86 citations counted in INSPIRE as of 22 Jan 2019

%\cite{deSoto:2007ht}
\bibitem{deSoto:2007ht}
  F.~de Soto and C.~Roiesnel,
  %``On the reduction of hypercubic lattice artifacts,''
  JHEP {\bf 0709} (2007) 007
  doi:10.1088/1126-6708/2007/09/007
  [arXiv:0705.3523 [hep-lat]].
  %%CITATION = doi:10.1088/1126-6708/2007/09/007;%%
  %56 citations counted in INSPIRE as of 22 Jan 2019

%\cite{Becirevic:1999hj}
\bibitem{Becirevic:1999hj}
  D.~Becirevic, P.~Boucaud, J.~P.~Leroy, J.~Micheli, O.~Pene, J.~Rodriguez-Quintero and C.~Roiesnel,
  %``Asymptotic scaling of the gluon propagator on the lattice,''
  Phys.\ Rev.\ D {\bf 61} (2000) 114508
  doi:10.1103/PhysRevD.61.114508
  [hep-ph/9910204].
  %%CITATION = doi:10.1103/PhysRevD.61.114508;%%
  %82 citations counted in INSPIRE as of 22 Jan 2019

%\cite{Blossier:2014kta}
\bibitem{Blossier:2014kta}
  B.~Blossier {\it et al.} [ETM Collaboration],
  %``Renormalization of quark propagator, vertex functions, and twist-2 operators from twisted-mass lattice QCD at $N_f$=4,''
  Phys.\ Rev.\ D {\bf 91} (2015) no.11,  114507
  doi:10.1103/PhysRevD.91.114507
  [arXiv:1411.1109 [hep-lat]].
  %%CITATION = doi:10.1103/PhysRevD.91.114507;%%
  %11 citations counted in INSPIRE as of 22 Jan 2019

%\cite{Boucaud:2018xup}
\bibitem{Boucaud:2018xup}
  P.~Boucaud, F.~De Soto, K.~Raya, J.~Rodríguez-Quintero and S.~Zafeiropoulos,
  %``Discretization effects on renormalized gauge-field Green’s functions, scale setting, and the gluon mass,''
  Phys.\ Rev.\ D {\bf 98} (2018) no.11,  114515
  doi:10.1103/PhysRevD.98.114515
  [arXiv:1809.05776 [hep-ph]].
  %%CITATION = doi:10.1103/PhysRevD.98.114515;%%
  %2 citations counted in INSPIRE as of 22 Jan 2019

%\cite{Rothe:1992nt}
\bibitem{Rothe:1992nt}
  H.~J.~Rothe,
  %``Lattice gauge theories: An Introduction,''
  World Sci.\ Lect.\ Notes Phys.\  {\bf 43} (1992) 1
   [World Sci.\ Lect.\ Notes Phys.\  {\bf 59} (1997) 1]
   [World Sci.\ Lect.\ Notes Phys.\  {\bf 74} (2005) 1]
   [World Sci.\ Lect.\ Notes Phys.\  {\bf 82} (2012) 1].
  %%CITATION = 00327,43,1;%%
  %135 citations counted in INSPIRE as of 24 Jan 2019

%\cite{Capitani:2002mp}
\bibitem{Capitani:2002mp}
  S.~Capitani,
  %``Lattice perturbation theory,''
  Phys.\ Rept.\  {\bf 382} (2003) 113
  doi:10.1016/S0370-1573(03)00211-4
  [hep-lat/0211036].
  %%CITATION = doi:10.1016/S0370-1573(03)00211-4;%%
  %108 citations counted in INSPIRE as of 24 Jan 2019

%\cite{Alkofer:2004it}
\bibitem{Alkofer:2004it}
  R.~Alkofer, C.~S.~Fischer and F.~J.~Llanes-Estrada,
  %``Vertex functions and infrared fixed point in Landau gauge SU(N) Yang-Mills theory,''
  Phys.\ Lett.\ B {\bf 611} (2005) 279
   Erratum: [Phys.\ Lett.\ B {\bf 670} (2009) 460]
  doi:10.1016/j.physletb.2008.11.068, 10.1016/j.physletb.2005.02.043
  [hep-th/0412330].
  %%CITATION = doi:10.1016/j.physletb.2008.11.068, 10.1016/j.physletb.2005.02.043;%%
  %198 citations counted in INSPIRE as of 17 Apr 2019


%\cite{Taylor:1971ff}
\bibitem{Taylor:1971ff}
  J.~C.~Taylor,
  %``Ward Identities and Charge Renormalization of the Yang-Mills Field,''
  Nucl.\ Phys.\ B {\bf 33} (1971) 436.
  doi:10.1016/0550-3213(71)90297-5
  %%CITATION = doi:10.1016/0550-3213(71)90297-5;%%
  %808 citations counted in INSPIRE as of 25 Jan 2019

%\cite{Ball:1980ax}
\bibitem{Ball:1980ax}
  J.~S.~Ball and T.~W.~Chiu,
  %``Analytic Properties of the Vertex Function in Gauge Theories. 2.,''
  Phys.\ Rev.\ D {\bf 22} (1980) 2550
   Erratum: [Phys.\ Rev.\ D {\bf 23} (1981) 3085].
  doi:10.1103/physrevd.23.3085.2, 10.1103/PhysRevD.22.2550
  %%CITATION = doi:10.1103/physrevd.23.3085.2, 10.1103/PhysRevD.22.2550;%%
  %182 citations counted in INSPIRE as of 29 Jan 2019

%\cite{Fischer:2018sdj}
\bibitem{Fischer:2018sdj}
  C.~S.~Fischer,
  %``QCD at finite temperature and chemical potential from Dyson-Schwinger equations,''
  arXiv:1810.12938 [hep-ph].
  %%CITATION = ARXIV:1810.12938;%%
  %3 citations counted in INSPIRE as of 30 Jan 2019


%\cite{Adler:1981sn}
\bibitem{Adler:1981sn}
  S.~L.~Adler,
  %``An Overrelaxation Method for the Monte Carlo Evaluation of the Partition Function for Multiquadratic Actions,''
  Phys.\ Rev.\ D {\bf 23} (1981) 2901.
  doi:10.1103/PhysRevD.23.2901
  %%CITATION = doi:10.1103/PhysRevD.23.2901;%%
  %136 citations counted in INSPIRE as of 23 Feb 2019

%\cite{Adler:1987ce}
\bibitem{Adler:1987ce}
  S.~L.~Adler,
  %``Overrelaxation Algorithms for Lattice Field Theories,''
  Phys.\ Rev.\ D {\bf 37} (1988) 458.
  doi:10.1103/PhysRevD.37.458
  %%CITATION = doi:10.1103/PhysRevD.37.458;%%
  %65 citations counted in INSPIRE as of 23 Feb 2019

%\cite{Kennedy:1985nu}
\bibitem{Kennedy:1985nu}
  A.~D.~Kennedy and B.~J.~Pendleton,
  %``Improved Heat Bath Method for Monte Carlo Calculations in Lattice Gauge Theories,''
  Phys.\ Lett.\  {\bf 156B} (1985) 393.
  doi:10.1016/0370-2693(85)91632-6
  %%CITATION = doi:10.1016/0370-2693(85)91632-6;%%
  %231 citations counted in INSPIRE as of 23 Feb 2019

%\cite{Wolff:2003sm}
\bibitem{Wolff:2003sm}
  U.~Wolff [ALPHA Collaboration],
  %%``Monte Carlo errors with less errors,''
  Comput.\ Phys.\ Commun.\  {\bf 156} (2004) 143
   Erratum: [Comput.\ Phys.\ Commun.\  {\bf 176} (2007) 383]
  doi:10.1016/S0010-4655(03)00467-3, 10.1016/j.cpc.2006.12.001
  [hep-lat/0306017].
  %%%%CITATION = doi:10.1016/S0010-4655(03)00467-3, 10.1016/j.cpc.2006.12.001;%%%%
  %%178 citations counted in INSPIRE as of 24 Jan 2018

%\cite{Mandula:1990vs}
\bibitem{Mandula:1990vs}
  J.~E.~Mandula and M.~Ogilvie,
  %``Efficient gauge fixing via overrelaxation,''
  Phys.\ Lett.\ B {\bf 248} (1990) 156.
  doi:10.1016/0370-2693(90)90031-Z
  %%CITATION = doi:10.1016/0370-2693(90)90031-Z;%%
  %97 citations counted in INSPIRE as of 25 Feb 2019

%\cite{Cucchieri:1995pn}
\bibitem{Cucchieri:1995pn}
  A.~Cucchieri and T.~Mendes,
  %``Critical slowing down in SU(2) Landau gauge fixing algorithms,''
  Nucl.\ Phys.\ B {\bf 471} (1996) 263
  doi:10.1016/0550-3213(96)00177-0
  [hep-lat/9511020].
  %%CITATION = doi:10.1016/0550-3213(96)00177-0;%%
  %65 citations counted in INSPIRE as of 28 May 2018

%\cite{Mandula:1987rh}
\bibitem{Mandula:1987rh}
  J.~E.~Mandula and M.~Ogilvie,
  %``The Gluon Is Massive: A Lattice Calculation of the Gluon Propagator in the Landau Gauge,''
  Phys.\ Lett.\ B {\bf 185} (1987) 127.
  doi:10.1016/0370-2693(87)91541-3
  %%CITATION = doi:10.1016/0370-2693(87)91541-3;%%
  %269 citations counted in INSPIRE as of 25 Feb 2019

%\cite{Bonnet:1999bw}
\bibitem{Bonnet:1999bw}
  F.~D.~R.~Bonnet, P.~O.~Bowman, D.~B.~Leinweber, D.~G.~Richards and A.~G.~Williams,
  %``Improved Landau gauge fixing and discretization errors,''
  Nucl.\ Phys.\ Proc.\ Suppl.\  {\bf 83} (2000) 905
  doi:10.1016/S0920-5632(00)91841-3
  [hep-lat/9909110].
  %%CITATION = doi:10.1016/S0920-5632(00)91841-3;%%
  %7 citations counted in INSPIRE as of 20 Mar 2019


%\cite{Giusti:1996kf}
\bibitem{Giusti:1996kf}                                 
  L.~Giusti,
  %``Lattice gauge fixing for generic covariant gauges,''
  Nucl.\ Phys.\ B {\bf 498} (1997) 331
  doi:10.1016/S0550-3213(97)00273-3
  [hep-lat/9605032].
  %%CITATION = doi:10.1016/S0550-3213(97)00273-3;%%
  %39 citations counted in INSPIRE as of 12 Mar 2019

%\cite{Cucchieri:2009kk}
\bibitem{Cucchieri:2009kk}
  A.~Cucchieri, T.~Mendes and E.~M.~S.~Santos,
  %``Covariant gauge on the lattice: A New implementation,''
  Phys.\ Rev.\ Lett.\  {\bf 103} (2009) 141602
  doi:10.1103/PhysRevLett.103.141602
  [arXiv:0907.4138 [hep-lat]].
  %%CITATION = doi:10.1103/PhysRevLett.103.141602;%%
  %53 citations counted in INSPIRE as of 12 Mar 2019


%\cite{Bicudo:2015rma}
\bibitem{Bicudo:2015rma}
  P.~Bicudo, D.~Binosi, N.~Cardoso, O.~Oliveira and P.~J.~Silva,
  %``Lattice gluon propagator in renormalizable $\xi$ gauges,''
  Phys.\ Rev.\ D {\bf 92} (2015) no.11,  114514
  doi:10.1103/PhysRevD.92.114514
  [arXiv:1505.05897 [hep-lat]].
  %%CITATION = doi:10.1103/PhysRevD.92.114514;%%
  %43 citations counted in INSPIRE as of 19 Apr 2019

%\cite{Cucchieri:2018doy}
\bibitem{Cucchieri:2018doy}
  A.~Cucchieri, D.~Dudal, T.~Mendes, O.~Oliveira, M.~Roelfs and P.~J.~Silva,
  %``Faddeev-Popov Matrix in Linear Covariant Gauge: First Results,''
  Phys.\ Rev.\ D {\bf 98} (2018) no.9,  091504
  doi:10.1103/PhysRevD.98.091504
  [arXiv:1809.08224 [hep-lat]].
  %%CITATION = doi:10.1103/PhysRevD.98.091504;%%
  %5 citations counted in INSPIRE as of 19 Apr 2019

%\cite{Teper:1998te}
\bibitem{Teper:1998te}
  M.~J.~Teper,
  %``SU(N) gauge theories in (2+1)-dimensions,''
  Phys.\ Rev.\ D {\bf 59} (1999) 014512
  doi:10.1103/PhysRevD.59.014512
  [hep-lat/9804008].
  %%CITATION = doi:10.1103/PhysRevD.59.014512;%%
  %291 citations counted in INSPIRE as of 19 Apr 2019

%\cite{Golterman:2014ksa}
\bibitem{Golterman:2014ksa}                   
  M.~Golterman, K.~Maltman and S.~Peris,
  %``New strategy for the lattice evaluation of the leading order hadronic contribution to $(g−2)\mu$,''
  Phys.\ Rev.\ D {\bf 90} (2014) no.7,  074508
  doi:10.1103/PhysRevD.90.074508
  [arXiv:1405.2389 [hep-lat]].
  %%CITATION = doi:10.1103/PhysRevD.90.074508;%%
  %33 citations counted in INSPIRE as of 19 Apr 2019

%\cite{Shintani:2010ph}
\bibitem{Shintani:2010ph}
  E.~Shintani, S.~Aoki, H.~Fukaya, S.~Hashimoto, T.~Kaneko, T.~Onogi and N.~Yamada,
  %``Strong coupling constant from vacuum polarization functions in three-flavor lattice QCD with dynamical overlap fermions,''
  Phys.\ Rev.\ D {\bf 82} (2010) no.7,  074505
   Erratum: [Phys.\ Rev.\ D {\bf 89} (2014) no.9,  099903]
  doi:10.1103/PhysRevD.82.074505, 10.1103/PhysRevD.89.099903
  [arXiv:1002.0371 [hep-lat]].
  %%CITATION = doi:10.1103/PhysRevD.82.074505, 10.1103/PhysRevD.89.099903;%%
  %61 citations counted in INSPIRE as of 19 Apr 2019

%\cite{Cucchieri:1997dx}
\bibitem{Cucchieri:1997dx}
  A.~Cucchieri,
  %``Gribov copies in the minimal Landau gauge: The Influence on gluon and ghost propagators,''
  Nucl.\ Phys.\ B {\bf 508} (1997) 353
  doi:10.1016/S0550-3213(97)80016-8, 10.1016/S0550-3213(97)00629-9
  [hep-lat/9705005].
  %%CITATION = doi:10.1016/S0550-3213(97)80016-8, 10.1016/S0550-3213(97)00629-9;%%
  %155 citations counted in INSPIRE as of 20 Mar 2019

%\cite{Sternbeck:2005tk}
\bibitem{Sternbeck:2005tk}
  A.~Sternbeck, E.-M.~Ilgenfritz, M.~Muller-Preussker and A.~Schiller,
  %``Towards the infrared limit in SU(3) Landau gauge lattice gluodynamics,''
  Phys.\ Rev.\ D {\bf 72} (2005) 014507
  doi:10.1103/PhysRevD.72.014507
  [hep-lat/0506007].
  %%CITATION = doi:10.1103/PhysRevD.72.014507;%%
  %146 citations counted in INSPIRE as of 20 Mar 2019

%\cite{August:2013jia}
\bibitem{August:2013jia}
  D.~August and A.~Maas,
  %``On the Landau-gauge adjoint quark propagator,''
  JHEP {\bf 1307} (2013) 001
  doi:10.1007/JHEP07(2013)001
  [arXiv:1304.4423 [hep-lat]].
  %%CITATION = doi:10.1007/JHEP07(2013)001;%%
  %9 citations counted in INSPIRE as of 25 Apr 2019



\end{thebibliography}
\end{document}